%% file: tesisJPWD.tex
\documentclass[chap,11pt]{usthesis} 
\usepackage{epsfig,amssymb,multirow,amsmath, bm} 

\begin{document}

\include{titlepage}

\include{abstract}
\tableofcontents  
\listoftables       
\listoffigures
\include{acknowledgements}
\include{introduction}
\include{chapRel}
\include{chapTOV}
\include{chapNucmatter}
\include{chapRQM}
\include{chapMechanics}
\include{chapQFT}
\include{chapQHD}
\include{chapFSUG}
\include{chapNS}
\include{chapResults}
\include{chapConclu}

\appendix
\include{Appendix}

\include{biblio}
\end{document}

%% file: titlepage.tex
\thesistitle{Relativistic mean-field theory applied to the study of neutron star properties}
\author{Jacobus Petrus Willem Diener}         
\degree{Master~of~Science} 
\supervisor{Doctor B. I. S. van der Ventel} 
\cosupervisor{Professor G. C. Hillhouse}
\submitdate{March 2008}  
\declarationdate{11 February 2008}
\copyrightyear{2008}

\titlepage 
\declaration

%% file: abstract.tex
\specialhead{ABSTRACT}
Nuclear physics can be applied in various ways to the study of neutron stars. This thesis reports on one such application, where the relativistic mean-field approximation has been employed to calculate the equations of state of matter in the neutron star interior. In particular the equations of state of nuclear and neutron star matter of the NL3, PK1 and FSUGold parameter sets were derived. A survey of available literature on neutron stars is presented and we use the derived equations of state to reproduce the properties of saturated nuclear matter as well as the mass-radius relationship of a static, spherical symmetric neutron star. Results are compared to published values of the properties of saturated nuclear matter and to available observational data of the mass-radius relationship of neutron stars.
\newpage
\begin{center}
{\large {\bfseries OPSOMMING}}
\end{center}
Kernfisika kan op vele maniere aangewend word binne die studie van neutronsterre. 
Een so 'n toepassing is die gebruik van die relatiwistiese gemiddelde-veld benadering om die toestandsvergelyking van neutronstermaterie af te lei. 
Die afleiding van die toestandsvergelyking van kern- en neutronstermaterie vir onderskeidelik die NL3, PK1 en FSUGold parameter stelle vorm die basis van die tesis. 'n Oorsig oor die beskikbare literatuur aangaande neutronsterre word gebied en van di$\acute{\mbox{e}}$ gepubliseerde resultate word herbereken. In die besonder word die genoemde toe-standsvergelykings gebruik om die eienskappe van versadigde kernmaterie te bepaal, asook om die massa-radius verhoudings van statiese, sferies-simmetriese neutronsterre mee te bereken. Die resultate 
word vergelyk met gepubliseerde waardes vir die eienskappe van versadigde kernmaterie en die waargenome massa-radius verhoudings van neutron sterre.

%% file: acknowledgements.tex
\specialhead{ACKNOWLEDGEMENTS}
I would like to acknowledge the contribution of the following people and institutions:
\begin{itemize}
	\item My supervisor, Dr B. I. S. van der Ventel for his guidance.
	\item My co-supervisor, Prof. G. C. Hillhouse for always challenging me.
	\item My parents, brothers and friends for who they are and for the role they have played in making me who I am.
	\item The NRF, Harry Crossley Foundation and the University of Stellenbosch for their financial support.
	\item To my heavenly Father for making neutron stars and the other interesting stuff in nature.
\end{itemize}

%% file: introduction.tex
\chapter{Introduction}\label{intro}
\begin{figure}
	\centering
		\includegraphics[width=0.75\textwidth]{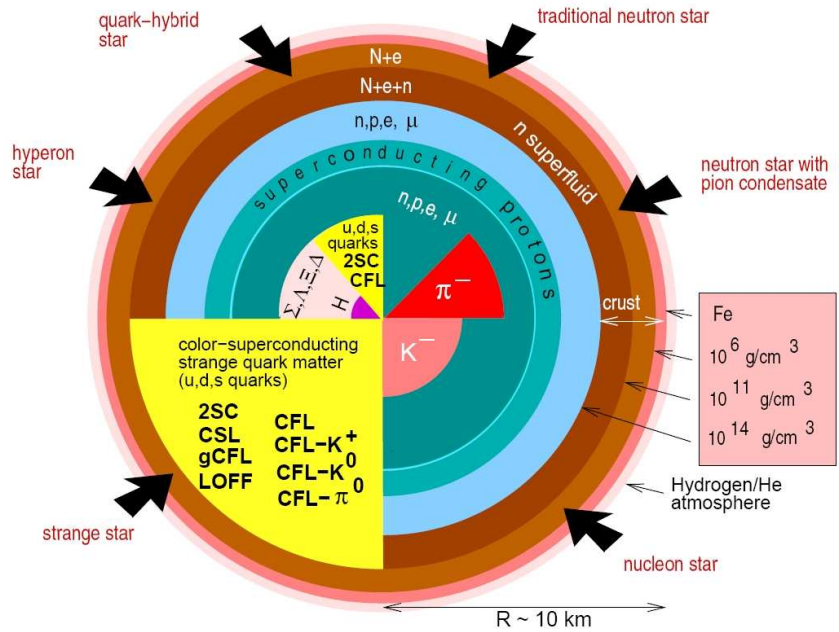}
	\caption{A representation of different ideas regarding the composition of the interior of neutron stars, from Ref. \cite{weberSQM}}
	\label{fig:weberNSinterior}
\end{figure}

Judging from current experimental activities, one of the pertinent questions that nuclear \mbox{physicists} are trying to answer is: What was the state of matter, at a time when the universe was very hot and dense? High-energy collisions of heavy-ions, such as the ones performed in the PHENIX experiment at the Relativistic Heavy-Ion Collider (RHIC) and planned at CERN in the ALICE experiment, aim to study such hot, dense nuclear matter \cite{PHENIX, ALICE}. More specifically the experiments at RHIC aim to study the state of matter, called \textsl{quark-gluon plasma}, which is believed to be the state of matter shortly after the Big Bang \cite{PHENIX}.\\
\\
Another question, that is currently investigated in earnest by nuclear physicists, is if the structure of normal nuclear matter (protons and neutrons) breaks down under conditions of low temperature and extreme pressure, i.e.\ whether the ground state of nuclear matter consists out of matter other than nuclear matter. To answer this question, 
nuclear physicists must look further afield than laboratory experiments. In laboratory experiments, where either two beams of high energy particles are collided or high energy particles are shot onto a target, a highly energetic, but short-lived state of matter is created, which is rather far removed from the ground state of matter \cite{Links}. 
A physical system is in its ground state when the system is in its lowest energy configuration. The ground state is usually achieved after the system had some time to equilibrate. $^{56}$Fe is the nucleus with the greatest (negative) binding energy and therefore has the lowest ground state of all the nuclei. Questions have been raised whether $^{56}$Fe is the ultimate ground state of matter since W. Baade and E. Zwicky proposed in 1934 that some supernovae are driven by the binding energy of a neutron star \cite{csg}, i.e. that a massive star explodes due to the release of energy in forming a neutron star. 
On average neutron stars have a mass of about 1.4 times the mass of our sun \cite{Lattimer07}. The gravitational binding energy of a neutron star is about 10$\%$ of its mass, while the nuclear binding energy of $^{56}$Fe is about 9 MeV/nucleon, which is about 1$\%$ of the mass of a $^{56}$Fe nucleus \cite{csg}. Thus the gravitational binding energy of a neutron star consisting of $^{56}$Fe would be about ten times larger than the nuclear binding energy. Since the star is in a stable long-lived state, one can assume that the matter is in a state other than that of $^{56}$Fe. As such nature has provided us with a laboratory to study cold, dense matter, which we currently cannot synthesise in the laboratory. (By cold is meant that the thermal energy of the particles is very small compared to the fermi energy and thus thermal excitations are assumed not to take place \cite{csg, webertxt}.)\\
\\
Neutron stars are believed to be formed in the core-collapse supernovae of massive stars \cite{csg, shapiro}.
A normal star is stabilised against gravitational collapse by the thermal pressure due to the energy release of nuclear fusion processes in the star. Fusion in the core of the star would proceed the fastest due to the higher pressure at the centre of the star. Once these fusion processes have reached the formation of $^{56}$Fe, the last exothermic fusion phases, it will shut down and the core will start to cool and contract under its own gravity \cite{csg}. As the core contracts densities where electrons become relativistic will quite easily be reached. The energy of the core can thus be reduced by the capture of electrons by the protons (inverse beta-decay), thus making the core more neutron-rich \cite{shapiro}. The crushing effect of gravity will be halted by the short-ranged repulsion of the strong nuclear force.\\ 
Matter falling onto the core, due to the low pressure created by the contracting core, would rebound off the stiffened core, creating a shockwave that travels outward from the centre of the star, which stalls after some hundred kilometres, due to energy losses as the wave travels through the interior of the star. This stalled shockwave creates an accreting front as matter from the outer parts of the star collapses towards 
the centre of the star. Through poorly understood mechanisms the binding energy of the neutron star (the core of the collapsing star) gets transferred to the accreting shock front, which causes the rest of the star to explode in a supernova. \cite{csg} \\
\\
In 1967 Jocelyn Bell observed a pulsating radio-source in outer space that had \mbox{characteristics} unlike any other radio-source \cite{hewish}. Initially the origin of these radio-pulses was unclear (little green men were not ruled out), but eventually the source was explained to be a rapidly rotating neutron star, today known as a pulsar \cite{NS1, NS2}. Today more than 1100 pulsars, the commonly observed state of neutron stars, are known \cite{csg}. \\
\\
As with the question regarding the ultimate ground state of matter, the constituents of the interior of neutron stars are also not known. As the average densities of neutron stars are comparable to that of nuclei, it is assumed that neutron stars consist (at least in some part) of baryonic matter (such as protons and neutrons) and therefore can be viewed as giant nuclei, but with a mass number of $10^{57}$ \cite{shapiro}! The main difference between nuclei and neutron stars is that while nuclei are bound by the nuclear strong force, neutron stars are bound by gravity.
The fact that gravity is attractive on all scales implies that the neutron star must have some form of internal pressure to counteract the effect of gravity, otherwise no stable neutron stars would exists, only black holes (objects that have collapsed under their own gravity). Baade and Zwicky envisioned that a neutron star is supported against gravitational collapse by the nucleon \mbox{degeneracy} pressure: the pressure that is due to Pauli's Exclusion Principle that states that \textsl{no two identical particles can occupy the same energy state.} Thus the pressure is given by particles that all want to occupy the lowest energy state, but these states are filled from the bottom and once they are full particles have to occupy higher lying states \cite{csg}. Due to the central role of \mbox{gravity,} neutron stars can only be adequately described using the general theory of relativity \cite{csg}. In Chapter \ref{chap:Rel} necessary concepts in relativity will be introduced to formulate such a description, which will be derived in Chapter \ref{chap:TOV}.\\
\\
But that would not be the full story. To be able to have such an adequate, relativistic description of neutron stars information regarding the relation between the pressure and the energy density (equation of state) of matter in the interior of the neutron star is needed. If we assume that the neutron star consists of baryonic matter, models of nuclear matter can be used to provide the equation of state of the neutron star interior \cite{csg, webertxt, shapiro}. 
Fig. \ref{fig:weberNSinterior} shows a schematic presentation of various assumptions about the interior of neutron stars (and as such for the ground state of matter).\\
\\
These different models for the interior of the neutron star can be tested by comparing calculated properties, specifically the mass-radius relationship fo a neutron star, to the observed values. By this process suitable models (and equations of state derived from these models) can be identified for the description of cold, dense matter. In general these different models of nuclei and nuclear matter are complex and difficult to solve exactly and therefore these models have to be \mbox{approximated.} The focus of this work will be to derive the equation of state of neutron star matter from a specific model of nuclei and nuclear matter, called quantum hadrodynamics (QHD), through the application of relativistic mean-field theory. This will be done by making a survey of some of the available literature on this subject as well as to try to reproduce some of the published values of certain properties of nuclear matter and neutron stars. 
The aim of this work is to consolidate some of the current knowledge in the theoretical study of neutron stars. Since these theoretical studies can only be validated through good agreement with observed properties of neutron stars, close co-operation between theorists and astronomers are crucial in the study of neutron stars. The converse is also true: to explain observational results theoretical modelling is needed to predict certain observed properties. Only through this interplay between theoretical and observational science can the understanding of our universe be advanced. This work therefore also aims to include some of the observational results to constrain the theoretical calculations. \\
\\
It is foreseen that much more will be learned about neutron stars with the commission of the Square Kilometre Array (SKA). The SKA will be the world's biggest radio-telescope, with an \mbox{effective} collecting area of one square kilometre, which will make the SKA fifty times more \mbox{powerful} than the current most powerful radio-telescope \cite{SKA1, SKA}. To explain the observations made by the SKA knowledge of the theoretical modelling of neutron stars would be crucial. 
\section{South African perspective}
This work is applicable within the South African context since South Africa is currently \mbox{developing} the 
meerKAT radio-telescope, formerly known as the Karoo Array Telescope (KAT), and has also been shortlisted to host SKA \cite{SKA}. It is foreseen that the meerKAT will be the largest radio-telescope in the world, until the SKA is commissioned, thus South Africa will have a strategic advance in, amongst others, the study of pulsars and neutron stars.\\
\\
KAT-7, a seven-dish engineering test bed, will be commissioned towards the end of 2009 and it is foreseen that the full fifty-dish array will be completed by 2012 \cite{SKA}.\\
\\
The development of the meerKAT could not only be of importance to radio-astronomers but also to nuclear physicists studying dense matter. This work can be seen as one example of how nuclear physics can be advanced by using data gathered from outer space.\\
\section{Conventions and notations}
In this thesis the following conventions and notations will be used.
\subsection{Vectors, fields and operators}
Three dimensional vectors will be expressed in bold font, e.g. ${\bm k}$, or using Latin indices, $k^i$. Four dimensional vectors will be denoted by the Lorentz (Greek) indices, e.g. $\mu$. In the case of the position vector, $x^\mu$, it is given by $$x^{\mu} = (x^{0},x^{1},x^{2},x^{3})\,.$$
In most instances the Lorentz index in the four-vector will be suppressed, after it has been \mbox{initially} defined with the index.\\
 \\
The explicit dependence of a field on spacetime coordinates will be suppressed, especially in later chapters, i.e.
$$
	\phi_\mu(x^\nu) = \phi_\mu(x) = \phi_\mu\,.
$$
If the field or the vector is constant it will be defined as such and denoted by a subscript, usually zero e.g. $\phi_0$. \\
\\
Derivatives with regards to contravariant, $x^\mu$, or covariant vectors, $x_\mu$, are defined as
$$
	\partial_\mu \equiv \frac{\partial}{\partial x^\mu} = 
	\left(\frac{\partial}{\partial x^0},\frac{\partial}{\partial x^1},
	\frac{\partial}{\partial x^2},\frac{\partial}{\partial x^3}\right) = 
	\left(\frac{\partial}{\partial x^0},{\bm\nabla}\right)\,,
$$
and
$$
	\partial^\mu \equiv \frac{\partial}{\partial x_\mu} = 
	\left(\frac{\partial}{\partial x_0},\frac{\partial}{\partial x_1},
	\frac{\partial}{\partial x_2},\frac{\partial}{\partial x_3}\right) = 
	\left(\frac{\partial}{\partial x^0},-{\bm\nabla}\right)\,,
$$
where the properties of the metric tensor (\ref{metrictensor}) have been used.\\
\\
The time-derivative is sometimes expressed as
$$
	\frac{\partial \phi}{\partial t} = \dot{\phi}\,.
$$
\subsection{Natural units}
In this work the notion of ``natural units'', where $$\hbar = c = 1,$$ will be adopted. This will be used since some equations are very cumbersome if all the factors of $\hbar$ and $c$ are included. Where the inclusion of $\hbar$ or $c$ is of specific interest or meaning they will still be initially included. \\
\\
This convention has the added benefit that it establishes a simple relation between mass and length scale, through the relation
$$
	\hbar c = 1 = 197.33\ \mbox{MeV}\,\mbox{fm}.
$$
In natural units the relativistic energy-mass relationship,
$$
	E^{2} = {\bm p}^{2}c^{2} + m^{2}c^{4},
$$
reduces to 
$$ E^{2} = {\bm p}^{2} + m^{2},$$ and the units of mass and energy are the same.
Table \ref{tab:constants} contains the constants and conversion factors (in natural units) used in this work, that are not explicitly defined elsewhere.
\section{Matrices}
The $N\times N$ identity matrix will be denoted by ${\bm 1}$. If the value of $N$ is known  or specific, it will be denoted by a subscript, i.e.\ for $N= 2$,
$$
	{\bm 1}_2 = {\left[\begin{array}{cc}
	1&0\\
	0&1\end{array}\right].}
$$
\begin{table*}[hht]
	\centering
		\begin{tabular}{lll}
				\hline
				Name&Symbol&Value\\
				\hline\hline
				Solar mass & \ $M_{\odot}$& $1.98892\times10^{30}$ kg or 
				\\&& $1.1155\times10^{60}$ MeV\\
				Gravitational constant &\  $G$ & $6.6726\times10^{-11}$ m$^3\,$kg$^{-1}\,$s$^{-2}$ or\\&&
				 $1.325\times10^{-42}$ fm/MeV\\
				Length & 1 fm& $1\times10^{-15}$ m\\
				Energy & 1 MeV & \\
				Energy density & 1 MeV/fm$^{3}$ & $1.7827\times10^{12}$ g/cm$^{3}$\\
				Pressure & 1 MeV/fm$^{3}$ & $1.6022\times10^{33}$ dyne/cm$^{2}$\\
				\hline	
		\end{tabular}
	\caption{Constants used in this thesis.}
	\label{tab:constants}
\end{table*}

%% file: chapRel.tex
\chapter{Relativity}\label{chap:Rel}
Due to the central role that gravity plays in binding neutron stars, any description of these objects must include the effects of gravity. In this work gravitational effects will be included by using a description that is consistent with the special and general theories of relativity.
\section{Introduction}
One of Albert Einstein's phenomenal achievements was to recognise that mass and energy are related. As such both mass and energy are gravitational sources 
and not only mass, as is implied in Newtonian theory.\\
\\
Since neutron stars are large, ultra-dense objects with large gravitational fields and high particle energies, relativistic effects 
need to be considered in the description of these objects. The aim of this chapter is to give a brief introduction of the concepts and the necessary tools in the general and special theories of relativity that are applied in the description of neutron stars.
\section{Special theory of relativity}\label{secSpesRel}
The principle of relativity states that the same mechanical laws apply in any frame of \mbox{reference.} This principle was
 already clearly formulated by Galileo Galilei in the 1600's for \mbox{classical} \mbox{mechanics} (i.e. mechanics governed by Newton's
 Laws), but Albert Einstein, in the special theory of relativity, extended this principle to include all physical
 laws. \cite{schutz} \\
\\
The special theory of relativity (commonly referred to as special relativity) is grounded upon two postulates \cite{griff}:\\
\\
{\bf The postulate of relativity:} That the same physical laws apply
 in any inertial (uniformly \mbox{moving}) frame of reference.\\
{\bf The speed of light is universal:} The speed of light in a vacuum (denoted by $c$) is the same for all inertial
 observers,
 regardless of the motion of the source.\\
\\
The line element in special relativity, known as the proper time $d\tau$, is given by
\begin{eqnarray}
d\tau^{2} = \eta_{\mu\nu}dx^{\mu}dx^{\nu} = c^{2}dt^{2}-dx^{2}-dy^{2}-dz^{2}\label{minkowski}.
\end{eqnarray}
It is the preserved interval that defines Minkowski space. $x^{\mu}$ refers to a spacetime point and is expressed in Cartesian coordinates as 
\begin{eqnarray}
x^{\mu} = (ct,x,y,z). \label{fourvector}
\end{eqnarray}
From the expression of the line element (\ref{minkowski}), the metric tensor of Minkowski space is given by
\begin{eqnarray}
\eta_{\mu\nu} = {\left[\begin{array}{cccc}
1&0&0&0\\
0&-1&0&0\\0&0&-1&0\\0&0&0&-1\end{array}\right].}\label{metrictensor}
\end{eqnarray}\\
\\
Special relativity does not include any gravitational effects and therefore
 spacetime is flat and the metric tensor (\ref{metrictensor}) is fixed. Frames of reference in which the expressions of special relativity are valid 
are called Lorentz (inertial) frames.
These frames are related by {\em Lorentz \mbox{transformations}}. Due to the postulate of relativity the physical expression of any observable should not differ when observed from different Lorentz frames. 
Because any two inertial frames of reference are related by a Lorentz transformation it means that the equations describing a physical observable must be invariant under Lorentz transformations. This property is known as {\em covariance} and if an equation is expressed in a covariant manner the form of the equation does not change when transformed from one Lorentz frame to another. 
To be able to write equations in a
 covariant manner tensor analyses is used. \cite{schutz}
\section{General theory of relativity}
The special theory of relativity gives a very accurate description of our world within a Lorentz frame, but neglects one of the most prominent forces that govern the large scale behaviour in our universe, namely gravity. The distinction between gravity and other forces is that gravity acts on all matter, regardless of their internal structure or composition: Particles with wholly different physical properties will follow the same free path in a gravitational field, given only that their initial \mbox{velocities} are the same. The general theory of relativity (commonly referred to as simply general relativity) expands the locally very accurate special \mbox{theory} of relativity to a global scale, by including gravitational effects. \cite{schutz}\\ 
\\
Special relativity would unfortunately not suffice if gravitational effects are included since no global Lorentz frame can be constructed in a non-uniform gravitational field 
\cite{schutz}. This is \mbox{because} Euclid's Parallel \mbox{Postulate} holds in Minkowski space, i.e. in Minkowski space parallel worldlines stay parallel if extended infinitely. In a non-uniform gravitational field particles do not travel along straight lines, but along curved trajectories: if two particles are released from the same height above the earth's surface, but some distance apart, they will both fall towards the centre of the earth. Their trajectories would be parallel on a local scale but overall they will follow a curved path \cite{csg}. The spacetime on a global scale and in the presence of gravitational sources is one in which Euclid's Parallel Postulate does not hold, i.e. a spacetime that is flat on a local scale but overall curved, as described by Riemannian geometry \cite{schutz}.\\
%
%
%
%
\\
The absence of a global Lorentz frame might spell trouble for special relativity: if no Lorentz frame with regards to an event can be found, special relativity will be a very \mbox{interesting} but unphysical theory. Luckily within a free-falling frame in a non-uniform gravitational field the effect of the gravitational field would not be noted. Thus a free-falling frame of reference would constitute a Lorentz frame with regards to the event. Einstein used this fact to relate the special and general theories of relativity through the Equivalence Principle:\\
\\
{\bf The Equivalence Principle:} Physical laws that describe how the forces of nature behave in a gravitational field have the same (covariant) form as in the special theory of relativity when these laws are expressed in a frame of reference, free-falling with the event, in the gravitational field.\\
\\
The Equivalence Principle is analogous to the theorem in differential geometry that states that a flat space tangent to the manifold can be imposed on any point in a differentiable manifold. The spacetime of general relativity constitutes such a manifold and therefore a Lorentz frame can be imposed at any point. \cite{schutz}\\
\\
To translate from special to general relativity the metric tensor, $\eta^{\mu\nu}$, has to be replaced by the general metric, $g^{\mu\nu}$, and normal derivatives by covariant derivates (see Sec. \ref{secRieman} for details of $g^{\mu\nu}$ and covariant differentiation) \cite{csg}.
%
%
%
\section{Riemannian spaces}\label{secRieman}
Albert Einstein recognised the similarities between gravitational physics and Riemannian spaces and used the mathematics describing these spaces to describe gravitational phyics.\\
\\
In this section the main points of Riemannian spaces will be introduced.
\subsection{Geodesics}
A line in curved space is defined as straight (i.e. the shortest distance between two points) if the tangent vector to the line is parallel-transported \cite{schutz}. These lines are known as geodesics.\\
\\
The metric tensor of curved space is denoted by $g_{\mu\nu}$ and is defined in terms of the natural basis vector ($\underline g_\mu$) of the space
$$
	\underline g_\mu = \frac{\partial s}{\partial x^\mu},
$$
where $s$ is a small displacement along the coordinate curve. The metric tensor is given by the dot product of two basis vectors
$$
	g_{\mu\nu} = \underline g_\mu \cdot \underline g_\nu\ .
$$
In Minkowski space the metric tensor ($\eta_{\mu\nu}$) is fixed and given by (\ref{metrictensor}). In curved space the metric tensor is not necessarily constant, but dependant on the geometry of the space. \cite{toeg} \\
\\
Geodesics, just as straight lines in a Euclidean space, can also be defined as the extreme value of the integral of the proper time interval ($d\tau$) between the two spacetime points. The line element, the proper time, in general relativity is given by
\begin{eqnarray}
d\tau^{2} = g_{\mu\nu}dx^{\mu}dx^{\nu}.\label{lineGR} 
\end{eqnarray}
As is shown by N.K. Glendenning in Ref.\,\cite{csg}, geodesics are thus described by the Geodesic equation,
\begin{eqnarray}
\frac{d^{2}x^{\lambda}}{d\tau^{2}} + \Gamma^{\lambda}_{\mu\nu}\frac{dx^{\mu}}{d\tau}\frac{dx^{\nu}}{d\tau} = 0
\end{eqnarray} 
where $\Gamma^{\lambda}_{\mu\nu}$ is the affine connection. It also shown in Ref.\,\cite{csg} that the affine connection is equal to the Christoffel symbol of the second kind.
\subsection{Christoffel symbols}
The Christoffel symbol (of the second kind), $\Gamma^{\alpha}_{\nu\mu}$, is defined in terms of the relation of the derivative of the natural base vector to the natural base vectors as \cite{toeg},
\begin{eqnarray}
	\frac{\partial g_\nu}{\partial x^\mu} \equiv \Gamma^{\alpha}_{\nu\mu}\underline g_\alpha\ .\label{CS}
\end{eqnarray}
As shown in Ref.\,\cite{stephani}, the Christoffel symbols can be expressed in terms of the metric tensor 
\begin{eqnarray}
	\Gamma^{\alpha}_{\nu\mu} = \frac{1}{2}g^{\alpha\beta}(g_{\nu\beta,\mu} + g_{\mu\beta,\nu} - g_{\nu\mu,\beta}),\nonumber
\end{eqnarray}	
and so it can be seen that the Christoffel symbols are related to the curvature of the space through the change in the metric.\\
\\
The Christoffel symbol of the first kind, $\Gamma_{\kappa\nu\mu}$, is related to the Christoffel symbol of the second kind through \cite{csg}:
$$
	\Gamma_{\kappa\nu\mu} = g_{\kappa\alpha}\Gamma^{\alpha}_{\nu\mu}.
$$
The Christoffel symbols of the second kind are symmetric in their covariant (lower) components \cite{toeg}, i.e.
$$
	\Gamma^{\alpha}_{\nu\mu} = \Gamma^{\alpha}_{\mu\nu}
$$
and so the Christoffel symbols of the first kind are symmetric in the last two covariant components.
\subsection{Covariant derivatives}\label{seccovarderiv}
The partial derivative of a vector $v$, with $v = v^\mu \underline g_\mu$, is:
\begin{eqnarray}
	\frac{\partial\; v^\mu \underline g_\mu}{\partial x^\nu} 
	&=&  \frac{\partial v^\mu }{\partial x^\nu}\underline g_\mu + 
		\frac{\partial \underline g_\mu }{\partial x^\nu}v^\mu \nonumber\\
	&=&  v^\mu_{,\nu}\underline g_\mu + \underline g_{\mu,\nu}v^\mu \nonumber\\
	&=&  v^\mu_{,\nu}\underline g_\mu + v^\mu\Gamma^{\alpha}_{\mu\nu}\underline g_\alpha \ \ \ \ \ \ 
		\mbox{from (\ref{CS})} \nonumber\\
	&=& (v^\mu_{,\nu} + v^\kappa\Gamma^{\alpha}_{\kappa\nu})\underline g_\mu \ \ \ \ \ \ \ 
		\mu \rightleftharpoons \kappa \nonumber\\	
	&=& (v^\mu_{;\nu})\underline g_\mu\ , \nonumber
\end{eqnarray}
and the covariant derivative is defined as \cite{toeg}:
\begin{eqnarray}
	v^\mu_{;\nu} = v^\mu_{,\nu} + v^\kappa\Gamma^{\alpha}_{\kappa\nu}\ . \label{covarderiv}
\end{eqnarray}
The covariant derivative transforms covariantly, i.e. the covariant derivative of a tensor is once again a tensor \cite{stephani}.
\subsection{Riemann-Christoffel curvature tensor}
The Riemann-Christoffel curvature tensor (or just curvature tensor) describes the curvature of space. It is given by 
\begin{eqnarray}
R^{\rho}_{\sigma\mu\nu} = \Gamma^{\rho}_{\sigma\nu,\mu} - \Gamma^{\rho}_{\sigma\mu,\nu} +
 \Gamma^{\alpha}_{\sigma\nu}\Gamma^{\rho}_{\alpha\mu} - \Gamma^{\alpha}_{\sigma\mu}\Gamma^{\rho}_{\alpha\nu}.
\end{eqnarray}
The curvature tensor is derived by evaluating the parallel-transport of a tangent vector along a small closed loop on a curved surface. The vector field is defined on the surface and a vector ($V^{\mu}$) 
parallel-transported along the loop. Once the starting point is reached again, the vector is compared to the original vector. The difference between the two vectors is proportional to the curvature and the curvature tensor \cite{schutz}.\\
A more intuitive explanation might be given when it is considered that in flat spacetime the metric tensor is constant and the order of covariant differentiation does not matter. But in curved spacetime the order of covariant differentiation is important as two successive covariant differentiations do not commute, as is shown in Ref.\,\cite{schutz}:
\begin{eqnarray}
	\left[ \nabla_{\alpha},\nabla_{\beta}\right]V^{\mu} = R^{\mu}_{\nu\alpha\beta}V^{\nu}\nonumber
\end{eqnarray}
where the commutator is given by the curvature tensor. The commutator is analogous to computing the change in a vector along a closed path: the change in the vector is first computed in one direction and then in another and the changes in the reverse order is subtracted. 
\cite{schutz}\\
\\
The curvature tensor is zero if and only if the space is flat \cite{stephani}.
\subsection{Bianchi identities}
From the properties of the curvature tensor it is shown in Ref.\,\cite{schutz} 
that the following equation holds in any frame:
\begin{eqnarray}
	R_{\alpha\beta\mu\nu;\lambda} + R_{\alpha\beta\lambda\mu;\nu} + R_{\alpha\beta\nu\lambda;\mu} = 0\ .\label{BI}
\end{eqnarray}
This is known as the {\em Bianchi identities}. These identities can be used to determine the metric tensor if the curvature is known \cite{stephani}.
\subsection{Ricci tensor and scalar curvature}
The {\em Ricci tensor}, $R_{\mu\nu}$, is a contraction of the curvature tensor:
\begin{eqnarray}
 R_{\alpha\beta} = R^{\,\nu}_{\,\alpha\nu\beta} = R_{\beta\alpha}.
\end{eqnarray}
The Ricci tensor can also be formed by contracting indices, other than the first and the third, 
but the symmetry of the curvature tensor implies that these contractions will either vanish or at most add a minus sign to the Ricci tensor. The Ricci tensor is symmetric. \cite{schutz}\\
\\
%
%
%
%
The {\em scalar curvature}, $R$, is defined as the contraction of the metric tensor and the Ricci tensor \cite{csg}:
\begin{eqnarray}
R = g^{\mu\nu}R_{\mu\nu} = g^{\mu\nu}R^{\sigma}_{\mu\sigma\nu}\nonumber.
\end{eqnarray}
\subsection{Einstein tensor}
If the Bianchi identity is contracted twice \cite{schutz}:
\begin{eqnarray}
		g^{\alpha\mu}\left[R_{\alpha\beta\mu\nu;\lambda} + R_{\alpha\beta\lambda\mu;\nu} 
		+ R_{\alpha\beta\nu\lambda;\mu}\right]
		= R_{\beta\nu;\lambda} - R_{\beta\lambda;\nu} + R^{\nu}_{\beta\nu\lambda;\mu} = 0\nonumber
\end{eqnarray}
\mbox{and}
	\begin{eqnarray}
		g^{\beta\nu}\left[ R_{\beta\nu;\lambda} - R_{\beta\lambda;\nu} + R^{\mu}_{\beta\nu\lambda;\mu}\right]
		= R_{;\lambda} - R^{\nu}_{\lambda;\nu} - R^{\mu}_{\lambda;\mu} = 0,\nonumber
\end{eqnarray}
then the last equation can be rewritten as:
\begin{eqnarray}
	{(2R^{\mu}_{\lambda} - \delta^{\mu}_{\lambda}R)}_{;\mu} = 0. \label{rewrite2ContractRCT}
\end{eqnarray}
By defining $G^{\mu\nu}$ as 
\begin{eqnarray}
	G^{\mu\nu} \equiv R^{\mu\nu} - \frac{1}{2}g^{\mu\nu}R \label{EinsteinTensor}
\end{eqnarray}
it is clear from Eq. (\ref{rewrite2ContractRCT}) that it has 
vanishing covariant divergence, i.e.
\begin{eqnarray}
	{(G^{\mu\nu})}_{;\mu} = 0\ .
\end{eqnarray}
$G^{\mu\nu}$ is known as the Einstein tensor and it plays a fundamental role in general relativity (see Sec. \ref{secEFT}). From the symmetry of the Ricci and metric tensors it is clear that the Einstein tensor is also symmetric.

%
%
\section{Energy-momentum tensor}
\label{sec:EnergyMomentumTensor}
The energy-momentum tensor ($T^{\mu\nu}$) describes the internal properties of an energy-mass \mbox{distribution} (i.e. matter) and is the source of the curvature of spacetime. To be able to give an accurate, covariant description of the internal properties of matter within a relativistic framework, not only the energy distribution, but also the distribution of momentum within the matter needs to be considered.\\
\\
The energy-momentum tensor is defined in terms of the flux of the 4-momentum across a constant surface. The element, $T^{\mu \nu}$, can be described as the $\mu$-component of the momentum flux across the $x^{\nu}$-surface. Thus the following components can be explicitly named:
\begin{center}
	\begin{tabular}{l}
		$T^{00}$ = energy density,\\
		$T^{0i}$ = energy flux across the $x^{i}$ surface,\\
		$T^{i0}$ = $i$ momentum density,\\
		$T^{ij}$ = flux of the $i$ momentum across the $j$ surface.\\
	\end{tabular}
\end{center}	
If we evaluate the energy-momentum tensor in a momentarily comoving frame of reference all elements of the matter would be momentarily static (no spatial momentum), but that would not imply that there is no transfer of energy. Energy might be transferred by heat conduction and therefore the $T^{0i}$ terms of the energy-momentum tensor would be non-zero. The $T^{i0}$ components would also be non-zero 
for the energy that is being transferred will carry momentum. The $T^{0i}$ components would be equal to the $T^{i0}$ components since the energy flux is the energy density times the speeds at which it flows. In the relativistic frame, mass is equal to energy so the energy flux is equal to the mass density times the speeds at which it flows, which is the density of momentum.\\
The spatial components of the energy-momentum tensor ($T^{ij}$) are symmetric. If they were not it would be imply that the elements are whirling around inside the fluid, in the \mbox{momentarily} comoving frame of reference. 
Thus $T^{\mu\nu}$ is in general symmetric and symmetry of the tensor in one frame of reference would imply that it is symmetric in all frames of reference.\\
The spatial components of the energy-momentum tensor represent the forces between adjacent elements in the matter: the off-diagonal spatial elements represent the viscosity (or any other forces that are parallel to the interface between elements) and the diagonal elements, the normal forces (i.e. the pressure). \cite{schutz} \\
\\
In general the energy-momentum tensor for static, spherically symmetric perfect fluid (no viscosity or heat conduction), moving with a velocity ${\bf v}$ is \cite{csg, walecka}:
\begin{eqnarray}
	T^{\mu\nu} = -P\eta^{\mu\nu} + (P + \epsilon)u^\mu u^\nu\label{genEMT},
\end{eqnarray}
where
\begin{itemize}
	\item $\epsilon$ is the energy density, 
	\item $P$ is the pressure, and,
	\item $u^\mu$ is the four-velocity,
\begin{eqnarray}
	u^\mu &=& \frac{dx^\mu}{d\tau} \nonumber\\
		&=& \sqrt{1-{\bm v}^2}\,\big(1,v^1,v^3,v^3\big)\,, \label{u4velo}
\end{eqnarray}
and therefore $u^\mu u_\mu = 1$ 
	\cite{csg}.
\end{itemize}
The conservation of mass and energy is expressed in terms of the energy-momentum tensor by the fact that the energy-momentum tensor has vanishing divergence, i.e.
\begin{eqnarray}
	T_{\mu\nu, \kappa} = 0.\label{divEMTSpesRel}
\end{eqnarray}
In general relativity Eq. (\ref{divEMTSpesRel}) generalises to \cite{stephani}
\begin{eqnarray}
	T_{\mu\nu; \kappa} = 0.\label{divEMT}
\end{eqnarray}

%
%
\section{Einstein's field equations} \label{secEFT}
Physical theories are mostly categorised by defining equations, which in the case of general relativity are the
 Einstein field equations. The Einstein field equations are given by solutions of the Einstein tensor. For 
 spacetime inside a distribution of mass and energy (i.e. a star), Einstein's field equations are given by
\begin{eqnarray}
G^{\mu\nu} = \kappa T^{\mu\nu} \label{EFeqn},
\end{eqnarray}
where $T^{\mu\nu}$ is the energy-momentum tensor and $\kappa$ is a constant that is determined by comparing 
general relativity to
 Newtonian mechanics in the Newtonian limit \cite{csg}. The energy-momentum tensor is a symmetric, divergenceless tensor that
 is constructed from the mass-energy properties of the medium. \\
\\
The Einstein field equations (\ref{EFeqn}) do not only govern spacetime within a mass-energy distribution, but also the arrangement 
of mass-energy distribution as well as its dynamics. Thus spacetime is acted upon by a
 mass-energy distribution, as these are sources of gravity. On the other hand, spacetime influences the mass-energy
 distribution through the curvature of spacetime. Thus spacetime and matter co-determine the universe within which we
 live.
\\

%% file: chapTOV.tex
\chapter{Tolman-Oppenheimer-Volkoff equations}\label{chap:TOV}
By using the concepts defined in Chapter \ref{chap:Rel} a relativistic description for a neutron star will be derived in this chapter.
\section{Introduction}
As a first approximation a star (such as a neutron star) is assumed to be a static, spherical symmetric 
fluid (gas) in hydrostatic equilibrium \cite{csg}: the star is bound by an external pressure that compresses the star (gravity) while the star is stabilised against gravitational collapse by the internal pressure in the star due to some repulsive force. \\
\\
If a neutron star is assumed to be a static spherical symmetric fluid in hydrostatic \mbox{equilibrium} it would have to be described by a relativistic equation for hydrostatic equilibrium.\\
\\
This problem was first studied by R. C. Tolman \cite{tolman} and by J. R. Oppenheimer and G. M. Volkoff \cite{OV} in 1939. They derived an equation to study hydrostatic equilibrium in a relativistic environment by describing a neutron star assuming that it consists of a neutron gas at high density \cite{shapiro}:
\begin{eqnarray}
	\frac{dP(r)}{dr} = -\frac{G\epsilon(r)M(r)}{c^{2}r^{2}}\left[1+\frac{P(r)}{\epsilon(r)}\right]
\left[1+\frac{4\pi r^{3}P(r)}{M(r)c^2}\right]
\left[1-\frac{2GM(r)}{c^{2}r}\right]^{-1}\label{TOV}
\end{eqnarray}
where 
\begin{eqnarray}
	\frac{dM(r)}{dr} = \frac{4\pi \epsilon(r)r^{2}}{c^2}\label{TOVM}
\end{eqnarray}
and 
\begin{itemize}
	\item $M(r)$ is the enclosed mass of the star,
	\item $\epsilon(r)$ is the energy density, and
	\item $P(r)$ is the internal pressure of the star.
\end{itemize}
Eq. (\ref{TOV}) is known as the Tolman-Oppenheimer-Volkoff (TOV) equation \cite{ns4u}.\\
\\
The TOV equation expresses how the pressure decreases from the centre of the star to the edge (where $P = 0$) in terms of the energy density and the pressure. These two quantities are related through the equation of state of the matter in the interior of the star. The larger the star, the higher the pressure in the centre of the star will be, since it is gravity that compresses the star. Since the pressure appears on the right-hand side of Eq. (\ref{TOV}), an increase in the central pressure will increase the pressure gradient [the left-hand side of Eq. (\ref{TOV})]. 
Thus the more massive the star, the smaller the radius at which $P = 0$, since the pressure gradient is steeper. Thus there exists a critical value for the mass of the star, above which the star will collapse under its own gravity \cite{csg}. This critical value is known as the maximum mass of the star and will differ for different equations of state of the matter in the neutron star interior.

\section {Derivation of the TOV equation}
The following derivation is taken almost entirely from Ref. \cite{csg}. (The derivation in Ref. \cite{csg} is more elaborate and hints to perform some of the calculations are also given.)\\
\\
To derive the TOV equations from the Einstein field equation (\ref{EFeqn}) it suffices to consider a static,
 isotropic star. The line element for static, isotropic spacetime can be given by:
\begin{eqnarray}
d\tau^{2} = e^{2\nu(r)}dt^{2}-e^{2\lambda(r)}dr^{2} - r^{2}d\theta^{2} - r^{2}\sin^2\theta d\phi^{2}\label{line}
\end{eqnarray}
where $\nu(r)$ and $\lambda(r)$ are functions that need to be determined and $x^{\mu}$ is the spacetime point described in natural units by
\begin{eqnarray}
x^{\mu} = \left[t, r, \theta, \phi \right]. \nonumber
\end{eqnarray}
\\
From the expression for the line element (\ref{line}) the components of the metric tensor can be read off as
\begin{subequations}
	\begin{eqnarray}
		g_{00} &=& e^{2\nu(r)}\\ g_{11} &=& -e^{2\lambda(r)}\label{metricGR}\\ 
		g_{22} &=&  -r^{2}\\ g_{33} &=& -r^{2}\sin^2\theta\\ 
		g^{\mu\nu} &=& g_{\mu\nu} = 0 \;\; (\mbox{for }\mu \neq \nu).
	\end{eqnarray}
\end{subequations}	
Since $g_{\mu\alpha}g^{\alpha\nu} = \delta^{\nu}_{\mu}$, $g_{\mu\nu}$ is given by $g_{\mu\mu} = \left({g^{\mu\mu}}\right)^{-1}$ (no summation over indices).\\
\\
Using the symmetry properties of the Christoffel symbols, the components of the Ricci tensor in static isotropic spacetime are:
\begin{subequations}\label{riccitov}
	\begin{eqnarray}
		R_{00} &=& (-\nu'' + \lambda'\nu' - \nu'^{2} - \frac{2\nu'}{r})e^{2(\nu-\lambda)}, \\
		R_{11} &=& \nu'' - \lambda'\nu' + \nu'^{2} - \frac{2\lambda'}{r},\\
		R_{22} &=& (1 + r\nu'- r\lambda')e^{2(\lambda)} + 1,\ \mbox{and},\\
		R_{33} &=& R_{22}\sin^{2}\theta,
	\end{eqnarray}
\end{subequations}	
where the primed index refers to differentiation with respect to $r$.\\
\\
For the construction of the TOV equation it is easier to work with mixed tensors, therefore the Einstein field equations
 (\ref{EFeqn})
 are written as
\begin{eqnarray}
	G^{\nu}_{\mu} &=& g_{\mu\alpha}G^{\alpha\nu} \nonumber\\
	&=& g_{\mu\alpha}R^{\alpha\nu} - \frac{1}{2}g_{\mu\alpha}g^{\alpha\nu}R \nonumber\\
	&=& R^{\nu}_{\mu} - \frac{1}{2}\delta^{\nu}_{\mu}R \nonumber\\
	&=& \kappa T^{\nu}_{\mu}\label{mixedTensor}\,,
\end{eqnarray}
and thus, using the properties of the Ricci tensor (\ref{riccitov}), the components of the Einstein tensor become:
\begin{subequations}\label{mixed}
	\begin{eqnarray}
		r^{2}G^{0}_{0} &=& e^{-2\lambda}(1 - 2r\lambda') - 1 =  -\frac{d}{dr} \left[ r(1 - e^{-2\lambda}) \right] 	
		\label{mixeda}\\
		r^{2}G^{1}_{1} &=& e^{-2\lambda}(1 + 2r\nu') - 1 \\
		G_{2}^{2} &=& e^{-2\lambda}\left(\nu'' + \nu'^{2} - \lambda'\nu'  + \frac{\nu' - \lambda'}{r}\right)\\
		G_{3}^{3} &=& G_{2}^{2}.
	\end{eqnarray}
\end{subequations}	
From (\ref{genEMT}) and the equivalence principle, the energy-momentum tensor 
is 
\begin{eqnarray}
	T^{\mu\nu} = -P(r) g^{\mu\nu} + \big(P(r) + \epsilon(r)\big)u^{\mu}u^{\nu} \nonumber.
\end{eqnarray}
Since the 
the star is static ($u^{i} = 0$) and therefore from (\ref{u4velo}), 
$g^{\mu\nu}u^{\mu}u^{\nu} = 1$, $u^0$ is given by
$$
		u^0 = \frac{1}{\sqrt{g_{00}}}\,.
$$
From the equivalence principle, a comoving Lorentz frame can be defined \footnote{This is a valid assumption, since the change in the metric from the centre of the star to its boundary is not significant over the spacing between a few nucleons \cite{csg}. [The change in the metric can be estimated by considering the espression for $g_{11}$ (\ref{g11}).]}, and therefore the energy-momentum tensor can be expressed as
\begin{eqnarray}
	T^{0}_{0} = \epsilon(r)\  \mbox{and} \ T^{i}_{i} = -P(r) .\label{energymoment}
\end{eqnarray}\\
From (\ref{mixeda}), (\ref{mixedTensor}) and (\ref{energymoment})
\begin{eqnarray}
	r^{2}G^{0}_{0} &=& -\frac{d}{dr} \left[ r(1 - e^{-2\lambda(r)})\right] \nonumber\\
	&=& \kappa r^{2}T^{0}_{0} \nonumber\\
	&=& \kappa r^{2}\epsilon(r)\label{exp4lambda}\,,
\end{eqnarray}\\
and (\ref{exp4lambda}) can be solved to obtain
\begin{eqnarray}
	e^{-2\lambda(r)} &=& 1- \frac{\kappa}{r}\int^{r}_{0}\epsilon(r)r^{2}dr \nonumber\\
	&=& 1- \frac{\kappa}{4\pi r}M(r)\label{defM(r)},
\end{eqnarray}
where $M(r)$ is defined as
\begin{eqnarray}
M(r) = 4\pi \int^{r}_{0}\epsilon(r')r'^{2}dr'. \label{grav mass}
\end{eqnarray}
Eq. (\ref{grav mass}) is an expression for the gravitational mass i.e. the mass-energy that generates a gravitational field included, up to radius $r$, in the star. The relativistic expression (in natural units) that is used to describe the relation between the energy density ($\epsilon$) and the mass density ($\rho$) is:
$$
	\epsilon (r) = \rho(r).
$$
Using (\ref{defM(r)}) 
the following expression for (\ref{metricGR}) can be obtained:
\begin{eqnarray}
	g_{11} = -e^{2\lambda(r)} = -\left( 1 - \frac{\kappa M(r)}{4\pi r} \right)^{-1}.\label{g111}
\end{eqnarray}
For (\ref{g111}) to agree with Newtonian mechanics in the Newtonian limit, $\kappa$ 
should be defined as 
\begin{eqnarray}
	\kappa = -8\pi G, \label{kappa def}
\end {eqnarray}
 with $G$ the Gravitational constant \cite{csg}. 
 Thus (\ref{g111}) becomes
\begin{eqnarray}
	g_{11} = -e^{2\lambda(r)} = -\left( 1 - \frac{2GM(r)}{r} \right)^{-1}.\label{g11}
\end{eqnarray}\\
Using (\ref{kappa def}) and (\ref{energymoment}) the expression for the Einstein field equations (\ref{mixed}),
 becomes:
\begin{subequations}
	\begin{eqnarray}
		G^{0}_{0} &=& e^{-2\lambda}\left(\frac{1}{r^{2}} - \frac{2\lambda'}{r}\right) - \frac{1}{r^{2}} =  -8\pi G\epsilon(r)
		 \label{G00}\\
		G^{1}_{1} &=& e^{-2\lambda}\left(\frac{1}{r^{2}} + \frac{2\nu'}{r}\right) 
		- \frac{1}{r^{2}}	 =  -8\pi GP(r)\label{G11}\\
		G_{2}^{2} &=& e^{-2\lambda}\left(\nu'' + \nu'^{2} - \lambda'\nu'  + \frac{\nu' - \lambda'}{r}\right) 
		= -8\pi GP(r)\label{G22}\\
		G_{3}^{3} &=& G_{2}^{2} 
	\end{eqnarray}
\end{subequations}	
By manipulating (\ref{G00}), (\ref{G11}) and (\ref{g11}) and substituting it into (\ref{G22}), the following expression for
 $P'(r)$ is obtained: 
\begin{eqnarray}
P'(r) = \frac{dP}{dr} = -\frac{G\left[\epsilon(r)+P(r)\right]
\left[M(r)+4\pi r^{3}P(r)\right]}
{r\left[1-2GM(r)\right]}.\label{TOV2}
\end{eqnarray}
Eq. (\ref{TOV2}) is the TOV equation in natural units.
\section{Solving the TOV equation}
To obtain sensible information from the TOV equation the coupled differential equations, Eqs (\ref{TOV}) and (\ref{TOVM}), must be solved. To be able to do this the initial conditions (at $r = 0$) for $P(r)$ and $M(r)$ must be known.\\
\\
Since the expression for $M(r)$ (\ref{grav mass}) refers to the enclosed mass in the star and $r=0$ refers to the centre of the star, the initial values for both $M(r)$ and $r$ are taken to be zero (or very small). The choice for the initial $P$ depends on the inferred central baryon density of the neutron star. 
\\\\
Using the initial values for $P$, $r$ and $M$ the 
pressure ($P(r')$) at an incremental increase in $r$, namely $r' = r + dr$ can be calculated using Eq. (\ref{TOV}). Using the equation of state, the corresponding energy density, $\epsilon(r')$, 
can be calculated. 
Using (\ref{TOVM}) the value for $M(r')$ can be obtained. This cycle is repeated until $r' = R$ at which the $P(r')$ is zero, is reached. This denotes the boundary of the star and $R$ is the radius of the star. $M(R)$ is the mass enclosed by the radius of the star, i.e. the mass of the star.\\
\\
Each different assumption of the central density will correspond to a unique mass and radius \mbox{relationship} of the neutron star. Therefore by varying the initial central density, a whole sequence of possible masses and radii of neutron stars, corresponding to a specific equation of state, can be generated.\\
\\
As was expounded upon at the beginning of this chapter, each neutron star sequence has a maximum mass and as such each equation of state of the neutron star interior implies a different maximum mass. The equations of state with higher maximum masses are referred as stiff, i.e. the pressure increases rapidly with an increase in density. The equations of state with lower maximum masses are referred to as softer equations of state \cite{csg}.

%% file: chapNucmatter.tex
\chapter{Modelling dense nuclear matter}\label{DNM}
The challenge in the description of matter at high densities, such as that in neutron stars, is to develop a model that not only describes matter at high densities, but also the properties of matter 
observed at normal densities. Additionally, relativistic effects become much more \mbox{pronounced} in dense systems as the particles attain energies that are comparable to their rest mass. Therefore it is neccesary that a description of matter at high densities must incorporate the general properties of quantum mechanics, Lorentz covariance, electromagnetic gauge \mbox{invariance} and \mbox{microscopic} causality within a many-body system \cite{walecka}. The only framework that can describe a relativistic, quantum-mechanical, many-body system in such a way is relativistic quantum field theory based on a local, Lorentz-invariant Lagrangian density \cite{walecka, recentprogress}.\\
\\
Quantum chromodynamics (QCD) describes the interaction between quarks via the exchange of gluons. This fundamental theory is an obvious candidate to describe dense matter systems, but unfortunately this theory has computational difficulties at such scales \cite{walecka}. It is also cumbersome to describe nuclear phenomena in terms of quarks and gluons since quark degrees of freedom are not observed in nuclear experiments but, hadronic degrees of freedom. Hadrons are 
particles that consist of a number of quarks and are further subdivided into baryons and mesons: baryons are particles that contain three quarks (such as protons and neutrons)
, while mesons contain a quark - anti-quark pair. A description of the interaction between two nucleons (particles in the nucleus)
, based on the exchange of mesons was first introduced by Hideki Yukawa in 1935 \cite{machleidt} and 
gained much acceptance after the actual discovery of mesons. A complete discussion of the mesonic theory of the nucleus is given in Ref.\,\cite{machleidt}.\\
\\
A quantized field theoretical description of nuclei and nuclear matter, called quantum hadrodynamics (QHD), which is based on hadronic degrees of freedom was introduced by John Walecka in 1974 \cite{walecka1}. It should be noted that QHD is not a fundamental theory, but an effective one since hadrons are composite particles.\\
\\
In QHD, as in QCD, one soon runs into computational difficulties and to obtain solutions from the description certain approximations have to be made. 
The coupling constants in QHD are large and thus a perturbation expansion in terms of the coupling constants will not suffice. Instead the system can be approximated in the relativistic mean-field. This approximation will be discussed in Sec. \ref{RMF}. The saturation properties of nuclei and nuclear matter, that any description of dense matter should reproduce, are discussed in Section \ref{satprop}. In Chapter \ref{chapRQM} a brief overview of relativistic quantum mechanics is given. This chapter is followed by a brief discussion of the necessary tools and concepts in classical field theory that is used in quantum field theory in Chapter \ref{mechanics}. Quantum field theory is discussed in Chapter \ref{QFT}, after which QHD is formulated in Chapters \ref{chapQHD1} and \ref{chapFSUG}.
%
%
%
\section{Properties of symmetric nuclear matter constraining nuclear models}\label{satprop}
Symmetric nuclear matter, or just nuclear matter, is an idealised system that stems from one of the original models of the nucleus, the liquid-drop model. The properties of nuclear matter are inferred from the experimentally observed properties of finite nuclei \cite{csg}, but, since it cannot be directly observed, there seem to be some disagreement as to what the exact values of certain properties of nuclear matter are. In this work the values of the properties of nuclear matter given in Ref.\,\cite{csg} will be taken as the observed values.
\subsection{Saturation density}
Nuclear matter is a saturated system, due the characteristics of the strong interaction. The short-ranged, strong nuclear interaction is the dominant interaction between nucleons. 
It is essentially attractive, which is necessary to form stable nuclei, but repulsive at short distance ($\leq 0.4$ fm). Since the strong force acts only over a short distance, the interaction is limited to nearest neighbours in a dense system. Therefore at a certain density the central density of the system will not increase any further, even as more nucleons are added to the system. 
The density at which this occurs is referred to as the saturation density. At saturation density the pressure of the system is zero and the system will remain in this state if left undisturbed. \cite{csg}\\
\\
The density of saturated nuclear matter given as 0.153 fm$^{-3}$ in Ref.\,\cite{csg} and 0.16 fm$^{-3}$ in Ref.\,\cite{webertxt}.
\subsection{Binding energy}
In general terms the binding energy of a system is the energy expended or required to form a system. In the case of a stable system the binding energy is negative, hence energy was donated to the surrounding environment in forming the system, leaving the system at a lower energy state than the sum of all the individual parts of the system.\\
\\
At saturation density the binding energy of the system will be at a minimum since at this density 
the system is in its most stable (lowest energy) state, compared to other densities. The binding energy of nuclear matter is given as -16.3 MeV/nucleon in Ref.\,\cite{csg} and -16.0 MeV/nucleon in Ref.\,\cite{webertxt} 

\subsection{Symmetry energy}\label{asym}
Stable nuclei with low proton ($Z$) number prefer a nearly equivalent neutron ($N$) number. As $Z$ increases the (repulsive) Coulomb interaction between the protons also increases. Stable nuclei then diverge from being $N=Z$ nuclei to nuclei with a $N$ larger than $Z$, as can been seen on a Segre diagram
. The influence of this preference is accounted for by the symmetry energy. The symmetry energy coefficient, $a_{4}$, stems from liquid-drop model of the nucleus, and refers to the contribution made by the isospin assymetry to the energy of the nucleus \cite{waleckatext}. \\
\\
The symmetry energy coefficient is given by:
\begin{eqnarray}
	a_{4} = \frac{1}{2}\left(\frac{\partial^{2}}{\partial t^{2}}\frac{\epsilon}{\rho}\right)_{t=0}\ \ \ \ \  \left(t\equiv\frac{\rho_{n} - \rho_{p}}{\rho}\right) \label{a4}.
\end{eqnarray}
The value of $a_4$ is estimated to be between 31 and 33 MeV according to Ref.\,\cite{Lattimer07}, while Refs \cite{csg} and \cite{webertxt} state the value of $a_4$ to be 32.5 MeV (without any uncertainty).
\subsection{Compression modulus}\label{comp}
The compression modulus defines the curvature of the equation of state at saturation \cite{csg} and is related to the high density behaviour of the equation of state. If the energy density rapidly increases with an increase in pressure the equation of state is referred to as stiff. With a soft equation of state the energy density increases more gradually with an increase in the pressure \cite{csg}. \\
The compression modulus, $K$, defined as 
\begin{eqnarray}
	K \equiv 9\left[\rho^{2}\frac{d^{2}}{d\rho^{2}}\left(\frac{\epsilon}{\rho}\right)\right]_{\rho=\rho_{0}}.
\end{eqnarray}
The value of $K$ has been estimated to 234 MeV (with some uncertainty) \cite{csg}. Ref.\,\cite{webertxt} state the value of $K$ to be 265 MeV.

%% file: chapRQM.tex
\chapter{Relativistic quantum mechanics}\label{chapRQM}
This chapter is a very concise introduction to relativistic quantum mechanics, introducing only concepts necessary within the scope of this work. A much more thorough discussion can be found in Refs \cite{A+H} to \cite{landau}.
%
%
%
%
%

\section{Introduction}
Relativistic quantum mechanics is the merger of quantum mechanics and special relativity. One possible candidate for a relativistic quantum mechanical description could be the Schr\"{o}dinger equation. The Schr\"{o}dinger equation can be motivated by applying the canonical quantization to the non-relativistic energy-momentum relationship \cite{A+H},
$$
	E = \frac{{\bm p}^{2}}{2m},
$$
where the physical quantities are replaced by operators, i.e.
\begin{subequations}\label{canquant}
	\begin{eqnarray}
		E \rightarrow i\hbar\frac{\partial}{\partial t}\\
		{\bm p}\rightarrow -i\hbar \nabla,
	\end{eqnarray}
\end{subequations}	
and by letting the operators operate on the wave function $\psi(t, \bm{x} )$, the Schr\"{o}dinger equation is obtained:
$$
	i\hbar\frac{\partial}{\partial t}\psi(t, {\bm x}) = \frac{-\hbar^2 \nabla^2}{2m}\psi(t, {\bm x}).
$$\\
Relativity requires that the equations of motion in one inertial frame are valid in all inertial frames, meaning that all equations describing the motion of a particle must transform covariantly under Lorentz transformations. The Schr\"{o}dinger equation does not transform covariantly and therefore another approach is needed to describe relativistic particles in a quantized theory \cite{drell}. An obvious candidate would be to apply the canonical quantization to the relativistic mass-energy relationship.
%
%
%
%
%
%
\section{Klein-Gordon equation}
Imposing the canonical quantization procedure to the relativistic energy-momentum relationship,
$$ E^{2} = {\bm p}^{2} + m^{2},$$
results in the Klein-Gordon equation \cite{A+H}:
\begin{eqnarray}
	-\frac{\partial^{2}}{\partial t^{2}}\phi(x^\mu) = \left(-\nabla^{2} + m^{2}\right)\phi(x^\mu),\nonumber 
\end{eqnarray}
where $\phi(x^\mu)$ is some wave function that depends on the four-vector $x^\mu = (t, \bm{x})$.\\
\\
The Klein-Gordon equation can be expressed in a covariant manner as:
\begin{eqnarray}
	\left(\partial_{\mu}\partial^{\mu} + m^{2}\right)\phi(x^\nu) = 0 \label{KGE}.
\end{eqnarray}
\\
Since $$\partial_{\mu}\partial^{\mu} = \frac{\partial^2}{\partial^2t} - \nabla^2$$ transforms like a scalar under Lorentz transformations \cite{csg}, the wave function, $\phi$, has to describe a scalar particle for Eq. (\ref{KGE}) to be a covariant equation. Therefore plane-wave solutions of the form
\begin{eqnarray}
	\phi(x^\mu) \propto e^{-iEt + i{\bm p\cdot \bm x}} = e^{-ip\cdot x},
\end{eqnarray}
with $p^\mu = (E, \bm p)$ the momentum four-vector and $p\cdot x = p_{\mu}x^{\mu} = Et-{\bm p\cdot \bm x}$, can be solutions of the free particle Klein-Gordon equation \cite{A+H}.\\
Substituting this wave function into the Klein-Gordon equation (\ref{KGE}), the relativistic energy-momentum relation, $$	E^{2} = {\bm p}^{2} + m^{2},$$ is obtained. As such, the possible energies are:
$$
	E = \pm\sqrt{{\bm p}^{2} + m^{2}}\,.
$$
Thus the Klein-Gordon equation allows for positive as well as negative energy solutions. \\
\\
The Klein-Gordon equation has the conserved current $j^{\mu}$ (i.e. $\partial_\mu j^{\mu} = 0$), with $j^{\mu} = (\rho,{\bm j})$, where	
\begin{eqnarray}
	\rho &=& i(\phi^{*}\partial_{0}\phi - \phi\partial_{0}\phi^{*})\nonumber\\
			{\bm j} &=& -i(\phi^{*}\nabla\phi - \phi\nabla\phi^{*}).\nonumber
\end{eqnarray}
If the time-like component of the conserved current ($\rho$) is interpreted as the probability density, the negative energy solution would imply that the probability density will not be positive definite \cite{capri}.\\
\\
Historically the possibility of negative energy solutions and the probability not being positive definite led to the Klein-Gordon equation being abandoned as a physical description \cite{A+H}. However, the Klein-Gordon equation was redeemed when a physical interpretation was given to the negative energy solutions (see Sec.\,\ref{negE}).
%
%
%
%
%
%
%
%
%
%
\section{Dirac equation}
P. A. M. Dirac wanted to formulate an equation based on the relativistic energy-momentum relationship that is linear in spatial and time derivatives in an effort to overcome the difficulties of the Klein-Gordon equation, especially the positive definiteness of the probability density \cite{capri}. \\
\\
One way to obtain the Dirac equation, using the relativistic energy-momentum relationship, is by making the following substitution \cite{guidry}:
$$
 E= \sqrt{p^2 + m^{2}} \rightarrow {\bm \alpha\cdot \textbf{p}} + \beta m,
$$
where {$\bm \alpha$} and $\beta$ need to be determined, but are assumed to be constants, independent of spacetime and to commute with position and momentum operators \cite{landau}.\\
Making the substitutions of the canonical quantization (\ref{canquant}), the above expression reduces to the Dirac-equation
\begin{eqnarray}
	 i\frac{\partial \psi(x)}{\partial t} = (-i{\bm \alpha\cdot\nabla } + \beta m)\psi(x)\,.  \label{DE} 
\end{eqnarray}
As is discussed in Ref.\,\cite{drell} the properties of the Dirac equation (\ref{DE}) can be derived by considering certain requirements: Since Eq. (\ref{DE}) must be invariant under spatial rotations, ${\bm \alpha}$ and $\beta$ cannot simply be numbers. If it is considered that the probability density, $\psi^*\psi$, must be the time component of a conserved four-vector if integrated over all space at constant time, Dirac proposed that the wave function must be analogous to a $N$-component spin wave function,
\begin{eqnarray}
	\psi = \left[
						\begin{array}{c}
								\psi_1\\\vdots\\
								\psi_N	 
							\end{array}	
				\right].		
\end{eqnarray}
The column vectors satisfying the Dirac equation are called Dirac spinors. Therefore $\bm{\alpha}$ and $\beta$ must be $N\times N$ matrices. Since the Dirac equation (\ref{DE}) must satisfy the relativistic energy-momentum relationship, the following properties can be derived:
\begin{subequations}\label{covargamma}
		\begin{eqnarray}
				\left\{\alpha_{i},\beta \right\}= 0													&&	  i=1,2,3\\
				\left\{\alpha_{i},\alpha_{j}\right\} = 2\delta_{ij}{\bm 1} &&		i,j = 1,2,3.
		\end{eqnarray}
\end{subequations}
Since the Hamilton operator, in this case
$$
	{\bm \alpha\cdot \textbf{p}} + \beta m,
$$
must be hermitian it implies that $\bm{\alpha}$ and $\beta$ must also be hermitian \cite{landau}. Thus $\bm{\alpha}$ and $\beta$ both must be traceless, even dimensional matrices with eigenvalues of $\pm 1$.\\
For $N$ = 2 the conditions of (\ref{covargamma}) are satisfied by the Pauli-matrices and the identity matrix, but not all these matrices are traceless. Thus the lowest dimensionality that satisfies all conditions for $\bm{\alpha}$ and $\beta$ is $N$ = 4 \cite{drell,capri}. One representation of these matrices is known is the Pauli-Dirac representation, given by
:
\begin{subequations}
	\begin{eqnarray}
	\alpha_i &=& 
				\left[
						\begin{array}{cc}
							0      &   \sigma_i\\
							\sigma_i &   0				 
						\end{array}	
					\right]	\label{pd1}\\		
		\beta &=& 
				\left[
					\begin{array}{cc}
						{\bm 1}_2 &   0\\
						0       &   -{\bm 1}_2				 
					\end{array}	
				\right]
				 = 
				 \left[
					\begin{array}{cccc}
						1 &   0&0&0\\
						0       & 1&0&0\\
						0&0&-1&0\\
						0&0&0&-1
					\end{array}	
				\right],\label{pd2}
	\end{eqnarray}
\end{subequations}
	where
\begin{subequations}\label{pauli-spin}
	\begin{eqnarray}
			\sigma_{1} &=& 
					\left[
						\begin{array}{cc}
								0 & 1\\
								1 & 0				 
						\end{array}	
					\right]	,		\label{pauli1}\\
			\sigma_{2} &=&
					\left[\begin{array}{cc}
								0   & -i\\
								i  &  0				 
							\end{array}	
						\right],\ \mbox{and},\label{pauli2}\\
			\sigma_{3} &=&
					\left[\begin{array}{cc}
								1  &   0\\
								0  &  -1				 
							\end{array}	
						\right]\label{pauli3}			
	\end{eqnarray}
\end{subequations}
are the Pauli-matrices.\\ 
\\
In the non-relativistic limit the Dirac equation reduces to the Pauli equation describing spin-$\frac{1}{2}$ \cite{drell}. Thus the free particle solutions of the Dirac equation can be found in analogy to that of a non-relativistic spin-$\frac{1}{2}$ particle, where the wave functions consist of a two component spinor and a plane-wave function \cite{A+H}. A possible solution to the Dirac equation (\ref{DE}) 
is \cite{A+H} 
\begin{eqnarray}
	\psi(x) = \omega e^{-ip\cdot x}\label{freeDEsol}
\end{eqnarray}
where $x^\mu$ and $p^\mu$\, are spacetime and momentum four-vectors and $\omega$ is a four component column vector written as
$$
	\omega = 
		\left(
			\begin{array}{c}
				\phi\\
				\chi
			\end{array}	
		\right)
$$with $\phi$ and $\chi$ two component spinors.\\
Solving for $\psi$, Eq. (\ref{DE}) reduces to:
\begin{eqnarray}
	E	\left(
			\begin{array}{c}
				\phi\\
				\chi
			\end{array}	
		\right)
	= 
		\begin{tabular}{rl}
			$\left(
				\begin{array}{cc}
					m{\bm 1}_2      				&   	{\bm \sigma}\cdot{\bm p}\\
					{\bm \sigma}\cdot{\bm p}    &   	-m{\bm 1}_2			 
				\end{array}	
			\right)$
		&
			$\left(
				\begin{array}{l}
					\phi\\
					\chi
				\end{array}	
			\right)$
	\end{tabular}\nonumber 
\end{eqnarray}
thus obtaining coupled equations for $\phi$ and $\chi$:
\begin{subequations}
	\begin{eqnarray}
		(E - m)\,\phi =  {\bm \sigma}\cdot{\bm p}\,\chi\label{eerste}\\
		(E + m)\,\chi = {\bm \sigma}\cdot{\bm p}\,\phi\ .\label{tweede}
	\end{eqnarray}
\end{subequations}
\\
From Eq. (\ref{tweede}), $\omega$ is given by:
\begin{eqnarray}
	\omega  =  
			\left(
				\begin{tabular}{c}
					$\phi$\\
					$\frac{{\bm \sigma}\cdot{\bm p}}{E + m}\phi	$
				\end{tabular}	\nonumber
			\right).
\end{eqnarray}
When the solution for $\chi$ [from Eq. (\ref{tweede})] is substituted into Eq. (\ref{eerste})
, it can be shown that \cite{A+H}:
$$
	(E - m)(E + m)\phi = {\bm p}^2\phi,
$$
which implies that
$$
	E = \pm\sqrt{{\bm p}^2 + m^2},
$$
thus allowing positive and negative energy solutions to the Dirac equation \cite{A+H}.\\
\\
To be able to write the Dirac equation in a covariant form, the $\gamma$-matrices are defined as:
\begin{subequations}\label{gammamatrices}
	\begin{eqnarray}
		\gamma^{0} &=& \beta\\
		\gamma^{i} &=& \beta\alpha^{i} = \gamma^{0}\alpha^{i}.
	\end{eqnarray}
\end{subequations}
The properties of the $\gamma$-matrices in the Pauli-Dirac representation are derived from the properties of $\bm{\alpha}$ and $\beta$ (\ref{covargamma}) and that of the Pauli-spin matrices (\ref{pauli-spin}):
\begin{subequations}
		\begin{eqnarray}
				\{\gamma^{\mu}, \gamma^{\nu}\} &=& 2\eta^{\mu\nu}{\bm 1}_4\,,\nonumber\\
				(\gamma^{i})^{2} &=& -{\bm 1}_4\ \mbox{for} \ i=1,2,3\,,\nonumber\\
				(\gamma^{0})^{2} &=& {\bm 1}_4 \,,\nonumber
		\end{eqnarray}
\end{subequations}
and the adjoints of the $\gamma$-matrices are
\begin{eqnarray}
	\gamma^{\mu\dagger} = \gamma^{0}\gamma^{\mu}\gamma^{0}.\nonumber
\end{eqnarray}
The Dirac equation (\ref{DE}) can be written in a covariant form, by multiplying it from the left by $\gamma^{0}$, as:
\begin{eqnarray}
	(i\gamma^{\mu}\partial_{\mu} - m)\psi(x)= 0\,. \label{covarDE}
\end{eqnarray}\\
Eq. (\ref{covarDE}) will only be covariant under a Lorentz transformation,
$$	x^\mu \stackrel{\lambda}{\longrightarrow} (x^\nu)' = \Lambda^{\nu}_{\mu}x^\mu\,,$$
if the wave function, $\psi(x)$ transforms as
$$	\psi(x) \stackrel{S(\Lambda)}{\longrightarrow} \psi'(x') = S(\Lambda)\psi(x)\,,$$
where $S(\Lambda)$ is some matrix \cite{capri}.\\ 
\\
As shown in Refs \cite{drell} and \cite{capri}, $S(\Lambda)$ must satisfy:
$$
	S^{-1}(\Lambda)\gamma^{\mu}\Lambda^{\nu}_{\mu}S(\Lambda) = \gamma^{\nu}\,.
$$
Defining the Dirac adjoint $\bar{\psi}$ as
\begin{eqnarray}
	\bar{\psi} = \psi^{\dagger}(x)\gamma^{0},\label{Dadjoint}
\end{eqnarray}
it is also shown in Ref.\,\cite{drell} that $\bar{\psi}$ transforms under a Lorentz transformation as
$$
	\bar{\psi'}(x') = \bar{\psi}(x)S^{-1}(\Lambda)\,.
$$\\
The Dirac equation has a conserved current, $j^{\mu}$, 
$$	j^{\mu} = \bar{\psi}\gamma^{\mu}\psi, $$ 
of which the time-like component, $\rho$, is given by
$$	\rho = \bar{\psi}\gamma^{0}\psi = \psi^\dagger\psi$$
so that
$$	\rho = \sum^4_{i = 1}\left|\psi_i\right|^2>0\,.$$
Thus the probability density of the Dirac equation is positive definite, but negative energy solutions are still allowed \cite{A+H}.
\section{Interpretation of negative energy solutions of the Klein-Gordon and Dirac equations}\label{negE}
There are two interpretations of the negative energy solutions. First P. A. M. Dirac put forward an interpretation of the negative energy solution of the Dirac equation for fermions and then R. P. Feynman added another interpretation that could be applied to fermions as well as bosons (scalar particles, described by the Klein-Gordon equation) \cite{guidry}.
\subsection{Dirac's interpretation}
The Dirac description states that there are positive and negative energy states available to a free spin-$\frac{1}{2}$ particle, such as an electron. These energy levels are symmetric around the zero \mbox{energy} state. To prevent positive energy particles from spontaneously decaying to negative \mbox{energy} states Dirac postulated that in the vacuum state all the negative energy states are filled (the Dirac sea \cite{A+H}). Since all negative energy states are filled, Pauli's Exclusion Principle prevents the decay of a positive energy electron to a negative energy state. \cite{A+H}\\
\\
An electron occupying a negative energy state can be excited to a positive energy state, leaving a hole in the Dirac vacuum. Due to the absence of a negatively charged electron the hole will behave as a positively charged particle with a positive energy with regards to the Fermi sea, predicting the existence of a positron \cite{A+H}.\\
\\
It should be noted that due to this interpretation the Dirac equation no longer describes a single particle state, but a many-particle state, thus necessitating a many-particle description such as quantum field theory to describe relativistic particles \cite{drell}.
\subsection{Feynman's interpretation}
R.P. Feynman interpreted the negative energy state as positive energy particle propagating backwards in time or as an anti-particle propagating forward in time \cite{guidry}.\\
\\
The Feynman's interpretation vindicated the Klein-Gordon equation. Since the Klein-Gordon equation describes bosons, which do not obey the Pauli Exclusion Principle, the Dirac interpretation could not be applied to explain the negative energy solutions of the Klein-Gordon equation.
%
%
%
%
%
%
%

%% file: chapMechanics.tex
\chapter{Lagrangian mechanics and field theory}\label{mechanics}
\section{Lagrangian mechanics}
H. Goldstein \textsl{et al.} explain Lagrangian mechanics in Ref.\,\cite{goldstein} as a description of a physical system in terms of the degrees of freedom of the system. Lagrangian mechanics does not contain any new physics, but is simply an alternative expression of the physical laws governing the equations of motion of objects. Since the description is in terms of the degrees of freedom of the system, the
 formulation is not restricted to just describing classical particles, but can also describe discrete or continuous systems in the classical or relativistic regimes. Lagrangian \mbox{mechanics} is a desirable description since all the information pertaining to the system is contained in one function: the Lagrangian.\\
\\ 
The Lagrangian (L) is defined by $$L = T - V$$ where,
\begin{itemize}
	\item $T$ is the kinetic energy of the system and
	\item $V$ the potential energy of the system,
\end{itemize}
and can thus be seen as an expression of the energy of a system. Each degree of the $N$ degrees of freedom of the system is described by a generalised coordinate $q_{\alpha}$, with $\alpha = 1,2,\ldots,N$. The Lagrangian is given in terms of the generalised independent coordinates and the time-derivates of these, $\dot{q}_{\alpha}$ and time, $t$.  \\
\\
The generalised momentum (conjugate momentum) of each generalised coordinate is $p_\alpha$ with
\begin{eqnarray}
	p_\alpha = \frac{\partial L}{\partial \dot{q}_\alpha}.\label{conmoment1}
\end{eqnarray}
The evolution of the system is governed by Hamilton's principle. It 
states that the system will evolve over a time interval $[t_1, t_2]$ in such a way that the action ($I$) is stationary. The action is the line integral of the Lagrangian between time $t_1$ and time $t_2$. If the action is stationary the variation of the action will be equal to zero, i.e.
$$0 = \delta I = \delta \int_{t_1}^{t_2}L(q,\dot q, t)dt\,.$$
\\
The equation of motion of a specific degree of freedom described by the coordinate, $q_{\alpha}$, is given by the Euler-Lagrange equation \cite{goldstein}:
$$\frac{d}{dt}\left(\frac{\partial L}{\partial \dot q_{\alpha}}\right) - \frac{\partial L}{\partial q_{\alpha}} = 0\,.$$ 
The Euler-Lagrange equation follows from the application of Hamilton's principle with regards to a specific coordinate \cite{goldstein}.
%
%
%
%
%
%
\section{Hamiltonian formulation}
The Hamiltonian formulation is an alternative way to express Lagrangian mechanics. The Hamiltonian formulation has found various extension in physics and is therefore also briefly discussed.\\
\\
The essential difference between the Lagrangian and the Hamiltonian formulation is that in the former the Lagrangian is expressed in terms of the generalised coordinates, the time-derivatives of the generalised coordinates and time itself, ($q, \dot{q}, t$). In the Hamiltonian formulation a new quantity, the Hamiltonian, $H$, is defined and expressed in terms of the generalised coordinates, the conjugate momenta of the generalised coordinates and time, ($q, p, t$). Thus on a mathematical level, the switch from the Lagrangian to the Hamiltonian formulation can be seen as a Legendre transformation \cite{goldstein}. \\
\\
The Hamiltonian is defined by (with the sum over $\alpha$ implied) \cite{goldstein, jose}:
\begin{eqnarray}
	H(q, p, t) \equiv \dot{q}_\alpha p_\alpha - L(q, \dot{q}, t).
\end{eqnarray}
Anagolous to the interpretation of $L$, 
$H$ can be seen as an expression of the total energy of the system. If $L$ is $L = T - V$ and if the momentum times the speed of each degree of freedom summed over the whole system is twice the kinetic energy ($T$) of the system, $H$ is given as $H = T + V$ \cite{jose}. 
%
%
%
%
%
%
%
\section{Field theory}
In Ref.\,\cite{goldstein} a field is defined as set of one or more functions of space and time. These functions can be used to describe the displacement and velocity of points in a (continuous) system \cite{jose}. The formalism of previous section can be applied to describe continuous systems using fields, by replacing 
the discrete coordinates, $q_\alpha$, by fields, $\phi_\alpha$. Accordingly the fields $\phi_\alpha$ with $\alpha = 1,2,\ldots,N$, describe the degrees of freedom of the system.\\
\\
In general, all the fields described in this work will be in 4-dimensional Minkowski space [see Eq. (\ref{minkowski}) for the metric of this space] and thus the fields will depend on spatial and time coordinates, i.e.
$$\phi_\alpha(t,x,y,z) = \phi_\alpha(t,{\bm x}) = \phi_\alpha(x^\mu).$$
\\
In field theory 
the Lagrangian would be defined in terms of the Lagrangian density ($\cal L$):	
\begin{eqnarray}
	L &=& \int {\cal L}\,d{\bm x}\nonumber\\
	&=& \int{\cal L}\left(\phi_\alpha(t,{\bm x}),\frac{d\phi_\alpha(t,{\bm x})}{dt}, 
			\frac{d\phi_\alpha(t,{\bm x})}{d{\bm x}}, {\bm x},t\right)d{\bm x}\nonumber\\
	&=& \int{\cal L}(\phi_\alpha(x^\mu),\frac{\partial \phi_\alpha(x^\mu)}{\partial x^\nu},x^\mu)\,d{\bm x}\nonumber\\
	&=& \int{\cal L}(\phi_\alpha(x),\partial_\nu(\phi_\alpha(x)),x)\,d{\bm x}\,.\label{densitylagrange}
\end{eqnarray}
The Euler-Lagrange equations can be generalised (see for instance Refs \cite{goldstein} and \cite{jose}) to directly apply to the Lagrangian density, i.e.
\begin{eqnarray}
	\partial_\nu\left(\frac{\partial {\cal L}}{\partial (\partial_\nu \phi_\alpha)}\right) - 
	\frac{\partial {\cal L}}{\partial \phi_\alpha} = 0\label{EL}.
\end{eqnarray}
For the field $\phi_\alpha$ the conjugate momentum of (\ref{conmoment1}) is defined as $\pi^\alpha$ \cite{goldstein}
\begin{eqnarray}
	\pi^\alpha = \frac{\partial {\cal L}}{\partial \dot{\phi}_\alpha}\ .\label{conjugmom}
\end{eqnarray}
The Hamiltonian is defined as:
\begin{eqnarray}
	H &=& \int {\cal H}\,d{\bm x}\nonumber\\
		&=& \int \left(\frac{\partial {\cal L}}{\partial \dot{\phi}_\alpha}\dot{\phi}_\alpha - 
		{\cal L}\right)d{\bm x}\nonumber\\
		&=& \int \left(\pi^\alpha\dot{\phi}_\alpha - {\cal L}\right)d{\bm x}\,,\label{hamil}
\end{eqnarray}
and thus the Hamiltonian density ${\cal H}$ is given by
\begin{eqnarray}
	\cal H = \pi^\alpha\dot{\phi}_\alpha - {\cal L}\,.\label{hamildensity}
\end{eqnarray}
%
%
%
%
%
%
%
%
%
\section{Noether's theorem}\label{secnoether}
Noether's theorem describes the relation between the invariance of the Lagrangian density under certain transformations and conserved quantities (currents).\\
\\
The transformation could be a coordinate transformation,
\begin{eqnarray}
	x^{\mu} \rightarrow x'^{\mu} = x^{\mu} + \delta x^{\mu}\label{codelta}
\end{eqnarray}
or a field transformation,
\begin{eqnarray}
		\phi_\alpha(x) \rightarrow \phi'_\alpha(x') = \phi_\alpha(x) + 
	\delta \phi_\alpha(x).\label{fielddelta}
\end{eqnarray}
The theorem states that
\begin{itemize}
	\item if the space is flat, and,
	\item if the coordinate/field transformation is continuous and 
				the transformation can be written in terms of the coordinates as
				\begin{eqnarray}
					\delta x^{\mu} = \epsilon_rX^{\mu}_{r}\label{cofunc},
				\end{eqnarray}
				and in terms of the fields as 
				\begin{eqnarray}
					\delta \phi_\alpha = \epsilon_r\Psi_{r\alpha}\label{fieldfunc},
				\end{eqnarray}
				where the functions $X^{\mu}_{r}$ and $\Psi_{r\rho}$ depend on the coordinate/field variables and $\epsilon$ 					is an infinitesimal parameter ($r$ runs from $1,2,\dots,R$), and,
	\item if the Lagrangian density is form- and scale invariant under these transformations, 			
\end{itemize}
then the following will hold \cite{goldstein}:
\begin{eqnarray}
	\frac{\partial}{\partial x^\nu}\left\{\left(\frac{\partial{\cal L}}{\partial\, 
	\frac{\partial \phi_\alpha(x^\sigma)}{\partial x^\nu} }
	\,\frac{\partial\phi_\alpha(x^\sigma)}{\partial x^\sigma}  - 
	{\cal L}\delta^{\nu}_{\sigma}\right)X^\sigma_r 
	- \frac{\partial{\cal L}}{\partial\, \frac{\partial \phi_\alpha(x^\sigma)}{\partial x^\nu}}\Psi_{\alpha r}
	\right\} = 0\ ,\label{noether}
\end{eqnarray}
so that there are $R$
conserved currents of the form: 
\begin{eqnarray}
	\left(\frac{\partial{\cal L}}{\partial\, \partial_\nu \phi_\alpha(x)}
	\,\partial_\sigma \phi_\alpha(x) - 
	{\cal L}\delta^{\nu}_{\sigma}\right)X^\sigma_r 
	- \frac{\partial{\cal L}}{\partial\, \partial_\nu \phi_\alpha(x)}\Psi_{\alpha r}\ .\nonumber
\end{eqnarray}
\\
Form invariance of the Lagrangian density under a transformation implies that 
	\begin{eqnarray}
		{\cal L'}\left( \phi'_\alpha(x'), \partial_\nu \phi'_\alpha(x'),x'\right) = 
		{\cal L}\left(\phi'_\alpha(x'), \partial_\nu \phi'_\alpha(x'),x'\right),\nonumber
	\end{eqnarray}
so that the effect of the transformation on the Lagrangian density would be
	\begin{eqnarray}
		{\cal L}\left( \phi_\alpha(x), \partial_\nu \phi_\alpha(x),x^{}_{}\right) \rightarrow 
		{\cal L}\left(\phi'_\alpha(x'), \partial_\nu \phi'_\alpha(x'),x'^{}_{}\right).\nonumber
	\end{eqnarray}
\\
Scale invariance implies that the magnitude of the action integral is also assumed to be invariant under the transformation, i.e.
	\begin{eqnarray}
		 I' &=& \int_{\Omega'}(dx'^4){\cal L}\left(\phi'_\alpha(x'^{\mu}), 
		 \partial_\nu \phi'_\alpha(x'^{\mu}),x'^{\mu}\right)\nonumber\\
		&=& \int_{\Omega}(dx^4){\cal L}
		\left( \phi_\alpha(x^{\mu}), \partial_\nu \phi_\alpha(x^{\mu}),x^{\mu}_{}\right)\,.\nonumber
	\end{eqnarray}\\
A proof of Noether's theorem is given in amongst other Refs \cite{goldstein} and \cite{greiner}.

%
%
%
%
%
%
%
%
%
\subsection{Energy-momentum tensor as conserved quantity}
A typical symmetry of the Lagrangian density is translational invariance. This means that the Lagrangian density describing the system is independent of the spatial and temporal origin of the system. With Noether's theorem one can show that translational invariance imply that the energy-momentum tensor is a conserved quantity (also see Sec. \ref{sec:EnergyMomentumTensor}) \cite{csg,goldstein,greiner}.\\
\\
Consider the following spacetime translation
\begin{eqnarray}
	x^\mu \rightarrow x'^\mu &=& x^\mu + \delta x^\mu\nonumber\\
	 &=& x^\mu + \epsilon_r h_r^\mu\nonumber\ ,
\end{eqnarray}	
where $h_r^\mu$ is an arbitrary constant and $\epsilon_r$ as in (\ref{cofunc}). Since the Lagrangian density (and therefore the fields) are translational invariant, the coordinate change would have no effect on the fields and $\delta \phi_\nu(x^{\mu})$ of Eq. (\ref{fielddelta}) would be zero and therefore also $\Psi_{r\nu}$ of Eq. (\ref{fieldfunc}).\\
\\
From Noether's theorem (\ref{noether}) it therefore follows that
\begin{eqnarray}
	0 &=& \frac{\partial}{\partial x^\nu}\left\{\left(\frac{\partial{\cal L}}{\partial(\partial_\nu \phi_\alpha)}
	(\partial_\sigma \phi_\alpha) - {\cal L}\delta^{\nu}_{\sigma}\right)
	h^\sigma_r\right\}\nonumber \\
	&=& \partial_\nu\left(T^\nu_\sigma h^\sigma\right).\nonumber
\end{eqnarray}
Since $h^\sigma$ is arbitrary, it follows that
\begin{eqnarray}
	\partial_\nu T^\nu_\sigma = 
	\frac{\partial}{\partial x^\nu}\left(\frac{\partial\cal{L}}{\partial (\partial_\nu \phi_\alpha)}\,
	\partial_\sigma \phi_\alpha  - \cal{L}\delta^\nu_\sigma \right) = 0\nonumber
\end{eqnarray}
and so the energy-momentum tensor, $T^{\mu\nu}$, is given by
\begin{eqnarray}
	T^{\mu\nu} &=& \frac{\partial\cal{L}}{\partial (\partial_\mu \phi_\alpha)}\,
	\partial^\nu \phi_\alpha - \cal{L}\eta^{\mu\nu}\ . \label{EMT}
\end{eqnarray}
In Chapters \ref{chapQHD1} and \ref{chapFSUG} the fields, $\phi_\alpha$, that will be considered will actually be field operators. Therefore the (ground state) expectation value have to be taken \cite{walecka}:
\begin{eqnarray}
	\left\langle T^{\mu\nu}\right\rangle &=& \left\langle \frac{\partial\cal{L}}{\partial (\partial_\mu \phi_\alpha)}\,
	\partial^\nu \phi_\alpha - \cal{L}\eta^{\mu\nu}\ \right\rangle \nonumber\\
	&=& -P\eta^{\mu\nu} + (P + \epsilon)u^\mu u^\nu\nonumber,
\end{eqnarray}
if a static, spherical symmetric fluid, moving with velocity $\textbf{v}$ is consider [from Eq. (\ref{genEMT})]. If $\textbf{v} = 0$, then 
\begin{subequations}\label{epsp13}
	\begin{eqnarray}
		\epsilon = \left\langle T^{00}\right\rangle,\\
		P = \frac{1}{3}\left\langle T^{ii}\right\rangle.
	\end{eqnarray}
\end{subequations}

%% file: chapQFT.tex
\chapter{Quantum field theory}\label{QFT}
Quantum field theory is the application of quantum mechanics to fields. One of the main motivations for this description is the single particle description of the very small scales (quantum mechanical) at high (relativistic) energies gives rise to negative energies, which in the interpretation of Sec.\,\ref{negE} gives rise to many-body states.\\
\\
Thus to describe this problem in terms of fields has some distinct advantages. The first being that field theory is a many-body description in which various degrees of freedom are handled with ease. The Lagrangian density is a Lorentz scalar and therefore invariant under any Lorentz boosts or rotations. Since the extreme value of the action is therefore also invariant, the Euler-Lagrange equation gives rise to Lorentz invariant equations of motion of the fields contained in the Lagrangian. A many-particle theory is also necessary in order to preserve causality as is shown in Ref.\,\cite{P+S}.\\
\\
The quantization procedure is very similar to that of normal quantum mechanics, except that in this case it is not the particles that are quantized, but fields describing the particles.
\section{Quantization of a field}
For the quantization of the fields, $\phi_r(x)$, the field and the conjugate fields, $\pi_s(x)$, are promoted to field operators, i.e. 
\begin{subequations}\label{fieldquant}
	\begin{eqnarray}
		\phi_r(x)\longrightarrow \hat{\phi}_r(x)\,,\\
		\pi_s(x)\longrightarrow \hat{\pi}_s(x).
	\end{eqnarray}
\end{subequations}
These operators are Heisenberg operators and the following equal-time commutation relationship are imposed \cite{M+S}:
\begin{subequations}\label{ETCR}
	\begin{eqnarray}
		\left[ \, \hat{\phi}_r(t,{\bm x}),\hat{ \pi}_s(t,{\bm x'})\,\right] = 
			i\hbar\delta_{rs}\,\delta^3({\bm x}-{\bm x'})\,,\label{ETCR1}\\
		\left[\,\hat{\phi}_r(t,{\bm x}), \hat{\phi}_r(t,{\bm x'})\,\right] = 
			\left[ \,\hat{\pi}_r(t,{\bm x}), \hat{\pi}_r(t,{\bm x'})\,\right] = 0\,.\label{ETCR2}
		\end{eqnarray}
\end{subequations}
These commutation relations are imposed to ensure that the correct spin statistics are satisfied \cite{P+S}.\\
\\
Since the field operators are time-dependent their dynamics are determined by Heisenberg's equation of motion \cite{greiner}. Heisenberg's equation of motion states that for an operator ${\cal O}(t)$ the time-evolution of the operator is given by
\begin{eqnarray}
	i\hbar\frac{d{\cal O}(t)}{dt} = [{\cal O}(t), H]\label{heisEQM},
\end{eqnarray}
where $H$ is the Hamiltonian (\ref{hamil}). \\
\\
The field operators are defined in the space defined by the state vectors of the particles described by the field. To obtain an explicit expression of these operators they are Fourier decomposed in terms of a complete set of wave functions in the space \cite{greiner}:
\begin{subequations}\label{opfourier}
	\begin{eqnarray}
		\hat{\phi}_r(t,{\bm x}) = \sum_{i}\hat{a}_{ri}(t)\,u_{ri}({\bm x})\,,\\
		\hat{\phi}^\dagger_r(t,{\bm x}) = \sum_{i}\hat{a}^\dagger_{ri}(t)\,u^{\ast}_{ri}({\bm x})\,.
		\end{eqnarray}
\end{subequations}
The operator properties and the time-dependence of $\hat{\phi}_r(t,{\bm x})$ are carried by the expansion coefficients, $\hat{a}_{ri}(t)$ \cite{greiner}.
\section{Relativistic Scalar fields}
From Chapter \ref{chapRQM}, relativistic, spin-0 (scalar) particles are described by the Klein-Gordon \mbox{equation.} The Lagrangian density that describes a real scalar field, $\phi(x)$, with particles of mass $m$, is (in natural units) \cite{greiner}:

\begin{eqnarray}
	{\cal L} &=& \frac{1}{2}\frac{\partial\phi(x^\nu)}{\partial x_{\mu}}\frac{\partial\phi(x^\nu)}{\partial x^{\mu}}
	 - \frac{1}{2}m^{2}\phi(x^\nu)^{2}\nonumber\\
	 &=& \frac{1}{2}\partial^{\mu}\phi(x)\partial_{\mu}\phi(x)
	 - \frac{1}{2}m^{2}\phi(x)^{2}.\nonumber
\end{eqnarray}
The equation of motion of the field is derived by applying the Euler-Lagrange equation (\ref{EL}), and is
\begin{eqnarray}
	\partial_{\mu}\partial^{\mu}\phi + m^{2}\phi = 0.\label{fieldKGE}
\end{eqnarray}
The conjugate field to $\phi(x)$ is defined analogous to the conjugate momentum of a classical field (\ref{conjugmom}) by
\begin{eqnarray}
	\pi \equiv \frac{\partial{\cal L}}{\partial\dot{\phi}} = \dot{\phi}\,,\label{conjugfield}
\end{eqnarray}
and the Hamiltonian density (\ref{hamildensity}) is
\begin{eqnarray}
	{\cal H} &=& \pi\dot{\phi} - {\cal L}\nonumber\\
	&=& \frac{1}{2}\left(\pi^2 + ({\bm \nabla}\phi)^2 +
	 m^2\phi^2\right).\nonumber
\end{eqnarray}
\subsection{Quantization}
In canonical field quantization the following substitutions are made
\begin{eqnarray}
	\phi(x)\longrightarrow \hat{\phi}(x)\,,\nonumber\\
	\pi(x)\longrightarrow \hat{\pi}(x)\,,\nonumber
\end{eqnarray}
and the operators have to satisfy the following \textsl{equal-time commutation relationships}:
\begin{subequations}\label{scalarETCR}
	\begin{eqnarray}
			\left[\hat{\phi}((t,{\bm x}),\ \hat{\pi}(t,{\bm x'})\right] &=& i\delta({\bm x} - {\bm x'})\\
			\left[\hat{\phi}(t,{\bm x}),\ \hat{\phi}(t,{\bm x'})\right] &=& 
			\Big[\hat{\pi}(t,{\bm x}),\ \hat{\pi}(t,{\bm x'})\Big]
			= 0
	\end{eqnarray}
\end{subequations}
It is shown in Ref.\,\cite{greiner} that the scalar field has to satisfy the above commutation relations, \mbox{otherwise} the principle of microcausality will be violated.\\
\\
The Hamiltonian of the quantized field is simply found to be \cite{greiner}
\begin{eqnarray}
	\hat{H} &=& \frac{1}{2}\int d{\bm p}\ \left(\hat{\pi}^2 + ({\bm \nabla}\hat{\phi})^2 +
	 m^2\hat{\phi}^2\right)\,.\label{KGEH}
\end{eqnarray}
\subsection{Expansion of operators}
The equation of motion of $\hat{\phi}(x)$ follows from Heisenberg's equation (\ref{heisEQM}) (keeping in mind that $\hat{\pi}(x) = \dot{\hat{\phi}}(x)$),
\begin{eqnarray}
	\ddot{\hat{\phi}}(t, {\bm x}) = ({\bm \nabla}^2 - m^2)\,\hat{\phi}(t, {\bm x})\,,\label{opKGE}
\end{eqnarray}
which is just the free Klein-Gordon equation (\ref{KGE}). Expanding the operator in terms of the plane wave basis for the operator space, the wave function in Fourier decomposition of the operator (\ref{opfourier}) becomes
$$ u_{\bm p}(t, {\bm x}) = N_p\ e^{i{\bm p}\cdot{\bm x}}$$ where $N_p$ is just a normalisation constant. Thus $\hat{\phi}(t, {\bm x})$ can be written as \cite{greiner}:
\begin{eqnarray}
	\hat{\phi}(t, {\bm x}) = \int d{\bm p}\ N_p\ e^{i{\bm p}\cdot{\bm x}}\ \hat{a}_{\bm p}(t).\label{scalar1}
\end{eqnarray}
Substituting the above equation back into Eq. (\ref{opKGE}), it is found, as shown in Ref.\,\cite{greiner}, that 
$$
	\ddot{\hat{a}}_{\bm p}(t) = -({\bm p}^2 + m^2)\,\hat{a}_{\bm p}(t)
$$ 
which means that solutions for $\hat{a}_{\bm p}(t)$ are given by
\begin{eqnarray}
	\hat{a}_{\bm p}(t) = \hat{a}^{(1)}_{\bm p}\,e^{-i\omega_pt} + \hat{a}^{(2)}_{\bm p}\,e^{+i\omega_pt},
\end{eqnarray}
where $\omega_p$ is $$\omega_p = \sqrt{{\bm p}^2 + m^2}\,,$$ and $\hat{a}^{(1)}_{\bm p}$ and 
$\hat{a}^{(2)}_{\bm p}$ are constants in time \cite{greiner}. Since the original $\phi(x)$ was a real-valued classical function, the operator should be hermitian ($\hat{\phi}^\dagger = \hat{\phi}$), and therefore one of the operators can be expressed in terms of the other as $$\left(\hat{a}^{(1)}_{\bm p}\right)^\dagger = \hat{a}^{(2)}_{-{\bm p}}\,.$$
\\
Thus the free particle solution for the quantized scalar field (\ref{scalar1}) becomes
\begin{eqnarray}
	\hat{\phi}(t, {\bm x}) = \int d{\bm p}\ N_p\left(\hat{a}_{\bm p}\,e^{i({\bm p}\cdot{\bm x}-\omega_pt)} +
	 \hat{a}^\dagger_{\bm p}\,e^{-i({\bm p}\cdot{\bm x}-\omega_pt)}\right)\,,\label{scalar2}
\end{eqnarray}
and the conjugate field operator, $\hat{\pi} = \dot{\hat{\phi}}$, as
\begin{eqnarray}
	\hat{\pi}(t, {\bm x}) = \int d{\bm p}\ N_p\,(-i\omega_p)\left(\hat{a}_{\bm p}\,e^{i({\bm p}\cdot{\bm x}-\omega_pt)} -
	 \hat{a}^\dagger_{\bm p}\,e^{-i({\bm p}\cdot{\bm x}-\omega_pt)}\right)\,.\label{scalar3}
\end{eqnarray}
As is shown in Ref.\,\cite{greiner}, by substituting the expressions for the field operator (\ref{scalar2}) and the conjugate field operator (\ref{scalar3}) back into the commutation relations (\ref{scalarETCR}) the commutation relations for $\hat{a}_p$ and $\hat{a}^\dagger_p$ can be obtained and are
\begin{subequations}\label{scalaropcom}
	\begin{eqnarray}
			\left[\hat{a}_{\bm p},\ \hat{a}^\dagger_{{\bm p}'}\right] &=& i\delta({\bm p} - {\bm p}')\,,\\
			\Big[\hat{a}_{\bm p},\ \hat{a}_{{\bm p}'}\Big] &=& 
			\Big[\hat{a}^\dagger_{\bm p},\ \hat{a}^\dagger_{{\bm p}'}\Big]= 0\,.
	\end{eqnarray}
\end{subequations}
Thus $\hat{a}^\dagger_{\bm p}$ and $\hat{a}_{\bm p}$ satisfy the commutation relationship for creation and annihilation operators for scalar particles (bosons) \cite{P+S}. Scalar particles are described by symmetric states. The states that the field operators will operate on are clearly symmetric from the above commutation relationships, since for an arbitrary state $\left|\Phi\right\rangle$
$$
	\hat{a}^\dagger_{\bm q}\,\hat{a}^\dagger_{\bm p}\left|\Phi\right\rangle = 
	\hat{a}^\dagger_{\bm p}\,\hat{a}^\dagger_{\bm q}\left|\Phi\right\rangle\,.
$$
The Hamiltonian is constructed from Eq. (\ref{KGEH}), by replacing the field operators by their \mbox{expansions} [Eqs (\ref{scalar2}) and (\ref{scalar3})]. After some algebra (as is shown in Ref.\,\cite{greiner}) the Hamiltonian is expressed as
\begin{eqnarray}
	\hat{H} &=& \frac{1}{2}\int d{\bm p}\ \omega_p\left(\hat{a}^\dagger_{\bm p}\,\hat{a}_{\bm p} + 
	\hat{a}_{\bm p}\hat{a}^\dagger_{\bm p}\right)\nonumber\\
	&=& \frac{1}{2}\int d{\bm p}\ \omega_p\left(2\hat{a}^\dagger_{\bm p}\,\hat{a}_{\bm p} + 
	\delta^3({\bm p}-{\bm p'})\right)\nonumber\\
	&=& \int d{\bm p}\ \omega_p\left(\hat{a}^\dagger_{\bm p}\,\hat{a}_{\bm p} + 
	\frac{1}{2}\delta^3({\bm p}-{\bm p'})\right)\,.\label{scalarhamil1}
\end{eqnarray}
If the above Hamiltonian operates on any state, even the vacuum, it will have infinite energy, due to the contribution of the last term in Eq. (\ref{scalarhamil1}). This contribution is attributed to the filled negative energy states that defines the vacuum \cite{greiner}. To remove this constant contribution, Hamiltonian is defined to contain the normal-ordered product of the operators. Normal-ordering entails that all creation operators have to stand on the left and all the annihilation operators on the right [when operators are switched around the commutation relationships (\ref{scalaropcom}) have to be taken into account]. The Hamiltonian, as the normal-ordered product of the field operators, is:
\begin{eqnarray}
	\hat{H} &=& \frac{1}{2}\int d{\bm p}\ {\bm :}\left(\hat{\pi}^2 + ({\bm \nabla}\hat{\phi})^2 +
	 m^2\hat{\phi}^2\right){\bm :}\nonumber\\
	&=& \int d{\bm p}\ \omega_p\left(\hat{a}^\dagger_{\bm p}\,\hat{a}_{\bm p}\right)\,.\nonumber
\end{eqnarray}
\section{Dirac fields}\label{secDEfields}
To describe a field of Dirac (spin-$\frac{1}{2}$) particles the ansatz is that the Lagrangian density for the field is given by
\begin{eqnarray}
	{\cal L} = i\psi^{\dagger}(x)\frac{\partial}{\partial t}\psi(x) + i\psi^{\dagger}(x)\alpha\cdot\nabla\psi(x)
							- m\psi^{\dagger}(x)\beta\psi(x),\nonumber
\end{eqnarray}
where $\psi(x)$ and the conjugate field $\psi^\dagger(x)$ [see (\ref{DEconjugate})] are classical fields describing the Dirac particles with mass $m$.
This Lagrangian density can be written covariantly using the definition of the $\gamma$-matrices (\ref{gammamatrices}) as well as the Dirac adjoint (\ref{Dadjoint}), as
\begin{eqnarray}
	{\cal L} = \bar{\psi}(x)\left[\,i\gamma^\mu \partial_\mu \psi(x) - m\right]\psi(x).\nonumber
\end{eqnarray}
Again the equation of motion can be derived using the Euler-Lagrange equation (\ref{EL}) and is
\begin{eqnarray}
	\left[ \,i\gamma^\mu \partial_\mu - m\right]\psi(x) = 0\,.\label{fieldDE}
\end{eqnarray}
The conjugate field is 
\begin{eqnarray}
	\pi = \frac{\partial{\cal L}}{\partial\dot{\psi}} = i\psi^\dagger\label{DEconjugate}
\end{eqnarray}
and the Hamiltonian density is
\begin{eqnarray}
	{\cal H} &=& \pi\dot{\phi} - {\cal L}\nonumber\\
		&=& \psi^\dagger(x)(-i{\bm \alpha}\cdot{\bm \nabla} + \beta m)\psi\,.\nonumber
\end{eqnarray}
\subsection{Quantization}
The Dirac-fields are quantized by replacing the Dirac-spinors by field operators,
\begin{eqnarray}
	\psi(x)\longrightarrow \hat{\psi}(x)\,,\nonumber\\
	\psi^{\dagger}(x)\longrightarrow \hat{\psi}^{\dagger}(x)\nonumber,
\end{eqnarray}
where the field operators satisfy the following \emph{equal-time anticommutation relationships},
\begin{subequations}\label{DEETCR}
	\begin{eqnarray}
		\left\{\hat{\psi}_{\alpha}(t,{\bm x}),\ \hat{\psi}_{\beta}^{\dagger}(t,{\bm x}')\right\} &=&
		 \delta_{\alpha\beta}\delta({\bm x} - {\bm x}'),\\
		\left\{\hat{\psi}_{\alpha}(t,{\bm x}),\ \hat{\psi}_{\beta}(t,{\bm x}')\right\} &=&
		 \left\{\hat{\psi}_{\beta}^{\dagger}(t,{\bm x}),\ \hat{\psi}_{\beta}^{\dagger}(t,{\bm x}')\right\} = 0.
	\end{eqnarray}
\end{subequations}
In contrast to the quantization of the scalar field, the anticommutation relationships of the Dirac field operators are defined so that Fermi-Dirac statistics is obeyed \cite{greiner}.
\subsection{Expansion of operators}
As is shown in Ref.\,\cite{greiner} the quantized field operator satisfies the equation for the Dirac field (\ref{fieldDE})
\begin{eqnarray}
	i\dot{\hat{\psi}}(t, {\bm x}) = (-i{\bm \alpha}\cdot{\bm \nabla} + \beta m)\hat{\psi}(t, {\bm x})\,,\nonumber
\end{eqnarray}
which is
\begin{eqnarray}
	(i\gamma^{\mu}\partial_{\mu} - m)\hat \psi(x)= 0\,,\label{fieldopDE}
\end{eqnarray}
when expressed in a covariant manner.\\
\\
The operator for the Dirac field can be expanded in terms of the solutions to the free Dirac equation (\ref{freeDEsol}). These solutions can be explicitly written (from Ref.\,\cite{greiner}) as
\begin{eqnarray}
	\psi^{(r)}_{\bm p}(t, {\bm x}) = \frac{1}{(2\pi)^{3/2}}\ \sqrt{\frac{m}{\omega_p}}\ w_r({\bm p})
	\ e^{-i\epsilon_r(\omega_pt - {\bm p} \cdot {\bm x})}\,,\label{7freeDE}
\end{eqnarray}
where $r$ runs for 1 to 4. $r = 1,2$ denotes the positive energy solutions ($E = \omega_p = \sqrt{{\bm p}^2 + m^2}$) and while the other two values denote the negative energy solutions ($E = -\omega_{p}$). Thus $\epsilon_r = 1$ for $r=1,2$ and $\epsilon_r = -1$ for $r = 3,4$. $w_r(\bm{p})$ are the Dirac spinors and from Ref.\,\cite{greiner} they have the following orthogonality and completeness properties:
\begin{subequations}\label{spinorprop}
	\begin{eqnarray}
		w^\dagger_{r'}(\epsilon_{r'}{\bm p})\ w_{r}(\epsilon_r{\bm p}) &=& \frac{\omega_p}{m}\ \delta_{rr'}\, ,\\
		\bar{w}_{r'}({\bm p})\ {w}_{r}({\bm p}) &=& \epsilon_r\ \delta_{rr'}\, ,\\
		\sum^4_{r = 1}w_{r\alpha}(\epsilon_r{\bm p})\ w^\dagger_{r\beta}(\epsilon_{r}{\bm p}) &=&
		\frac{\omega_p}{m}\ \delta_{\alpha\beta}\,,\\
		\sum^4_{r = 1}\epsilon_r\ w_{r\alpha}({\bm p})\ \bar{w}_{r\beta}({\bm p}) &=& \delta_{\alpha\beta}.
	\end{eqnarray}
\end{subequations}
The expansion of the field operator is thus given by
\begin{eqnarray}
	\hat{\psi}(t, {\bm x}) &=& \sum^4_{r = 1}\int d{\bm p}\ \hat a({\bm p},r)\ \psi^{(r)}_{\bm p}(t, {\bm x}) \nonumber\\
	&=& \sum^4_{r = 1}\int \frac{d{\bm p}}{(2\pi)^{3/2}}\sqrt{\frac{m}{\omega_p}}\ 
	\hat a({\bm p},r)\ w_r({\bm p})\ e^{-i\epsilon_r(p\cdot x)}\,,\label{DEsol1}
\end{eqnarray}
and that of the conjugate field operator by \cite{greiner}
\begin{eqnarray}
	\hat{\psi}^\dagger(t, {\bm x}) &=& \sum^4_{r = 1}\int d{\bm p}\ \hat a^\dagger({\bm p},r)\ 
	\psi^{(r)\dagger}_{\bm p}(t, {\bm x}) \nonumber\\
	&=& \sum^4_{r = 1}\int \frac{d{\bm p}}{(2\pi)^{3/2}}\sqrt{\frac{m}{\omega_p}}\ 
	\hat a^\dagger(\bm p,r)\ \bar{w}_r({\bm p})\,\gamma^0\,e^{+i\epsilon_r(p\cdot x)}\,.\label{DEsol2}
\end{eqnarray}
It can be shown (as in Ref.\,\cite{greiner}) that the operators $\hat a({\bm p},r)$ and $\hat a^\dagger(\bm p,r)$ satisfy the same anticommutation relationships imposed on the field operators (\ref{DEETCR}):
\begin{eqnarray}
	\left\{\hat{a}({\bm p},r),\ \hat{a}^\dagger({\bm p}',r')\right\} &=& 
	\delta({\bm p} - {\bm p'})\ \delta_{rr'},\nonumber\\
	\Big\{\hat{a}({\bm p},r),\ \hat{a}({\bm p'},r')\Big\} &=&
	\left\{\hat{a}^\dagger({\bm p},r),\ \hat{a}^\dagger({\bm p}',r')\right\} = 0.\nonumber
\end{eqnarray}
These anticommutation relationships describe anti-symmetric states which are occupied by fermions (spin-$\frac{1}{2}$ particles) \cite{P+S}, since for an arbitrary state $\left|\Phi\right\rangle$
$$
	\hat{a}^\dagger({\bm q},r')\,\hat{a}^\dagger({\bm p},r)\left|\Phi\right\rangle = 
	-\hat{a}^\dagger({\bm p},r)\,\hat{a}^\dagger({\bm q},r')\left|\Phi\right\rangle\,.
$$
The Hamiltonian of a quantized Dirac field is given by \cite{greiner}
\begin{eqnarray}
	\hat{H} = \int d{\bm p}\left(\sum^2_{r = 1}\omega_p\ \hat{a}^\dagger({\bm p},r)\ \hat{a}({\bm p},r) - 
	\sum^4_{r = 3}\omega_p\ \hat{a}^\dagger({\bm p},r)\ \hat{a}({\bm p},r)\right)\,.
\end{eqnarray}
This Hamiltonian can have an infinite (negative) value, for the last term adds the contribution of all the negative energy states ($r = 3,4$). In the Dirac interpretation of the negative energy states these states are all filled to form the vacuum. The only contribution of the negative energy states that should be considered is if an negative energy particle has been excited to a positive energy state, leaving a hole in the vacuum. To have a sensible interpretation of the Hamiltonian the contribution of the vacuum has to be removed, i.e.
\begin{eqnarray}
	\hat{H} &=& \int d{\bm p}\Big[\sum^2_{r = 1}\omega_p\ \hat{a}^\dagger({\bm p},r)\ \hat{a}({\bm p},r) + 
	\sum^4_{r = 3}\omega_p\ \left(1-\hat{a}^\dagger({\bm p},r)\ \hat{a}({\bm p},r)\right)\Big]\nonumber\\
	&=& \int d{\bm p}\Big[\sum^2_{r = 1}\omega_p\ \hat{a}^\dagger({\bm p},r)\ \hat{a}({\bm p},r) + 
	\sum^4_{r = 3}\omega_p\ \left(\hat{a}({\bm p},r)\ \hat{a}^\dagger({\bm p},r)\right)\Big]\,.\label{DEH}
\end{eqnarray}
For Eq. (\ref{DEH}) to be expressed as a product of normal-ordered operators, $\hat{a}({\bm p},r)$ with ($r = 1,2$) are the annihilation operators for particles and $\hat{a}^\dagger({\bm p},r)$ with ($r = 3,4$) are the annihilation operators for antiparticles (holes). Similarly $\hat{a}^\dagger({\bm p},r)$ with ($r = 1,2$) are the creation operators for particles and $\hat{a}({\bm p},r)$ with ($r = 3,4$) are the creation operators for antiparticles. To more easily distinguish between operators for particles and antiparticles, the spinors are renamed as:
\begin{eqnarray}
	w_1({\bm p}) &=& u(p,+s)\,,\nonumber\\
		w_2({\bm p}) &=& u(p,-s)\,,\nonumber\\
		w_3({\bm p}) &=& v(p,-s)\,,\nonumber\\
		w_4({\bm p}) &=& v(p,+s)\,,\nonumber
\end{eqnarray}
and the operators as:
\begin{eqnarray}
	\hat{a}({\bm p},1)&=&\hat{b}(p,+s)\,,\nonumber\\
 \hat{a}({\bm p},2)	&=& 	\hat{b}(p,-s)\,,\nonumber\\
\hat{a}({\bm p},3)	&=& 	\hat{d}^\dagger(p,-s)\,,\nonumber\\
\hat{a}({\bm p},4)	&=& 	\hat{d}^\dagger(p,+s) \,,\nonumber
\end{eqnarray}
where $s$ denotes the spin projections. The expression for the field operators therefore changes to (from Ref.\,\cite{greiner})
\begin{eqnarray}
	\hat{\psi}(t, {\bm x}) = \sum_{s}\int\frac{d{\bm p}}{(2\pi)^{3/2}}\sqrt{\frac{m}{\omega_p}}\ 
	\Big( \hat b(p,s)\ u(p,s)\ e^{-i(p\cdot x)} + \hat{d}^\dagger(p,s)\ v(p,s)\ e^{+i(p\cdot x)}\Big)\label{opDEsol}
\end{eqnarray}
and
\begin{eqnarray}
	\hat{\psi}^\dagger(t, {\bm x}) = \sum_{s}\int \frac{d{\bm p}}{(2\pi)^{3/2}}\sqrt{\frac{m}{\omega_p}}\ 
	\Big(\hat b^\dagger(p,s)\ \bar{u}(p,s)\ e^{+i(p\cdot x)}\ + \hat d(p,s)\ \bar{v}(p,s)\ e^{-i(p\cdot x)}\Big)
	\,.\label{conjopDEsol}
\end{eqnarray}
\section{Vector boson fields}
Vector bosons are spin-1 particles and the commonly known ones are the massless photon and the massive $W^{\pm}$ and $Z^0$ bosons \cite{greiner}. The $\omega$- and $\rho$-mesons are also spin-1 particles \cite{walecka}. \\
\\
The quantization of the fields describing the massive $\omega$- and $\rho$-mesons are of particular interest in this work, since these mesons are included in the description of nuclear matter studied in Chapters \ref{chapQHD1} and \ref{chapFSUG}. The quantization of massive vector boson is to a large extent based on the quantization of the photon, a massless vector boson, and  
therefore the quantization of the photon field will be discussed first.\\
\\
As is explained by Greiner \textsl{et al.} in Ref.\,\cite{greiner} the vector boson fields seem to be overdetermined since one would usually expect four degrees of freedom to be associated with a vector field (that of a scalar component, longitudinal and two transversal polarization directions). Photons, being massless, have only two observed degrees of freedom (two transversal polarization states), while massive vector bosons additionally also have a longitudinal polarization state. This reduction of the degrees of freedom has to be taken into account when quantizing the theory by imposing additional constraints on the field operators \cite{greiner}.
\subsection{Photon field}\label{photon}
The electromagnetic interaction between particles via the exchange of photons is described by Maxwell's equations. (For a complete description of the electromagnetic interaction see for instance \textsl{Introduction to Electrodynamics} by D. J. Griffiths \cite{griff}.)
Maxwell's equations can be expressed in a covariant manner by defining the field-strength tensor (or field tensor), $F^{\mu\nu}$,
\begin{eqnarray}
	\begin{tabular}{rl} 
		$F^{\mu\nu}$ = 
				$\left[
					\begin{array}{cccc}
						0&-E^1&-E^2&-E^3\\
						E^1&0&-B^3&B^2\\
						E^2&B^3&0&-B^1\\
						E^3&-B^2&B^1&0\\
					\end{array}	
				\right]$\nonumber
	\end{tabular}	
\end{eqnarray} 
as
\begin{subequations}
	\begin{eqnarray}
		\partial_\mu F^{\mu\nu} = j^\nu,\label{inhomoMax}\\
		\partial^\lambda F^{\mu\nu} + \partial^\nu F^{\lambda\mu} + \partial^\mu F^{\nu\lambda} = 0
	\end{eqnarray}
\end{subequations}	
where $j^\nu = (\rho, \textbf{j})$ is the electromagnetic four-current density.\\
\\
In terms of the electromagnetic vector potential, $A^\mu(x) = (A^0(x), {\bm A}(x))$, the field tensor can be expressed as \cite{greiner}
\begin{eqnarray}
	F^{\mu\nu} = \partial^\mu A^\nu(x) - \partial^\nu A^\mu(x).\nonumber
\end{eqnarray}
Eq. (\ref{inhomoMax}) then leads to a second-order wave equation and field equation for $A^\mu$
\begin{eqnarray}
	\partial_\mu\partial^\mu A^\nu(x) - \partial^\nu(\partial_\mu A^\mu(x)) = j^\nu(x), \label{Awave}
\end{eqnarray}
from which it can be seen (by taking the four-divergence) that the electromagnetic current is conserved, i.e.\ $\partial_\nu j^\nu(x) = 0$.\\
\\
Since the addition of any scalar function $\Lambda(t, {\bm x})$ to $A^\mu$ in the form of
\begin{eqnarray}
	A'^\mu(x) = A^\mu(x) + \partial^\mu\Lambda(x)\label{gaugeinvar}\,,
\end{eqnarray}
leaves $F^{\mu\nu}$ invariant and thus also the field equation (\ref{Awave}). Since Eq. (\ref{gaugeinvar}) describes a local gauge transformation, $A^{\mu}$ is therefore gauge invariant \cite{greiner}.\\
\\
The property of gauge invariance leads to technical complications in the theory \cite{greiner} and an additional constraint (``gauge'') needs to be imposed on the field to remedy the situation (in this case by reducing the degrees of freedom of the field) \cite{M+S}. There are various different conditions that can be imposed on the photon field, one of which is the Lorenz condition,
\begin{eqnarray}
	\partial_\mu A^\mu = 0. \label{Lorenzgauge}
\end{eqnarray}
By demanding that the Lorenz condition is satisfied, is known as the Lorenz gauge. (This condition is mostly misattributed to H. A. Lorentz \cite{Lorenzpaper}.)\\
\\
The photon field equation (\ref{Awave}) can be derived from the following Lagrangian density \cite{M+S},
\begin{eqnarray}
	{\cal L} = -\frac{1}{2}(\partial_\nu A_\mu)(\partial^\nu A^\mu) - j_\mu A^\mu \nonumber
\end{eqnarray}
and the conjugate field to the photon field is therefore
\begin{eqnarray}
	\pi_\mu = \frac{\partial {\cal L}}{\partial (\partial_0A^\mu)} = -\partial^0{A}_\mu.\label{conjugateA}
\end{eqnarray}
\subsection{Quantization}
The photon field is quantized by considering the field and the conjugate field as field operators and imposing the equal-time commutation relationships \cite{greiner}
\begin{subequations}\label{photonETCR}
	\begin{eqnarray}
		\left[ \, \hat{A}^\mu(t,{\bm x}),\hat{\pi}^\nu(t,{\bm x')}\,\right] = 
			i\eta^{\mu\nu}\,\delta^3({\bm x}-{\bm x'})\label{photonETCR1},\\
		\left[\,\hat{A}^\mu(t,{\bm x}), \hat{A}^\nu(t,{\bm x'})\,\right] = 
			\left[ \,\hat{\pi}^\mu(t,{\bm x}), \hat{\pi}^\nu(t,{\bm x'})\,\right] = 0\label{photonETCR2}
	\end{eqnarray}
\end{subequations}
where $\eta^{\mu\nu}$ is the metric as defined in (\ref{metrictensor}).\\
\\
Using the expression for the conjugate photon field (\ref{conjugateA}), Eq. (\ref{photonETCR1}) becomes
\begin{eqnarray}
	\left[ \,\hat{A}^\mu(t,{\bm x}),\partial^0\hat{A}^\nu(t,{\bm x'})\,\right] =	
	-i\eta^{\mu\nu}\,\delta^3({\bm x}-{\bm x'}).\label{photonETCR3}
\end{eqnarray}
The commutator of the divergence of the field operator, $\partial_\mu \hat{A}^\mu$, and the field operator 
can be evaluated, using Eqs (\ref{photonETCR}) and (\ref{photonETCR3}), and yields
\begin{eqnarray}
	\left[ \,\partial_\mu \hat{A}^\mu(t,{\bm x}),\hat{A}^\nu(t,{\bm x'})\,\right] &=&	
	\left[ \,\partial_0 \hat{A}^0(t,{\bm x}) + 
	\nabla\cdot \hat{{\bm A}}(t,{\bm x}),\hat{A}^\nu(t,{\bm x'})\,\right]\nonumber\\
	&=& -\left[ \,\hat{\pi}^0(t,{\bm x}),\hat{A}^\nu(t,{\bm x'})\,\right] + \nabla\cdot
	\left[ \,\hat{{\bm A}}(t,{\bm x}),\hat{A}^\nu(t,{\bm x'})\,\right]\nonumber\\
	&=& i\eta^{\nu0}\,\delta^3({\bm x}-{\bm x'}) \neq 0.	\nonumber
\end{eqnarray}
From the above it can be seen that $\partial_\mu \hat{A}^\mu \neq 0$ and thus the canonical quantization procedure and the Lorenz gauge are not compatible \cite{greiner}.\\
\\
If the Lorenz gauge is not imposed ``scalar'' photons can be created that arbitrarily add energy to the system. Since only transversal photons are observed, another constraint has to be imposed to be able to quantize the photon field. In the Gupta-Bleuler method only the state vectors in the Hilbert space, $\left|\Phi\right\rangle$, that satisfy
\begin{eqnarray}
	\left\langle \Phi\right|\partial^\mu\hat{A}_\mu\left|\Phi\right\rangle = 0,\label{gupta}
\end{eqnarray}
are admitted. By imposing condition (\ref{gupta}) on the states in the Hilbert space the photon field can be quantized by demanding that the photon field operators satisfy the conditions given in (\ref{photonETCR}). \cite{greiner}
\subsection{Massive vector bosons}
A massive neutral vector bosons (spin-1) field, $A^\mu(x)$, is described by the Lagrangian density \cite{greiner},
\begin{eqnarray}
	{\cal L} = -\frac{1}{4}F_{\mu\nu}F^{\mu\nu} + \frac{1}{2}m^2A_\mu(x) A^\mu(x) - j_\mu(x) A^\mu(x),\label{Lproca}
\end{eqnarray}
where $F_{\mu\nu}$ is once again the field tensor,
$$
	F_{\mu\nu} = \partial_\mu A_\nu(x) - \partial_\nu A_\mu(x)
$$
and $j_\mu(x)$ the current density.\\
\\
Using the Euler-Lagrange equation (\ref{EL}), the equation of motion of the field is calculated to be
\begin{eqnarray}
	\partial_\mu F^{\mu\nu} + m^2A^\nu(x) - j^\nu(x) = 0\label{proca1},
\end{eqnarray}
which is the Proca equation \cite{M+S}.\\
By taking the four-divergence of the Proca equation, one finds that
$$
	\partial_\nu A^\nu(x) = \frac{1}{m^2}\partial_\nu j^\nu(x).
$$
If one assumes that there are no sources, i.e. $j^\nu(x) = 0$, or that the source current is conserved, i.e. $\partial_\nu j^\nu(x) = 0$, the above condition on the field reduces to
\begin{eqnarray}
	\partial_\nu A^\nu = 0\nonumber,   
\end{eqnarray}
which is the Lorenz condition (\ref{Lorenzgauge}). Thus the Proca equation for massive vector bosons automatically satisfies the Lorenz condition under the assumptions mentioned and simplifies to
\begin{eqnarray}
	\partial_\mu\partial^\mu A^\nu(x) + m^2A^\nu(x)  = j^\nu(x).\label{proca}
\end{eqnarray}
The conjugate field (\ref{conjugfield}) is derived from the Lagrangian density (\ref{Lproca}) as
\begin{eqnarray}
	\pi^\mu = \frac{\partial {\cal L}}{\partial (\partial_0A_\mu)} = -F^{0\mu}.\label{conjugateP}
\end{eqnarray}
Since $F^{\mu\nu}$ has no diagonal elements, $\pi^0$ would be zero and $\pi^i = E^i$. Because the Proca equation automatically satisfies the Lorenz condition the time-component of the conjugate field is a dependent variable. This is to be excepted since by automatically satisfying the Lorenz condition the Proca equation only has three degrees of freedom \cite{greiner}. As a dependent variable the time-component of field is given in terms of the degrees of freedom ($E^i$) as
\begin{eqnarray}
	A^0 = -\frac{1}{m^2}\nabla\cdot{\bm E}.
\end{eqnarray}
\subsection{Quantization}
The canonical quantization of the massive vector field is achieved by imposing the following commutation relationships:
\begin{subequations}\label{procaETCR}
	\begin{eqnarray}
		\left[ \, \hat{A}^i(t,{\bm x}),\hat{E}^j(t,{\bm x')}\,\right] = 
			-i\delta_{ij}\,\delta^3({\bm x}-{\bm x'})\,,\label{procaETCR1}\\
		\left[\,\hat{A}^i(t,{\bm x}), \hat{A}^j(t,{\bm x'})\,\right] = 
			\left[ \,\hat{E}^i(t,{\bm x}), \hat{E}^j(t,{\bm x'})\,\right] = 0\,.\label{procaETCR2}
	\end{eqnarray}
\end{subequations}
It may seem that the quantization of the field might not be covariant, since it is only expressed in terms of the spatial components of the field operators. According to W. Greiner \textsl{et al.} a covariant quantization can be achieved by defining the time component of the field operator to be \cite{greiner}:
$$
	\hat{A}^0 = -\frac{1}{m^2}\nabla\cdot\hat{{\bm E}}.
$$
From (\ref{procaETCR}) this operator satifies the following commutation relationships
\begin{subequations}\label{proca0ETCR}
	\begin{eqnarray}
		\left[ \, \hat{{\bm A}}(t,{\bm x}),\hat{A}^0(t,{\bm x')}\,\right] &=& 
			i\frac{1}{m^2}\nabla\,\delta^3({\bm x}-{\bm x'})\label{proca0ETCR1}\nonumber,\\
		\left[\,\hat{A}^0(t,{\bm x')}, \hat{A}^0(t,{\bm x')}\,\right] &=& 0.\label{proca0ETCR2}\nonumber
	\end{eqnarray}
\end{subequations}

%% file: chapQHD.tex
\chapter{Quantum hadrodynamics}\label{chapQHD1}
\section{Introduction}
As mentioned in Chapter \ref{DNM}, J. D. Walecka introduced a description of nuclei and nuclear matter based on the exchange of mesons in 1974. This description will be referred to as quantum hadrodynamics or QHD.\\
\\
Since 
nuclei and nuclear matter are complex systems, there exist various models of nuclei and nuclear matter of which QHD is one. All models need some form of experimental input to constrain the model. In the case of QHD it is the coupling constants (or coupling strengths) between the different meson and nucleon fields that are the unknown parameters. These couplings are determined by fitting the calculated properties of nuclei and nuclear matter (such as the properties that were described in Sec. \ref{satprop}) to the experimentally observed (or inferred) values. In the last few decades various other parameter sets for QHD have been calculated by different authors by fitting various other observed properties of nuclei and/or nuclear matter. These different \mbox{parameter} sets differ amongst one another since they might include other meson fields and/or different couplings between the various fields and/or recalculated values of the couplings constants. 
\\\\
In Sec. \ref{secQHD1formalism} the basic philosophy and calculations will be introduced within the context of the original parameter set, QHD-I, as introduced by J. D. Walecka \cite{walecka1, walecka}.\\
\\
In Chapter \ref{chapFSUG} other parameter sets and variations based on QHD-I will be discussed.
\section{Formalism}\label{secQHD1formalism}
Quantum hadrodynamics I (QHD-I), also known as the $\sigma$ - $\omega$ model, is the original and \mbox{simplest} parameter set of QHD. It models the nuclear force by the exchange of neutral (isoscalar) scalar sigma ($\sigma$) mesons and neutral (isoscalar) vector omega ($\omega$) mesons. These mesons have been found to be the most important in describing the properties of nuclei and nuclear matter \cite{waleckatext}. The baryons included in QHD-1 are protons and neutrons. The scalar meson gives rise to a strong attractive central force and a spin-orbit force in the nucleon-nucleon interaction, while the vector meson gives rise to a strong repulsive central force and a spin-orbit force (with the same sign as the spin-orbit force of the scalar meson) \cite{machleidt}. Thus the nucleon-nucleon interaction is described by three fields, the baryon-, scalar meson- and vector meson-fields. No charged mesons are included in this parameter set (i.e. the electric properties of the baryons are not considered) and the masses of the proton and neutron are taken to be equal in QHD-I.\\
\\
The Lagrangian density of QHD-I (in natural units) is given by \cite{walecka}:
\begin{eqnarray}
		{\cal L} &=& \bar{\psi}(x)\Big[\gamma_{\mu}\Big(i\partial^{\mu}-g_{v}V^{\mu}(x)\Big) - 
		\Big(M-g_{s}\phi(x)\Big)\Big]\psi(x)\nonumber\\ &+&
		 \frac{1}{2}\Big(\partial_{\mu}\phi(x)\partial^{\mu}\phi(x) - m_{s}^{2}\phi^{2}(x)\Big) - 
		 \frac{1}{4}V_{\mu\nu}V^{\mu\nu} + \frac{1}{2}m_\omega^{2}V_{\mu}(x)V^{\mu}(x)\nonumber
\end{eqnarray}
where
\begin{itemize}
\item $V$ denotes the vector meson field,
\item $\phi$ denotes the scalar meson field,
\item $m_\omega$ and $m_s$ the different meson masses and $M$ denotes the nucleon mass,
\item $g_{v}$ and $g_{s}$ are the vector and scalar coupling constants and 
\item $V_{\mu\nu} = \partial_{\mu}V_{\nu}(x) - \partial_{\nu}V_{\mu}(x)$.
\end{itemize}
The Lagrangian density is a Lorentz scalar and has the dimension of length$^{-4}$.\\
\\
In constructing the Lagrangian density it was assumed that the neutral scalar mesons couple to the scalar density of the baryon field and that the neutral vector mesons couple to the conserved baryon current \cite{walecka}.\\
\\
Using the Euler-Lagrange equation (\ref{EL}) the equations of motion of the different fields can be derived:
\begin{subequations}\label{QHD1EQM}
	\begin{eqnarray}
		\partial_{\mu}\partial^{\mu}\phi(x) + m_{s}^{2}\phi(x) &=& g_{s}\bar{\psi}(x)\psi(x)\label{scalarEQM} \\
		\partial_{\mu}V^{\mu\nu} + m_\omega^{2}V^{\nu}(x) &=& g_{v}\bar{\psi}(x)\gamma^{\nu}\psi(x)\label{bosonEQM} \\
		\Big[\gamma_{\mu}\big(i\partial^{\mu}-g_{v}V^{\mu}(x)\big) - 
		\big(M-g_{s}\phi(x)\big)\Big]\psi(x) &=& 0\label{baryonEQM}.
	\end{eqnarray}
\end{subequations}
Eq.\,(\ref{scalarEQM}) is analogous to the equation of motion of the free scalar field, the Klein-Gordon equation (\ref{fieldKGE}), with the baryon scalar density, $\bar{\psi}(x)\psi(x)$, as the source term.\\
\\
Eq.\,(\ref{bosonEQM}) is analogous to the Proca equation (\ref{proca}), with the coupling of the vector field to the conserved baryon current as the source term.\\
\\
The equation of motion of the baryon field is the Dirac equation (\ref{fieldDE}) with modified mass (due to the scalar field) and interaction with the vector field introduced in a minimal fashion \cite{walecka}.\\
\\
Since the equations of motion (\ref{QHD1EQM}) are non-linear, coupled equations they are very difficult to solve and are therefore approximated. The approximation that will be used is the {\em relativistic mean-field approximation}.
\section{Relativistic mean-field theory}\label{RMF}
In other quantized theories, such as QCD, a perturbation expansion in the coupling strengths can be used to approximate complicated equations. Within the framework of QHD the coupling strengths are large and thus the higher-order terms in a perturbative approximation will diverge. Within the context of the study of dense nuclear systems an approximation that is increasingly valid as the density increases, is also preferred. In the relativistic mean-field approximation (RMF) within QHD, the meson field operators are replaced by their ground state ($\left|\Phi\right\rangle$) expectation values, which are classical fields, in the following way \cite{walecka}:
\begin{subequations}\label{MFT}
	\begin{eqnarray}
		\phi \longrightarrow \left\langle \Phi\left|\phi\right|\Phi\right\rangle = 
		\left\langle \phi\right\rangle = \phi_{0}\\
		V_{\mu} \longrightarrow \left\langle\Phi\left|V_{\mu}\right|\Phi\right\rangle  = \left\langle V_\mu\right\rangle 
		= \delta_{\mu 0}V_{0}.
	\end{eqnarray}
\end{subequations}
The mean-field approximation (\ref{MFT}) can be motivated if a system of $B$ baryons (which are at rest and at zero temperature), occupying a box of volume $V$ is considered. Since the system is static, the baryon flux, given by $\bar{\psi}(x)\gamma^{i}\psi(x)$, will be zero \cite{recentprogress}. If the baryon density, $B/V$, is increased the source terms on the right-hand side of Eqs (\ref{scalarEQM}) and (\ref{bosonEQM}) will become large. If the source terms are large enough the meson field operators can be approximated by their ground state expectation values, instead of evaluating the operators at every spacetime point in the box. For a stationary, uniform system $\phi_0$ and $V_0$ will be constants, independent of space and time. The spatial components of $\left\langle V_\mu\right\rangle $ will vanish, since, as mentioned above, the baryon flux is zero and the system is at rest. Another illustration for the vanishing spatial components of $\left\langle V_\mu\right\rangle $ is given in Ref. \cite{csg}. \\
\\
In the mean-field approximation the meson field operators are replaced by their ground state expectation values in order to simply the solution to the field equations. However, the nucleon field operators remain operators.
Since only the ground state expectation values of the meson field operators are considered in the approximation it follows that the baryon operators (the baryon sources to which the meson fields couple) must be evaluated by operating on the ground state. As such these baryon operators in the equations of motion of the meson fields (\ref{QHD1EQM}) are replaced by their normal-ordered ground state expectation values \cite{FSU1}:
\begin{subequations}\label{expectval}
	\begin{eqnarray}
		\bar\psi(x)\psi(x) &\longrightarrow &
		\left\langle \Phi\left|{\bm :}\bar\psi(x)\psi(x){\bm :}\right|\Phi\right\rangle
		= 	\left\langle \bar\psi\psi\right\rangle\\
		\bar\psi(x)\gamma^\mu\psi(x) &\longrightarrow& 
		\left\langle \Phi\left|{\bm :}\bar\psi(x)\gamma^\mu\psi(x){\bm :}\right|\Phi\right\rangle 
		= \left\langle\bar\psi\gamma^0\psi\right\rangle,
	\end{eqnarray}
\end{subequations}	
where the spatial components vanished due to the consideration of the ground state.
The normal-ordered expectation value is taken since the contribution of the filled negative energy baryon states (the vacuum) is ignored, since the vacuum has a (infinite!) constant energy. Thus only the positive energy baryon states are considered. This is known as the no-sea approximation \cite{NL3}.\\
\\
Since $\phi_{0}$ and $V_0$ are constants in the RMF approximation the equations of motion of the fields (\ref{QHD1EQM}) reduce to
\begin{subequations}\label{MFTQHD1EQM}
	\begin{eqnarray}
		 m_{s}^{2}\phi_{0} &=& g_{s}\left\langle\bar{\psi}\psi\right\rangle\label{MFTsigma} \\
		 m_\omega^{2}V_{0} &=& g_{v}\left\langle\bar{\psi}\gamma_{0}\psi\right\rangle\label{MFTomega}\\
		 \left[i\gamma_\mu\partial^\mu - g_v\gamma_0V_0- (M-g_s\phi_0)\right]\psi &=& 0\,. \label{MFTnucleon}
	\end{eqnarray}
\end{subequations}		
The RMF QHD-I Lagrangian density is:
	\begin{eqnarray}
			{\cal L}_{RMF} = \bar{\psi}\Big[i\gamma_{\mu}\partial^{\mu}-g_{v}\gamma^{0}V_{0} - (M-g_{s}\phi_{0})\Big]\psi -
		 \frac{1}{2}m_{s}^{2}\phi_{0}^{2} + \frac{1}{2}m_\omega^{2}V_{0}^{2}\,.\nonumber
	\end{eqnarray}
From the above Lagrangian density the RMF energy-momentum tensor of QHD-I can be constructed using the definition of the energy-momentum tensor (\ref{EMT}) as well as keeping Eq. (\ref{baryonEQM}) in mind, and is given by
	\begin{eqnarray}		 
		 (T^{\mu\nu})_{RMF} = i\bar{\psi}\gamma^{\mu}\partial_{\nu}\psi 
		 - \eta^{\mu\nu}\big(-\frac{1}{2}\,m_{s}^{2}\phi_{0}^{2}
		 +\frac{1}{2}\,m_\omega^{2}V_{0}^{2}\big)\,.\nonumber
	\end{eqnarray}
The energy density and the pressure, using (\ref{epsp13}) for QHD-1 is \cite{walecka}
\begin{subequations}\label{MFTeos}
	\begin{eqnarray}
		\epsilon &=& \left\langle T^{00}\right\rangle \nonumber\\ &=& 
		\left\langle i\bar{\psi}\gamma_{0}\partial_{0}\psi - \Big(-\frac{1}{2}\,m_{s}^{2}\phi_{0}^{2}
		 +\frac{1}{2}\,m_\omega^{2}V_{0}^{2}\Big)\right\rangle\nonumber\\
		&=&\left\langle i\bar{\psi}\gamma_{0}\partial_{0}\psi\right\rangle +\frac{1}{2}\,m_{s}^{2}\phi_{0}^{2}
		- \frac{1}{2}\,m_\omega^{2}V_{0}^{2} 
		\label{MFTeps}\,,\\
		P&=& \frac{1}{3}\left\langle T^{ii} \right\rangle\nonumber\\
		&=&	\frac{1}{3}\left\langle i\bar{\psi}\gamma^{i}\partial_{i}\psi 
		+\big(-\frac{1}{2}\,m_{s}^{2}\phi_{0}^{2}
		 +\frac{1}{2}\,m_\omega^{2}V_{0}^{2}\big)\right\rangle\nonumber\\
		 &=& \frac{1}{3}\left\langle i\bar{\psi}\gamma_{i}\partial_{i}\psi\right\rangle 
		 - \frac{1}{2}\,m_{s}^{2}\phi_{0}^{2} + \frac{1}{2}\,m_\omega^{2}V_{0}^{2} \label{MFTpres}\,.
	\end{eqnarray}
\end{subequations}

\section{Evaluation of expectation values}\label{secQHD1expect}
Since $\psi$ and $\bar{\psi}$ (the baryon field operators) are still operators, their expectation values have to be calculated to obtain sensible information for, amongst others, the energy density and the pressure. As mentioned previously, it is the ground state expectation values of the baryon field operators that have to be evaluated. 
\\\\
These expectation values can be evaluated by constructing an explicit expression for 
$\psi$. For the RMF approximation a uniform static system is assumed, so $\psi$ is expanded in terms of the momentum eigenstates \cite{walecka}. The momentum eigenstates are denoted by ${\bm k}$, which is the wave vector. The wave vector is related to the momentum (${\bm p}$) through the expression of the de Broglie wavelength \cite{griff2}, $$ {\bm p} = \hbar {\bm k}\,.$$ In natural units the momentum and the wave vector are equivalent. Solutions for the field operator are thus of the form [analogous to the free single-particle solution for a Dirac particle (\ref{7freeDE})],
\begin{eqnarray}
	\psi(x) = \psi({\bm k},s)\,e^{i{\bm k}\cdot {\bm x} - ie({\bm k})t}\,,\label{Kexpand}
\end{eqnarray}
where $\psi({\bm k},s)$ is the four component Dirac spinor ($s$ denotes the spin index) and $e({\bm k})$ the energy associated with specific state denote by ${\bm k}$ \cite{walecka}. Substituting Eq. (\ref{Kexpand}) into Eq. (\ref{MFTnucleon}) yields
\begin{eqnarray}
	\big(\,-\gamma_ik^i + \gamma_0e({\bm k}) - g_v\gamma_0V_0 - (M-g_s\phi_0)\big)\psi({\bm k},s) = 0\,.\nonumber
\end{eqnarray}
By multiplying the above equation by $\gamma_0$, it reduces to (having reverted back to the notation of the ${\bm{\alpha}}$- and $\beta$-matrices)
\begin{eqnarray}
	\Big(\beta^2\big(e({\bm k}) - g_vV_0\big)\Big)\psi({\bm k},s) = 
	\Big({\bm \alpha}\cdot{\bm k} + \beta m^*\Big)\psi({\bm k},s)\,,\label{MFTenergy1}
\end{eqnarray}
where
\begin{itemize}
	\item $m^*$ is the reduced nucleon mass, defined as 
			\begin{eqnarray}
				m^* \equiv (M-g_s\phi_0)\,.\label{mreddef}
			\end{eqnarray}
\end{itemize}
The presence of the scalar meson field thus shifts the mass of the nucleons \cite{walecka}.\\
\\
Squaring both sides and keeping the properties of the ${\bm \alpha}$- and $\beta$-matrices (\ref{covargamma}) in mind one finds that
\begin{eqnarray}
	e^{\pm}({\bm k}) = g_v V_0 \pm\sqrt{{\bm k}^2 + {m^*}^2}\,.\label{MFTenergy}
\end{eqnarray}
From Eq. (\ref{MFTenergy}) is it clear that positive and negative energy solutions, which are symmetric around $g_v V_0$, are still found. Defining positive and negative energy spinors $u({\bm k},s)$ and $v({\bm k},s)$ that in general satisfy the same orthogonality and completeness properties as in (\ref{spinorprop}), as well as the following normalisation condition \cite{walecka}
\begin{eqnarray}
	u^\dagger({\bm k},s)u({\bm k},s') = v^\dagger({\bm k},s)v({\bm k},s') = \delta_{ss'}\,,\nonumber
\end{eqnarray}
the general solution for the baryon field operator (\ref{MFTnucleon}) can be written as \cite{walecka}
\begin{eqnarray}
	\psi(x) = \sum_{s}\int \frac{d{\bm k}}{(2\pi)^{3/2}}
	\Big( \hat b({\bm k},s)\ u({\bm k},s)\ e^{i{\bm k}\cdot {\bm x} - ie^{(+)}({\bm k})t} 
	+ \hat d^\dagger({\bm k},s)\ v({\bm k},s)\ e^{-i{\bm k}\cdot{\bm x} -ie^{(-)}({\bm k})t}\Big)\,,\label{MFTpsi}
\end{eqnarray}
where the particle and anti-particle creation and annihilation have been included (using the same notation as in Sec. \ref{secDEfields}).\\
\\
The ground state $\left|\Phi\right\rangle$ is defined by $k_F$, called the Fermi-wavenumber: $k_F$ denotes the filled positive energy baryon momentum state below which all positive energy states are also filled. The ground state in the the no-sea approximation has the following properties:
\begin{eqnarray}
	\hat{d}({\bm k},s)\left|\Phi\right\rangle = 0 & & \forall\  \left|{\bm k}\right|\nonumber\\
	\hat{b}^\dagger({\bm k},s)\left|\Phi\right\rangle = 0 & & \forall\  \left|{\bm k}\right|<k_F\nonumber\\
	\hat{b}({\bm k},s)\left|\Phi\right\rangle = 0 & & \forall\  \left|{\bm k}\right|>k_F\nonumber
\end{eqnarray}
The expectation value in the expression for the RMF value of $V_0$ (\ref{MFTomega}) can be evaluated by using the construction of $\psi(x)$ (\ref{MFTpsi}) and taking the normal-ordered expectation value, as in Eq. (\ref{expectval}). In normal-ordering the expression the following should be kept in mind: The anti-commutation relationship of $\hat{b}({\bm k},s)$ and $\hat{d}^\dagger({\bm k},s)$, which is the same as the creation and annihilation operators of the general Dirac field given in Sec. \ref{secDEfields}, as well as the properties of the ground state. Then it can be shown that, after some algebra, $\left\langle \psi^\dagger\psi\right\rangle$ is given by
\begin{eqnarray}
	\left\langle \psi^\dagger\psi\right\rangle &=& \sum_s\frac{1}{(2\pi)^3}\int_0^{k_F} d^3k\nonumber\\
	&=& \sum_s\frac{1}{(2\pi)^3}\int_0^{k_F} dk\,4\pi k^2\nonumber\\
	&=& \frac{\gamma}{6\pi^2}{k_F}^3\nonumber\\ &=& \rho\,,\label{rhoB}
\end{eqnarray}
where $\gamma$ is the nucleon spin-degeneracy of the state. From (\ref{rhoB}) it is clear that $\left\langle \psi^\dagger\psi\right\rangle$ is the nucleon (baryon) number density at $\left|{\bm k}\right| = k_F$, which will be denoted by $\rho$.\\
\\
From (\ref{MFTeos}) the expressions for the energy density and the pressure for QHD-1 in the RMF approximation become
\begin{subequations}\label{MFTeos2}
	\begin{eqnarray}
		\epsilon &=& \left\langle{\psi}^\dagger\big(-i{\bm \alpha}\cdot{\bm \nabla} +
		\beta m^* + g_v V_0\big)\psi\right\rangle -
		 \frac{1}{2}\,m_\omega^{2}V_{0}^{2} 
		+\frac{1}{2}\,m_{s}^{2}\phi_{0}^{2}\,,\label{eMFTeos2}\\
		P&=& \frac{1}{3}\left\langle{\psi}^\dagger\big(-i{\bm \alpha}\cdot{\bm \nabla}\big)\psi\right\rangle -
			   \frac{1}{2}\,m_\omega^{2}V_{0}^{2} + \frac{1}{2}\,m_{s}^{2}\phi_{0}^{2}\,.\label{pMFTeos2}
	\end{eqnarray}
\end{subequations}
The expectation values in (\ref{MFTeos2}) can be evaluated in the same way as the one in Eq.\,(\ref{rhoB}). 
After some algebra [using in particular the properties of the spinors (\ref{MFTenergy1}) as well as Eq.\,(\ref{MFTenergy})], the expectation values in the expressions for the energy density and the pressure reduce to:
\begin{subequations}
	\begin{eqnarray}
		\left\langle{\psi}^\dagger\big(-i{\bm \alpha}\cdot{\bm \nabla} +
		\beta m^* + g_v V_0\big)\psi\right\rangle 
		&=& \frac{\gamma}{(2\pi)^3}\int^{k_F}_{0}d{\bm k}\,\sqrt{{\bm k}^2 + {m^*}^2} + g_v V_0 \rho\,,\label{expeps}\\
		\left\langle{\psi}^\dagger\big(-i{\bm \alpha}\cdot{\bm \nabla}\big)\psi\right\rangle 
	&	=& \frac{\gamma}{(2\pi)^3}\int^{k_F}_{0}d{\bm k}\,\frac{{\bm k}^2}{\sqrt{{\bm k}^2 + {m^*}^2}}\,.\label{exppres}
	\end{eqnarray}
\end{subequations}
The remaining expectation value, that of $\left\langle \bar{\psi}\psi\right\rangle$ in the equation of motion of the scalar meson field (\ref{MFTsigma}), can be evaluated in this explicit manner, but was not done in this work. Rather other considerations (as explained in Ref. \cite{walecka}) will be employed to yield the expectation value (see below and Sec. \ref{secQHD1eos}).\\
\\
Another method to evaluate the expectation values is given by N. K. Glendenning in Ref. \cite{csg}. It states that the expectation value of an operator ($\Gamma$) in the ground state can be given in terms of the expectation value of the single-particle state,
$$
	\left(\bar{\psi}\Gamma\psi\right)_{{\bm k},s},
$$
where
\begin{itemize}
	\item ${\bm k}$ denotes the momentum, and,
	\item $s$ the spin of the single-particle state.
\end{itemize}
The expectation value in the many-nucleon system is then \cite{csg}
\begin{eqnarray}
	\left\langle \bar{\psi}\Gamma\psi\right\rangle  = 
	\sum_s\int\frac{d{\bm k}}{(2\pi)^3}\left(\bar{\psi}\Gamma\psi\right)_{{\bm k},s}\,
	\Theta\big[\,\mu-e({\bm k})\big]\,,\label{NKGexpect}
\end{eqnarray}
where
\begin{itemize}
	\item $e({\bm k})= g_v V_0 + ({\bm k}^2 + {m^*}^2)^{\frac{1}{2}}$ is the positive single-particle energies 
	(\ref{MFTenergy}), since only the ground state is 
	considered, 
	\item $\mu$ is the chemical potential/Fermi energy \big(single-particle energy $e({\bm k}_F)$ \big), and ,
	\item $\Theta[\,\mu-e({\bm k})]$ is a step function with 
	$$
		\Theta[\,\mu-e({\bm k})] = \left\{\begin{array}{cc}
																			1&\mbox{if}\  \left|{\bm k}\right|\leq k_F\\
																			0&\mbox{if}\ \left|{\bm k}\right|> k_F
																		\end{array}\right..
	$$
\end{itemize}
All the possibilities for $\Gamma$ will in general appear in the Dirac Hamiltonian \cite{csg}. The Dirac Hamiltonian $H_D$ can be constructed from the equation of motion of the nucleon field (\ref{MFTnucleon}) and the expression of $\psi$ (\ref{Kexpand}) as
\begin{eqnarray}
	H_D = \gamma_0\big[{\bm\gamma}\cdot{\bm k} + g_v \gamma_0 V_0 + m^*\big]\,.
\end{eqnarray}
Taking the single-particle expectation value of $H_D$
\begin{eqnarray}
	\left(\psi^\dagger H_D\psi\right)_{{\bm k},s} &=& e({\bm k})\left({\psi}^\dagger\psi\right)_{{\bm k},s}\nonumber\\
	&=& e({\bm k})\,,\label{singexp}
\end{eqnarray}
where $\left({\psi}^\dagger\psi\right)_{{\bm k},s} =1$ (this can be seen from the expression for $\psi$ and the normalisation of the spinors). Taking the derivative of the left-hand side of Eq. (\ref{singexp}) with respect to any variable, $\zeta$, yields
\begin{eqnarray}
	\frac{\partial}{\partial\zeta}\,\left(\psi^\dagger H_D\psi\right)_{{\bm k},s} &=& 
	\left(\psi^\dagger\,\frac{\partial H_D}{\partial\zeta}\,\psi\right)_{{\bm k},s}
	+ e({\bm k})\,\frac{\partial}{\partial\zeta}\,\left(\psi^\dagger\,\psi\right)_{{\bm k},s}\nonumber\\
	& = & \left(\psi^\dagger\,\frac{\partial H_D}{\partial\zeta}\,\psi\right)_{{\bm k},s},\label{derivHD}
\end{eqnarray}
as $\psi(x)$ is an eigenfunction of $H_D$, the second term on the right vanishes \cite{csg}. Thus Eq. (\ref{derivHD}) [considering Eq. (\ref{singexp})] yields
\begin{eqnarray}
	\frac{\partial}{\partial\zeta}\,\left(\psi^\dagger H_D\psi\right)_{{\bm k},s} 
	&=&\frac{\partial}{\partial\zeta}\,e({\bm k})\nonumber\\
	 & = & \left(\psi^\dagger\,\frac{\partial H_D}{\partial\zeta}\,\psi\right)_{{\bm k},s}.\label{singlepatexp}
\end{eqnarray}
Any expectation value can therefore be obtained by using the general expression (\ref{NKGexpect}): taking the derivative of $H_D$ for the appropriate choice of $\zeta$ to yield $\Gamma$ and obtaining the expression for the single particle expectation value of $\Gamma$ from Eq. (\ref{singlepatexp}).\\
\\
Using this method the same results are obtained as with the first method, just with much less effort. The expectation value in the expression for the energy density (\ref{eMFTeos2}), when $\nabla$ operated on $\psi$ (\ref{Kexpand}), using Eqs (\ref{MFTenergy1}) and (\ref{MFTenergy}), is:
\begin{eqnarray}
	 \left\langle{\psi}^\dagger\big({\bm \alpha}\cdot{\bm k} + \beta m^* + g_v V_0\big)\psi\right\rangle  
	 &=&\left\langle{\psi}^\dagger\Big(\sqrt{{\bm k}^2 + {m^*}^2} + g_v V_0\Big)\psi\right\rangle\nonumber\\
	 &=& \left\langle{\psi}^\dagger e({\bm k})\psi\right\rangle\nonumber
\end{eqnarray}
and using the value of the single-particle expectation value of $H_D$ (\ref{singexp}), the expectation value in the energy density yields
\begin{eqnarray}
	\left\langle{\psi}^\dagger\Big(-i{\bm \alpha}\cdot{\bm \nabla} +
		\beta m^* + g_v V_0\Big)\psi\right\rangle  
		&=&  \sum_s\int\frac{d{\bm k}}{(2\pi)^3}\,e({\bm k})\,\Theta\big[\,\mu-e({\bm k})\big]\nonumber\\
		&=&  \sum_s\int\frac{d{\bm k}}{(2\pi)^3}\,\Big(\sqrt{{\bm k}^2 + {m^*}^2} + g_v V_0\Big)\,
		\Theta\big[\,\mu-e({\bm k})\big]\nonumber\\
		&=&  g_v V_0\,\rho + \frac{\gamma}{(2\pi)^3}\int^{k_F}_0d{\bm k}\,\sqrt{{\bm k}^2 + {m^*}^2}\,.\nonumber
\end{eqnarray}
The expectation value in the expression for the pressure (\ref{pMFTeos2}) is
\begin{eqnarray}
	\left\langle{\psi}^\dagger\Big(-i{\bm \alpha}\cdot{\bm \nabla}\Big)\psi\right\rangle  &=& 
	\left\langle{\psi}^\dagger\big({\bm \alpha}\cdot{\bm k}\big)\psi\right\rangle \nonumber\\
	&=& \left\langle\bar{\psi}\big({\bm \gamma}\cdot{\bm k}\big)\psi\right\rangle \nonumber\,,
\end{eqnarray}
and the single-particle expectation value of the operator ${\bm \gamma}\cdot{\bm k}$ is
\begin{eqnarray}
	\left(\bar{\psi}\,{\bm \gamma}\cdot{\bm k}\,\psi\right)_{{\bm k},s} 
	&=&  \left(\bar{\psi}\,{\bm \gamma}\,\psi\right)_{{\bm k},s}\cdot{\bm k}\nonumber \\
	&=&  \left(\bar{\psi}\,\frac{\partial H_D}{\partial {\bm k}}\,\psi\right)_{{\bm k},s}\cdot{\bm k}\nonumber \\
	&=& \frac{\partial e({\bm k})}{\partial{\bm k}}\cdot{\bm k}\nonumber \ \ \ \ \ \ \ \ \ \ \ \ \ \ \ \ \ \ \ \ 
	\mbox{using (\ref{singlepatexp})}\\
	&=& \frac{{\bm k}\cdot{\bm k}}{\sqrt{{\bm k}^2 + {m^*}^2}}\nonumber\,,
\end{eqnarray}
and using the general expression for the expectation value (\ref{NKGexpect})
\begin{eqnarray}
	\left\langle{\psi}^\dagger\big(-i{\bm \alpha}\cdot{\bm \nabla}\big)\psi\right\rangle  
		&=&  \sum_s\int\frac{d{\bm k}}{(2\pi)^3}\,\frac{{\bm k}^2}{\sqrt{{\bm k}^2 + {m^*}^2}}\,
		\Theta\big[\,\mu-e({\bm k})\big]\nonumber\\
		&=&  \frac{\gamma}{(2\pi)^3}\int^{k_F}_0d{\bm k}\,\frac{{\bm k}^2}{\sqrt{{\bm k}^2 + {m^*}^2}}\,.\label{NKGpresexp}
\end{eqnarray}
\\
A result for 
$\left\langle \bar{\psi}\psi\right\rangle$ can also be obtained in this manner: taking $M$ (the nucleon mass) as $\zeta$, the single-particle expectation value is 
\begin{eqnarray}
	\big(\bar{\psi}\psi\big)_{{\bm k},s} 
	&=&  \big(\psi^\dagger\,\gamma^0\,\psi\big)_{{\bm k},s}\nonumber \\
	&=&  \big(\psi^\dagger\,\frac{\partial H_D}{\partial M}\,\psi\big)_{{\bm k},s}\nonumber \\
	&=& \frac{\partial}{\partial M}\,e({\bm k})\nonumber\ \ \ \ \ \ \ \ \ \ \ \ \ \ \ \ \ \ \ \ 
	\mbox{using (\ref{singlepatexp})}\\
	&=& \frac{m^*}{\sqrt{{\bm k}^2 + {m^*}^2}}\,.\nonumber
\end{eqnarray}
Using the general expression for the expectation value (\ref{NKGexpect}) and the above result, $\left\langle \bar{\psi}\psi\right\rangle$ yields
\begin{eqnarray}
	\left\langle \bar{\psi}\psi\right\rangle 
	&=&  \sum_s\int\frac{d{\bm k}}{(2\pi)^3}\,\frac{m^*}{\sqrt{{\bm k}^2 + {m^*}^2}}\,
	\Theta[\,\mu-e({\bm k})]\nonumber\\
	&=&  \frac{\gamma}{2\pi^2}\int^{k_F}_0 dk\,\frac{ k^2\,m^*}{\sqrt{{k}^2 + {m^*}^2}}\,.\label{scalardensity}
\end{eqnarray}
\section{QHD-I parameter set}
The values of the particle masses used in this work for QHD-1 was taken from Ref. \cite{recentprogress} and are listed in Table \ref{tab:QHD1mass}.\\
\begin{table*}[ttb]
	\centering
		\begin{tabular}{cccc}
			\hline\hline
			$M_{\mbox{\footnotesize{proton}}}$& $M_{\mbox{\footnotesize{neutron}}}$& $m_s$& $m_\omega$\\
			\hline
			 939 & 939 & 520 & 783\\
			\hline
		\end{tabular}
	\caption{Particle masses (in MeV) for QHD-1 used in this work.}
	\label{tab:QHD1mass}
\end{table*}\\
The values for the coupling constants used in this work for QHD-1 were taken from Ref. \cite{recentprogress} (these values are preferred to those given in Ref. \cite{walecka}, as those in Ref. \cite{recentprogress} give a better description of the saturation properties of nuclear matter
). The couplings constants are given in terms of $C_s$ and $C_v$, which are defined as
\begin{eqnarray}
	C_s^2 = g_s^2\left(\frac{M^2}{m_s^2}\right)\ \ \ \mbox{and}\ \ \ C_v^2 =
	 g_v^2\left(\frac{M^2}{m_v^2}\right)\,.\nonumber
\end{eqnarray}
In Table \ref{tab:QHD1} the values for $C_s$ and $C_v$ and the corresponding values for $g_s$ and $g_v$, using the values of the particle masses in Table \ref{tab:QHD1mass}, are given.
\begin{table*}[bbb]
	\centering
		\begin{tabular}{cccc}
			\hline\hline
		  $C_s^2$ & $C_v^2$&$g_s^2$ & $g_v^2$\\
			\hline\\
			 357.4 & 273.8 & 109.6 &  190.4\\
			\hline
		\end{tabular}
	\caption{Coupling constants of the QHD-I parameter set.}
	\label{tab:QHD1}
\end{table*}
\section{Equation of state}\label{secQHD1eos}
Using the results for the expectation values obtained in Sec. \ref{secQHD1expect}, the RMF equations of motion for the fields (\ref{MFTQHD1EQM}) reduce to
\begin{subequations}
	\begin{eqnarray}
		 \phi_{0} &=& \frac{g_{s}}{m_{s}^{2}}\left\langle\bar{\psi}\psi\right\rangle 
		 = \frac{g_{s}}{m_{s}^{2}}\,\frac{\gamma}{2\pi^2}\label{QHD1scalar}
		 \int^{k_F}_0 dk\,\frac{ {k}^2\,m^*}{\sqrt{{k}^2 + {m^*}^2}}\,,\\
		 V_{0} &=& \frac{g_{v}}{m_\omega^{2}}\left\langle\psi^{\dagger}\psi\right\rangle 
		 = \frac{g_{v}}{m_\omega^{2}}\rho\,,\label{QHD1vector}
	\end{eqnarray}
\end{subequations}	
while the energy density and pressure, from (\ref{MFTeos}) and using Eq. (\ref{QHD1vector}) to simplify the expression for the energy density, are
\begin{subequations}
	\begin{eqnarray}
		\epsilon = \frac{1}{2}m_{s}^{2}\phi_{0}^{2} + 	\frac{1}{2}m_\omega^{2}V_{0}^{2} + 
								\frac{\gamma}{(2\pi)^{3}}\int^{k_{f}}_{0}d{\bm k}\,
								\sqrt{{\bm k} ^{2} 	+ (M - g_{s}\phi_{0})^{2}}\label{eps1}\,,\\
		P = -\frac{1}{2}m_{s}^{2}\phi_{0}^{2} + 	\frac{1}{2}m_\omega^{2}V_{0}^{2} + 
								\frac{1}{3}\left(\frac{\gamma}{(2\pi)^{3}}\int^{k_{f}}_{0}d{\bm k}\frac{{\bm k}^{2}}{\sqrt{{\bm k}^{2}
								 +
								 (M - g_{s}\phi_{0})^{2}}}\right)\,.\label{pres1}
	\end{eqnarray}
\end{subequations}	
The expression for $\phi_0$ could have also been obtained from the thermodynamic argument that a closed, isolated system such as the one assumed for the RMF approximation will minimise its energy. Therefore an expression for $m^*$ (and therefore $\phi_0$) can be obtained by minimising the energy density (\ref{eps1}) with regards to $m^*$. In this way the expression for $m^*$ is obtained,
\begin{eqnarray}
	m^* = M - \frac{g_{s}^2}{m_{s}^{2}}\,\frac{\gamma}{2\pi^2}
		 \int^{k_F}_0 dk\,\frac{ k^2\,m^*}{\sqrt{{k}^2 + {m^*}^2}}\,,\label{mreduced}
\end{eqnarray}
which is equivalent to Eq. (\ref{QHD1scalar}) when the definition for $m^*$ (\ref{mreddef}) is considered.\\
\\
Imposing spherical symmetry, the energy density (\ref{eps1}) and the pressure (\ref{pres1}) can be re-written as:
\begin{subequations}
	\begin{eqnarray}
		\epsilon &=& \frac{1}{2}m_{s}^{2}\phi_{0}^{2} + 	\frac{1}{2}m_\omega^{2}V_{0}^{2} + 
								\frac{\gamma}{2\pi^{2}}\int^{k_{f}}_{0}dk\,k^{2}\sqrt{k^{2} + {m^{*}}^{2}}\label{QHD1eps}\\
		P &=& -\frac{1}{2}m_{s}^{2}\phi_{0}^{2} + 	\frac{1}{2}m_\omega^{2}V_{0}^{2} + 
								\frac{1}{3}\left(\frac{\gamma}{2\pi^{2}}\int^{k_{f}}_{0}dk\frac{k^{4}}{\sqrt{k^{2} + 
								{m^{*}}^{2}}}\right)\label{QHD1pres}.						
	\end{eqnarray}
\end{subequations}
\section{Observables of nuclear matter in QHD-I}
\subsection{Symmetry energy}\label{QHD1symm}
To calculate the symmetry coefficient, $a_4$, given by (\ref{a4})
\[
		a_{4} = \frac{1}{2}\left(\frac{\partial^{2}}{\partial t^{2}}\frac{\epsilon}{\rho}\right)_{t=0}\ \ \ \ \  \left(t\equiv\frac{\rho_{n} -
		 \rho_{p}}{\rho}\right) ,
\]
the energy density must be expressed in terms of $t$. This can be done by considering the \mbox{expressions} for the proton ($\rho_p$), neutron ($\rho_n$) and baryon ($\rho$) number densities. These are given in terms of the different fermi momenta ($k_p$ for the protons and $k_n$ for the neutrons) of the species as
\begin{eqnarray}
	\rho_p &=& \frac{1}{\pi^{2}}\int^{k_{p}}_{0}dk\,k^{2} = \frac{k_p^3}{3\pi^2},\nonumber\\
	\rho_n &=& \frac{1}{\pi^{2}}\int^{k_{n}}_{0}dk\,k^{2} = \frac{k_n^3}{3\pi^2},\nonumber\\ \rho &=& \rho_p + \rho_n\nonumber.
\end{eqnarray}
In the case of symmetric nuclear matter ($t = 0$) the proton and neutron fermi momenta are equal, $k_p$ = $k_n$ = $k_F$, and the baryon density reduces to
\begin{eqnarray}
	\rho_0 = 2 \times \frac{1}{\pi^{2}}\int^{k_{F}}_{0}dk\,k^{2} = \frac{2k_F^3}{3\pi^2}\,\,.\label{rho0}
\end{eqnarray}
The only terms in the energy density (\ref{QHD1eps}) that will contribute to $a_4$ are the integrals over the filled baryon states. By defining $\epsilon_s$ as the isospin dependent part of the energy density, which is
\begin{eqnarray}
	\epsilon_s = 	\frac{1}{\pi^{2}}\int^{k_{n}}_{0}dk\,k^{2}\sqrt{k^{2} + {m^{*}}^{2}} + 
		\frac{1}{\pi^{2}}\int^{k_{p}}_{0}dk\,k^{2}\sqrt{k^{2} + {m^{*}}^{2}}\,,\nonumber
\end{eqnarray}
where the nucleon degeneracy, $\gamma$, has been explicitly included (as 2 for each baryon), the \mbox{expression} for $a_4$ (\ref{a4}) reduces to
\begin{eqnarray}
	a_{4} &=& \frac{1}{2}\left(\frac{\partial^{2}}{\partial t^{2}}\frac{\epsilon_s}{\rho}\right)_{t=0}\nonumber\\
		&=&\frac{1}{2}\left.\frac{\partial^{2}}{\partial t^{2}}
		\left(\frac{1}{\rho}\frac{1}{\pi^{2}}\int^{k_{n}}_{0}dk\,k^{2}\sqrt{k^{2} + {m^{*}}^{2}} + 
		\frac{1}{\rho}\frac{1}{\pi^{2}}\int^{k_{p}}_{0}dk\,k^{2}\sqrt{k^{2} + {m^{*}}^{2}}\right)\right|_{t=0}.\label{a4QHD1}
\end{eqnarray}
To evaluate the expression for $a_4$ 
the proton and neutron fermi momenta are expressed in terms of $t$, i.e.
\begin{eqnarray}
	k_n &=& \left[3\pi^2\frac{1}{2}\frac{k_p^3 + k_n^3}{3\pi^2}(1 + t)\right]^\frac{1}{3}\nonumber\\
			&=&  \left[\frac{k_p^3 + k_n^3}{2}(1 + t)\right]^\frac{1}{3}\label{kn}
\end{eqnarray}
and
\begin{eqnarray}
	k_p &=& \left(3\pi^2\frac{1}{2}\frac{k_p^3 + k_n^3}{3\pi^2}(1 - t)\right)^\frac{1}{3}\nonumber\\
			&=&  \left(\frac{k_p^3 + k_n^3}{2}(1 - t)\right)^\frac{1}{3}\,.\label{kp}
\end{eqnarray}
After substituting in the expressions for the different fermi momenta [(\ref{kn}) and (\ref{kp})] into the expression for $a_4$, as well as some algebra and remembering that $k_n$ and $k_p$ are equivalent at the evaluation point, (\ref{a4QHD1}) yields 
\begin{eqnarray}
		a_{4} &=& 
		 \frac{k_F^2}{6\sqrt{k_F^2 + m^{*2}}}\nonumber\, .
\end{eqnarray}
%
%
%

	

%
%
%
%
%
%
\section{Summary}
The aim of this chapter was to introduce the basic concepts of QHD and RMF within the context of the QHD-I parameter set. This chapter will serve as the foundation for the derivation of the equation of state of nuclear and neutron star matter in other, more advanced QHD parameter sets. These parameter sets will be discussed in the next chapter.

%% file: chapFSUG.tex
\chapter{Advanced QHD parameter sets}\label{chapFSUG}
\section{Introduction}
The coupling constants of QHD-I were fitted to reproduce the saturation of nuclear matter at a fermi wavenumber of 1.30 fm$^{-1}$ and binding energy of -15.75 MeV \cite{recentprogress}. Unfortunately this parameterisation produces too high a value of the nuclear matter compressibility ($K$). In 1977 J. Boguta and A. R. Bodmer proposed that self-couplings of the scalar meson field should be included in the Lagrangian density and the coupling constants re-adjusted to reproduce $K$ more accurately \cite{B+B}. These self-couplings of the scalar meson field are included in all other QHD parameter sets studied in this work. 
\\\\ 
Another addition is the inclusion of the charged (isovector) vector rho meson triplet ($\rho^0$, $\rho^\pm$) in all parameter sets discussed below. Since protons and neutrons practically only differ in terms of their isospin projections, the rho mesons are included to distinguish between these baryons and to give a better account of the symmetry energy \cite{csg}. (As these vector meson are charged, the reaction between a rho meson and a proton will differ from the reaction between a rho meson and a neutron.)\\ 
The rho mesons are introduced in a similar fashion as the other mesons in QHD: the free Lagrangian density describing the vector field is included in the QHD Lagrangian density as well as a coupling between the rho meson and the conserved isospin density \cite{csg, walecka}. 
\\\\
The {\bf NL-SH} parameter set is ``an expansion of the ideas of Boguta and Bodmer'' \cite{NLSH} and was formulated to reproduce the neutron radii of neutron-rich nuclei more accurately. This was to be achieved by fitting the coupling constants 
to reproduce the experimentally observed values of the root-mean-square neutron radii of neutron-rich nuclei \cite{NLSH}.\\
\\
In 1994 Y. Sugahara and H. Toki introduced a self-coupling in the omega meson field to the Lagrangian density and formulated the \textbf{TM1} and \textbf{TM2} parameter sets. This was done to obtain better agreement with other models for nuclei and nuclear matter. The parameters were fitted to obtain agreement with the observed properties of unstable nuclei \cite{TM1}. The TM1 \mbox{parameter} set is obtained when the fitting procedure is done to obtain agreement with light unstable nuclei and the TM2 parameter set for heavy unstable nuclei.\\
\\
According to Ref.\,\cite{NL3} the \textbf{NL3} parameter set was formulated by G. A. Lalazissis \textsl{et al.} to (amongst others) further improve the predicted value of the compressibility, $K$. The NL3 \mbox{parameter} set is based on the QHD-I Lagrangian density with the inclusion of rho mesons and self-couplings of the scalar field. This parameter set aimed to achieve a better description of \mbox{nuclei} with large isospin asymmetry, i.e. nuclei far from the valley of beta stability. The \mbox{parameters} were calculated by fitting the predicted values of various properties (such as binding energies, charge radii, and neutron radii of spherical nuclei) to the observed values of nuclei that has a large isospin asymmetry \cite{NL3}. \\
\\
As a further refinement to the TM1 parameter set, the \textbf{PK1} parameter set was obtained by fitting properties (such as binding energies and charge radii) of a wide range of heavy nuclei. This was done to, amongst others, better reproduce the nuclear symmetry energy and \mbox{compressibility} \cite{PK1}.\\
\\
B. Todd-Rutel and J. Piekarewicz introduced a coupling between the omega meson field and the rho meson field in an attempt to better describe the density dependence of the nuclear symmetry energy without changing the saturation properties of nuclear matter \cite{FSU}. The \textbf{FSUGold} parameter set was obtained by fitting the binding energies and charge radii of a variety of magic nuclei to the calculated properties.
\subsection{Photon field}
In general the photon (Coulomb) field is included in all the models mentioned above when studying the properties of finite nuclei. In this work the properties of infinite nuclear matter, specifically those which are applicable to the study of neutron star matter, will be studied. 
Since the Coulomb forces are much stronger than the gravitational forces on the nuclear scale, neutron stars are assumed to be charge neutral \cite{csg}, and therefore the contribution of the photon field is ignored in this treatment of nuclear matter.
\section{Lagrangian density}
The most general Lagrangian density, that describes the QHD-I, NL-SH, NL3, FSUGold, PK1, TM1 and TM2 parameter sets in QHD, is \cite{FSU1}:
\begin{eqnarray}\label{fsuglagrangian}
			{\cal L} &=& \bar{\psi}(x)\Big[\gamma^{\mu}\big(i\partial_{\mu} - g_{v}V_{\mu}(x)\big) 
			- \frac{g_\rho}{2}{\bm\tau}\cdot{\bm b}_\mu(x)
			- \big(M-g_{s}\phi(x)\big)\Big]\psi(x)\nonumber \\
			&+& \frac{1}{2}\partial_{\mu}\phi\partial^{\mu}\phi(x) - \frac{1}{2}m_s^{2}\phi^{2}(x) 
			- \frac{\kappa}{3!}\big(g_s\phi(x)\big)^3 - \frac{\lambda}{4!}\big(g_s\phi(x)\big)^4\nonumber \\
			&-& \frac{1}{4}V^{\mu\nu}V_{\mu\nu} + \frac{1}{2}m_\omega^{2}V^{\mu}(x)V_{\mu}(x) + 
			\frac{\zeta}{4!}\big(g_v^2V^\mu(x) V_\mu(x)\big)^2 \nonumber\\
			&-& \frac{1}{4}{\bm b}^{\mu\nu}{\bm b}_{\mu\nu} +
			 \frac{1}{2}m_{\rho}^{2}{\bm b}^{\mu}(x)\cdot{\bm b}_{\mu}(x)\nonumber\\
			&+& \Lambda_v\big(g_v^2V^\mu(x) V_\mu(x)\big)\big(g_\rho^2{\bm b}^{\mu}(x)\cdot{\bm b}_{\mu}(x)\big)\,,
		%
		%
\end{eqnarray}	
where the field tensors have been defined as
\begin{subequations}
	\begin{eqnarray}
		V_{\mu\nu} &=& \partial_\mu V_\nu(x) - \partial_\nu V_\mu(x) \nonumber\\
		{\bm b}_{\mu\nu} &=& \partial_\mu{\bm b}_{\nu}(x) - \partial_\nu{\bm b}_{\mu}(x)\nonumber
	\end{eqnarray}
\end{subequations}
with
\begin{itemize}
	\item $\psi(x)$ the isodoublet baryon field (only protons and neutrons are considered in this work),
	\item $\phi(x)$ the sigma (scalar) meson field,
	\item $V^\mu(x)$ the omega (vector) meson field, 
	\item ${\bf b}^\mu(x)$ is the Lorentz vector field denoting the three isospin components of the rho meson fields,
	$$
		{\bf b}^\mu = (b_1^\mu, b_2^\mu, b_3^\mu)\,.
	$$ 
	The charged rho meson fields ($\rho^\pm$) can be constructed in terms of the first two components of 
	${\bm b}^\mu$ as \cite{csg}
		\begin{eqnarray}
			b^\mu_\pm = \frac{1}{\sqrt{2}}(b^\mu_1 \pm b^\mu_2)\,, \label{oprho}
		\end{eqnarray}
		and,
	\item ${\bm \tau} = (\tau_1, \tau_2, \tau_3)$ the isospin operator. This operator is described in terms of the 
	Pauli 2$\times$2 spin-matrices
	 (\ref{pauli-spin}) as
		\[
			{\bm \tau}  = \left[
											\begin{array}{cc}
												{\bm \sigma} & 0\\
												0 & {\bm \sigma}				 
											\end{array}	
										\right].		
		\]
		It should be noted that since the nucleon field, $\psi$ is an isodoublet and therefore
		\[
			\psi = \left[
											\begin{array}{c}
												\psi_p\\
												\psi_n				 
											\end{array}	
										\right],\footnote{see \ref{sec:convent}}
		\]
		where is the $\psi_p$ is the proton field and $\psi_n$ is the neutron field, each consisting of the 4$\times$1 Dirac spinors, that the 				total dimension of $\psi$ is 8$\times$1. Therefore $\tau$ is in actual fact given by
			\begin{eqnarray}
				{\bm \tau} = {\bm \sigma} \otimes \left[\begin{array}[h]{cc}\textbf{1}_4 & 0\\0 &
				 \textbf{1}_4\end{array}\right]
			\end{eqnarray}
			and therefore the explicit expression for $\tau_3$, using explicit representation of $\sigma_3$ (\ref{pauli3}),
			 is
		
			\begin{eqnarray}
				\tau_3 
				= \left[\begin{array}[h]{cc}1 & 0\\0 & -1\end{array}\right] \otimes 
				\left[\begin{array}[h]{cc}\textbf{1}_4 & 0\\0 & \textbf{1}_4\end{array}\right]   
				= \left[\begin{array}[h]{cc}\textbf{1}_4 & 0\\0 & -\textbf{1}_4\end{array}\right]\label{tau3}
			\end{eqnarray}
\end{itemize}
Using the Euler-Lagrange equation (\ref{EL}) the equations of motion of the different fields are:
	\begin{subequations}\label{EQM}
	\begin{eqnarray}
			\partial_{\mu}\partial^{\mu}\phi + m_s^{2}\phi + \frac{\kappa}{2!}g_s^3\phi^2 + \frac{\lambda}{3!}g_s^4\phi^3  
			&=& g_{s}\bar{\psi}\psi\label{EQM1}\\
			\partial_{\mu}V^{\mu\nu} + m_\omega^{2}V^{\nu} + \frac{\zeta}{3!}g_v^4V_\nu^2 V^\nu + 
			2\Lambda_vg_v^2V^\nu g_\rho^2{\bm b}^{\mu}\cdot{\bm b}_{\mu}&=& g_{v}\bar{\psi}\gamma^{\nu}\psi\\
			\partial_{\mu}{\bm b}^{\mu\nu} + m_{\rho}^{2}{\bm b}^{\nu} + 
			2\Lambda_vg_v^2V^\nu V_\nu g_\rho^2{\bm b}^\nu&=&\frac{g_\rho}{2}\bar{\psi}\gamma^{\nu}{\bm \tau}\psi\\ 
			\Big[\gamma^{\mu}(i\partial_{\mu}-g_{v}V_{\mu} - \frac{g_\rho}{2}{\bm \tau}\cdot{\bf b}_\mu
			-(M-g_{s}\phi)\Big]\psi &=& 0.\label{EQM2}
	\end{eqnarray}
\end{subequations}	
\section{Conventions}\label{sec:convent}
In the literature two different conventions for expressing the Lagrangian density of QHD are used. These differences are obvious when the Lagrangian density and the values of the coupling constants of the NL3 parameter set in Ref.\,\cite{FSU} are compared to those in Ref.\,\cite{NL3}.\\
\\
The way in which the Lagrangian density (\ref{fsuglagrangian}) is expressed will be referred to as the \emph{Walecka} convention, which is used in the original formulation of QHD 
by J. D. Walecka (see Ref.\,\cite{walecka}). The other convention of expressing the Lagrangian density is mainly used by authors from Europe and the Orient. This formulation will be referred to as the \emph{Ring} convention.\\
\\
The differences are:
\begin{enumerate}
	\item In the Walecka convention the scalar field $\phi$ is explicitly positive, where as in the Ring convention it 
	is negative.
	\item The first four components in the nucleon field, $\psi$, of the Walecka convention refer to the proton field and 
	the last four to neutron field, whereas in the Ring convention the first four components of the nucleon field refer
	to the neutron field and the last four to the proton field.
	\item In the Walecka convention the coupling between the rho-field, ${\bm b}_\mu$ and the isospin density, $g_\rho$ 
	has an explicit factor of a half and in the Ring convention this factor is implied. \\
\end{enumerate}
\begin{table}[htb]
	\centering
		\begin{tabular}{ccc}
			\hline\hline
			Property & Walecka & Ring\\
			\hline
			Reduced nucleon mass ($m^*$)& $m^* = M - g_s\phi$& $m^* = M + g_s\phi$\\
			\\
			Coupling between the rho field and \\nucleon fields, 
			$g_\rho$ & $\bar{\psi}\gamma^\mu (g_\rho/2){\bm \tau}\cdot{\bm b}_\mu\psi$ & 
			$\bar{\psi}\gamma^\mu g_\rho{\bm \tau}\cdot{\bm b}_\mu\psi$\\
			\\
			Third-order self-coupling constant \\in the scalar meson field & $\kappa$ & $g_2$\\
			\\
			Fourth-order self-coupling constant \\in the scalar meson field & $\lambda$ & $g_3$\\
			\\
			Self-coupling constant in the omega\\meson field & $\zeta$ & $c_3$\\
			\\
			Coupling between the rho- and the omega\\meson fields & $\Lambda_v$ & Not applicable\\
			\\
			Nucleon field, $\psi$ & $\psi = \left[\begin{array}{c}\psi_p\\\psi_n\end{array}\right]$ & 
			$\psi = \left[\begin{array}{c}\psi_n\\\psi_p\end{array}\right]$\\
			\\
			\hline
		\end{tabular}
	\caption{Differences in Walecka and Ring expressions of the Lagrangian density of QHD parameter sets for nuclear matter.}
	\label{tab:convention}
\end{table}
{\bf In this work the Walecka convention will be used.} The expressions in the Ring convention will be given for certain properties, but will be identified as such. Both conventions are included in this work, because the different QHD parameter sets were studied in the specific context that they were defined in. It should also be noted that while certain terms will be present in one convention it might not be present in the other. This is due to the fact that different parameter sets include different couplings and as such certain couplings might only be used in parameter sets studied by authors using a specific convention.\\
\\
It should also be noted that the coupling constants are defined separately for each convention.\\
\\
The differences between the two conventions are contrasted in Table \ref{tab:convention}.
\subsection{Note on $m^*$}
The reduced mass (\ref{mreddef}),
\begin{eqnarray}
	m^* = (M - g_s\phi_0)\,,\label{mreduced1}
\end{eqnarray}
is only defined in terms of the generic baryon mass, $M$. The baryons considered in this work are protons and neutrons and their masses are usually taken to be equal (939 MeV). However, the PK1 parameter set distinguishes between the proton and the neutron mass (see Table \ref{tab:orientmasses}). This distinction should be taken (and is taken into account in this work) when calculating $m^*$ in the PK1 parameter set, but it is not explicitly shown in the expressions for $m^*$.
\section{Relativistic mean-field theory}
As in Chapter \ref{chapQHD1} the equations of motion for the different fields are solved in the relativistic mean-field approximation, which assumes that the ground state of a uniform, static, spherical symmetric system is described and therefore the meson fields can be replaced their expectation values \cite{FSU1}:
\begin{eqnarray}
		\phi(x) &\longrightarrow& \left\langle \phi(x)\right\rangle = \phi_{0}\nonumber\\
		V^{\mu}(x) &\longrightarrow& \left\langle V^\mu(x)\right\rangle = g^{\mu 0}V_{0}\nonumber\\
		{\bm b}^{\mu}(x) &\longrightarrow& \left\langle b^{\mu}_a(x)\right\rangle = g^{\mu 0}\delta_{a3}b_0\nonumber
\end{eqnarray}
The spatial components of ${\bm b}^\mu$ and $V^\mu$ vanish due to the consideration of a uniform, static ground state (as was explained in Chapter \ref{chapQHD1}). The ground state is also assumed to have definite spin and parity. Since the first two components of ${\bm b}^\mu$ can be written in terms of raising and lowering operators of the charged rho meson fields (\ref{oprho}), only the third component (that describes the neutral rho meson, $\rho^0$) has a non-vanishing expectation value in the RMF approximation \cite{csg}.\\
\\
Correspondingly the baryon sources in the equations of motion are replaced by their normal-order expectation values in the RMF grondstate, $\left|\Phi\right\rangle$,
\begin{eqnarray}
	\bar\psi(x)\psi(x) &\longrightarrow &
	\left\langle\Phi\right| {\bm :}\bar\psi(x)\psi(x){\bm :}\left|\Phi\right\rangle = 
	\left\langle \bar{\psi}\psi\right\rangle\nonumber\\
	\bar\psi(x)\gamma^\mu\psi(x) &\longrightarrow& 
	\left\langle\Phi\right| {\bm :}\bar\psi(x)\gamma^\mu\psi(x){\bm :}\left|\Phi\right\rangle = 
	\left\langle \bar{\psi}\gamma^0\psi\right\rangle\nonumber\\
	\bar\psi(x)\gamma^\mu\tau_a\psi(x) &\longrightarrow &
	\left\langle\Phi\right| {\bm :}\bar\psi(x)\gamma^\mu\tau_a\psi(x){\bm :}\left|\Phi\right\rangle =
	 \left\langle \bar{\psi}\gamma^0\tau_3\psi\right\rangle\nonumber.
\end{eqnarray}
In the RMF approximation, the equations of motion of the fields (\ref{EQM}) reduce to:
\begin{subequations}\label{MFTFSUEQM}
	\begin{eqnarray}
			g_s\phi_0 &=& \frac{g_{s}^2}{m_s^2}\left[\left\langle\bar\psi\psi\right\rangle - 
			\frac{\kappa}{2}(g_s\phi_0)^2 - \frac{\lambda}{6}(g_s\phi_0)^3\right]\label{FSUsigmaEQM}\\
			g_{v}V_0 &=& \frac{g_{v}^2}{m_\omega^2}\left[\left\langle \psi^{\dagger}\psi\right\rangle - 
			\frac{\zeta}{6}(g_vV_0)^3 -  2\Lambda_v(g_vV_0)(g_\rho b_0)^2\right]\label{FSUomegaEQM}\\
			g_{\rho}b_0 &=&
			 \frac{g_\rho^2}{m_\rho^2}\left[\frac{1}{2}\left\langle\psi^\dagger\tau_3\psi\right\rangle - 
			 2\Lambda_v(g_vV_0)^2(g_\rho b_0)\right]\label{FSUrhoEQM}\\
			 0 &=& 
			\left[i\gamma^{\mu}\partial_{\mu}-g_{v}\gamma^0V_{0} - \frac{g_\rho}{2}\tau_3\gamma^0b_0
			- (M-g_{s}\phi)\right]\psi\label{FSUbaryon}
	\end{eqnarray}	
\end{subequations}
In the Ring convention\footnote{The coupling between the rho- and the omega meson fields, $\Lambda_v$, is not included in any of the parameter sets studied within the Ring convention.} the equations of motion are given by:
\begin{subequations}\label{MFTRingEQM}
	\begin{eqnarray}
			m_s^2\phi_0 &=& -g_s\left\langle\bar\psi\psi\right\rangle -g_2\phi_0^2 - g_3\phi_0^3\label{PK1sigmaEQM}\\
			m_\omega^2V_0 &=& g_v\left\langle \psi^{\dagger}\psi\right\rangle - c_3 V_0^3\label{PK1omegaEQM}\\
			m_{\rho}^2b_0 &=& g_\rho\left\langle\psi^\dagger\tau_3\psi\right\rangle\label{PK1rhoEQM}\\
			0 &=& \left[i\gamma^{\mu}\partial_{\mu}-g_{v}\gamma^0V_{0} - g_\rho\tau_3b_0
			- (M+g_{s}\psi)\right]\psi \,.
	\end{eqnarray}
\end{subequations}
In the RMF approximation the Lagrangian density (\ref{fsuglagrangian}) reduces to 
	\begin{eqnarray}
		{\cal L}_{RMF} &=& \bar{\psi}(x)\Big[\gamma^{\mu}\big(i\partial_{\mu} - g_{v}V_{0}(x)\big) 
			- \frac{g_\rho}{2}\tau_3b_0
			- \big(M-g_{s}\phi_0\big)\Big]\psi(x)\nonumber \\
			&-& \frac{1}{2}m_s^{2}\phi^{2}_0 
			- \frac{\kappa}{3!}\big(g_s\phi_0\big)^3 - \frac{\lambda}{4!}\big(g_s\phi_0\big)^4
			+ \frac{1}{2}m_\omega^{2}V_0^2
			+ \frac{\zeta}{4!}\big(g_vV_0\big)^4 \nonumber\\
			&+& \frac{1}{2}m_{\rho}^{2}b_0^2 + \Lambda_v\big(g_vV_0\big)^2\big(g_\rho b_0\big)^2\,.	\label{FSUL}
	\end{eqnarray}
Using the RMF Lagrangian density and (\ref{epsp13}) [keeping Eq. (\ref{FSUbaryon}) in mind] 
the energy density and the pressure are given by 
\begin{subequations}\label{FSUeos1}
	\begin{eqnarray}
		\epsilon &=& \left\langle T^{00}\right\rangle \nonumber\\ &=& 
		\left\langle i\bar{\psi}\gamma_{0}\partial_{0}\psi\right\rangle - 
		\left\langle {\cal L}\right\rangle\label{FSUeps1}\,,\\
		P&=& \frac{1}{3}\left\langle T^{ii} \right\rangle\nonumber\\&=&
		\frac{1}{3}\left\langle i\bar{\psi}\gamma^{i}\partial_{i}\psi\right\rangle + \left\langle {\cal L}\right\rangle 
		\label{FSUpres1}\,,
	\end{eqnarray}
\end{subequations}
where
	$\left\langle {\cal L}\right\rangle$ is given by 
			\begin{eqnarray}
					\left\langle {\cal L}\right\rangle &=& - \frac{1}{2}m_s^{2}\phi^{2}_0 
					- \frac{\kappa}{3!}\big(g_s\phi_0\big)^3 - \frac{\lambda}{4!}\big(g_s\phi_0\big)^4
					+ \frac{1}{2}m_\omega^{2}V_0^2 + \frac{\zeta}{4!}\big(g_vV_0\big)^4 \nonumber\\
					&&+ \frac{1}{2}m_{\rho}^{2}b_0^2 + \Lambda_v\big(g_vV_0\big)^2\big(g_\rho b_0\big)^2.\label{varL}
			\end{eqnarray}
%
%
%
%
%
\section{Evaluation of expectation values}\label{secFSUexp}
The expectation values in the equations of motion of both conventions [Eqs (\ref{MFTFSUEQM}) and (\ref{MFTRingEQM})] and the equation of state (\ref{FSUeos1}), were evaluated using the method given by N.K. Glendenning in Ref.\,\cite{csg}, which is explained in Sec.\,\ref{secQHD1expect}.\\
\\
The calculation is essentially the same as in Sec.\,\ref{secQHD1expect}, with the only difference being that the expression for the Dirac Hamiltonian, $H_D$, and the positive single-particle energies \big($e({\bm k})$\big) are slightly different. In this case, $H_D$ is given by
\begin{eqnarray}
	H_D = \gamma_0\Big[{\bm\gamma}\cdot{\bm k} + g_v \gamma_0 V_0 + 
	\frac{1}{2}g_\rho\,\tau_3\,b_0 + m^*\Big]
\end{eqnarray}
and $e({\bm k})$ by
\begin{eqnarray}
	e({\bm k})= g_v V_0 + \frac{1}{2}g_\rho\,\tau_3\,b_0 +\sqrt{{\bm k}^2 + {m^*}^2}\label{ekfsu}
\end{eqnarray}
where $m^* = (M-g_s\phi_0)$ and the expansion for $\psi$ is taken to be as in (\ref{Kexpand}), with only the value for $e({\bm k})$ changing.\\
\\
Using the above it can be established that the expression of $\left\langle\psi^\dagger\psi\right\rangle$ will not change from that in Chapter \ref{chapQHD1}, except in this case the contribution of the baryons are evaluated separately (each baryon has 2 spin states, and $k_n$ and $k_p$ denote the fermi momenta of the different species),
\begin{eqnarray}
	\left\langle\psi^\dagger\psi\right\rangle &=& \sum_s\frac{1}{(2\pi)^3}\int_0^{k_p} d{\bm k} +
	 \sum_s\frac{1}{(2\pi)^3}\int_0^{k_n} d{\bm k}\nonumber\\
	&=& \frac{1}{3\pi^2}{k_p}^3 + \frac{1}{3\pi^2}{k_n}^3\nonumber\\
	&=& \rho_p + \rho_n \label{rho}.
\end{eqnarray}
\\
The same applies to the expression for $\left\langle\bar\psi\psi\right\rangle$,
\begin{eqnarray}
	\left\langle\bar\psi\psi\right\rangle	&=& 
	\frac{1}{\pi^2}\,\int^{k_p}_0 dk\,\frac{ k^2\,m^*}{\sqrt{{k}^2 + {m^*}^2}} 
	+ \frac{1}{\pi^2}\,\int^{k_n}_0 dk\,\frac{ k^2\,m^*}{\sqrt{{k}^2 + {m^*}^2}}
\end{eqnarray}
To evaluate $\left\langle\psi^\dagger\tau_3\psi\right\rangle$, a slightly different approach will be used. Since the baryon field can be expanded in terms of the different baryon components (in this case protons and neutrons), each described by a four component Dirac spinor, $\psi$ can be written as 
$$ 
	\psi = \left[\begin{array}{c}\psi_p\\\psi_n\end{array}\right]\,.
$$
$\tau_3$ is given by Eq. (\ref{tau3}) and therefore the effect of the operation of $\tau_3$ will be to separate the contribution of the baryons into a positive (proton) and negative (neutron) parts. The single particle expectation value will therefore be given, since the states are normalised, by
$$
	\big(\psi^\dagger\tau_3\psi\big) = \left\{\begin{array}{cc}
																				\big(\psi_p^\dagger\tau_3\psi_p\big)_p = 1\\
																				\big(\psi_n^\dagger\tau_3\psi_n\big)_n = -1
																		\end{array}\right.,
$$
and thus
\begin{eqnarray}
	\left\langle\psi^\dagger\tau_3\psi\right\rangle &=& \sum_s\frac{1}{(2\pi)^3}\int_0^{k_p} d{\bm k} -
	 \sum_s\frac{1}{(2\pi)^3}\int_0^{k_n} d{\bm k}\nonumber\\
	&=& \frac{1}{3\pi^2}{k_p}^3 - \frac{1}{3\pi^2}{k_n}^3\nonumber\\
	&=& \rho_p - \rho_n \label{iso} .
\end{eqnarray}
\\
Since the following expression still holds
\begin{eqnarray}
	\frac{\partial e({\bm k})}{\partial{\bm k}}\cdot{\bm k}
	&=& \frac{{\bm k}\cdot{\bm k}}{\sqrt{{\bm k}^2 + {m^*}^2}}\nonumber\,,
\end{eqnarray}
the expectation value in the expression of the pressure (\ref{FSUpres1}) does not change, and is therefore still given by (\ref{NKGpresexp})
\begin{eqnarray}
	\left\langle i\bar{\psi}\gamma^{i}\partial_{i}\psi\right\rangle = 
	\frac{1}{4\pi^3}\int^{k_p}_0d{\bm k}\,\frac{{\bm k}^2}{\sqrt{{\bm k}^2 + {m^*}^2}}+ 
	\frac{1}{4\pi^3}\int^{k_n}_0d{\bm k}\,\frac{{\bm k}^2}{\sqrt{{\bm k}^2 + {m^*}^2}} \label{presFSUexp}\,.
\end{eqnarray}
However, the expression for $e({\bm k})$ differs from the one in Chapter \ref{chapQHD1} [compare (\ref{MFTenergy}) and (\ref{ekfsu})]. The expectation value in the expression for the energy density is now given by
\begin{eqnarray}
	\left\langle i\bar{\psi}\gamma_{0}\partial_{0}\psi\right\rangle
	&=&  \sum_s\int\frac{d{\bm k}}{(2\pi)^3}\,e({\bm k})\,\Theta[\,\mu-e({\bm k})]\nonumber\\ 
	&=&  \sum_s\int\frac{d{\bm k}}{(2\pi)^3}\,\big(g_v V_0 + \frac{1}{2}g_\rho\,\tau_3\,b_0 +\sqrt{{\bm k}^2 
	+ {m^*}^2}\big)\,\Theta[\,\mu-e({\bm k})]\nonumber\\ 
	&=&  g_v V_0\rho + \frac{1}{2}g_\rho\,b_0\big(\rho_p-\rho_n\big) + 
	\sum_s\int\frac{d{\bm k}}{(2\pi)^3}\,\sqrt{{\bm k}^2 + {m^*}^2}\,\Theta[\,\mu-e({\bm k})]\nonumber\\ 
	&=&  g_v V_0\rho + \frac{1}{2}g_\rho\,b_0\big(\rho_p-\rho_n\big)\nonumber\\&& 
	+ \frac{1}{4\pi^2}\int^{k_p}_0d{\bm k}\,\sqrt{{\bm k}^2 + {m^*}^2} 
	+\frac{1}{4\pi^2}\int^{k_n}_0d{\bm k}\,\sqrt{{\bm k}^2 + {m^*}^2} \label{epsFSUexp} 
\end{eqnarray}

%
%
%
%
\section{Parameter sets}
The parameters for the QHD-I calculations done in this work are those given by B.D. Serot and J.D. Walecka in Ref.\,\cite{recentprogress}.\\
The Ring NL3 parameters are taken from the original article by G.A. Lalazissis \textsl{et al.} \cite{NL3}, while the parameters in the Walecka convention are from Ref.\,\cite{FSU}.\\
The parameters for the PK1 model are taken from the article by W. Long \textsl{et al.} \cite{PK1}, while the NL-SH and TM1 (as well as TM2) paramater sets are taken from the article by M.M. Sharma and M.A. Nagarajan \cite{NLSH} and the article by Y. Sugahara and H. Toki \cite{TM1}, respectively.\\
The parameters for the FSUGold model and the NL3 parameters in the Walecka convention are given by B. Todd-Rutel and J. Piekarewicz in Ref.\,\cite{FSU}.\\
\\
The values for the coupling constants and masses for models studied in the Walecka convention are given in Tables \ref{tab:westcoupling} and \ref{tab:westmasses} respectively, while those for the Ring convention are given in Tables \ref{tab:orientcoupling} and \ref{tab:orientmasses}.\\
\begin{table*}[hbt]
	\centering
		\begin{tabular}{cccccccc}
			\hline\hline
			Model & $g_s^2$ & $g_v^2$ & $g_\rho^2$& $\kappa$ (MeV) & $\lambda$ & $\zeta$ & $\Lambda_v$ \\
			\hline
			NL3  & 104.3871 & 165.5854 & 79.6000 & 3.8599 & -0.01591 & 0.00 & 0.00 \\
			S271  & 81.1071  & 116.7655 & 85.4357 & 6.6834 & -0.01580 & 0.00 & 0.00 \\
			Z271&  49.4401  & 70.6689  & 90.2110 & 6.1696 & +0.15634 & 0.06 & 0.00 \\
			FSUGold & 112.1996 & 204.5469 & 138.4701& 1.4203 & +0.0238  & 0.0600 & 0.0300 \\
			\hline
		\end{tabular}
	\caption{Coupling constants of different parameter sets defined in the Walecka convention. All coupling constants
	 are dimensionless, except for $\kappa$ which is given in MeV.}
	\label{tab:westcoupling}
\end{table*}
\begin{table*}[htb]
	\centering
		\begin{tabular}{cccccc}
			\hline\hline
			Model & $M_{\mbox{\footnotesize{proton}}}$& $M_{\mbox{\footnotesize{neutron}}}$& $m_s$& $m_\omega$& $m_\rho$ \\
			\hline
			NL3 & 939 & 939 & 508.1940 & 782.5  & 763\\
			S271 & 939 & 939 & 505.000 & 783  & 763\\
			Z271&939 & 939 & 465.000 & 783  & 763\\
			FSUGold & 939 & 939 & 491.5000 & 783  & 763\\
			\hline
		\end{tabular}
	\caption{Particle masses (in MeV) of the parameter sets defined in the Walecka convention.}
	\label{tab:westmasses}
\end{table*}
\begin{table*}[hbt]
	\centering
		\begin{tabular}{ccccccc}
			\hline\hline
			Model & $g_s$ & $g_v$ & $g_\rho$& $g_2$(fm$^{-1}$) & $g_3$ & $c_3$ \\
			\hline
			NL3 & 10.217 & 12.868 & 4.474 & -10.431 & -28.885 & 0.00\\
			PK1  & 10.3222 & 13.0131 & 4.5297 & -8.1688 & -9.9976 & 55.636\\
			TM1 & 10.0289 & 12.6139 & 4.6322 & -7.2325 & 0.6183 & 71.3075\\
			TM2 & 11.4694 & 14.6377 & 4.6783 & -4.4440 & 4.6076 & 84.5318\\
			NL-SH  & 10.444 & 12.945 & 4.383 & -6.9099 & -15.8337 & 0.00\\
			\hline
		\end{tabular}
	\caption{Coupling constants of different parameter sets defined in the Ring convention. All coupling constants are dimensionless, 
	except for $g_2$ which is given in fm$^{-1}$.}	\label{tab:orientcoupling}
\end{table*}
\begin{table*}[tth]
	\centering
		\begin{tabular}{cccccc}
			\hline\hline
			Model & $M_{\mbox{\footnotesize{proton}}}$& $M_{\mbox{\footnotesize{neutron}}}$& $m_s$& $m_\omega$& $m_\rho$ \\
			\hline
			NL3 & 939 & 939 & 508.194 & 782.501  & 763.0000\\
			PK1  & 939.5731 & 938.2796 & 514.0891 & 784.254 & 763\\
			TM1  & 938 & 938 & 511.198 & 783.0 & 770.0\\
			TM2  & 938 & 938 & 526.443 & 783.0 & 770.0 \\
			NL-SH  & 939 & 939 & 526.059 & 783.00 & 763.00\\
			\hline
		\end{tabular}
	\caption{Particle masses (in MeV) of the parameter sets defined in the Ring convention.}
	\label{tab:orientmasses}
\end{table*}

%
%
%
\newpage
\section{Equation of state}
The equations of motion of the different meson fields (\ref{MFTFSUEQM}) reduce to the following expressions when the results of Sec.\,\ref{secFSUexp} are taken into account. In the Walecka convention the equations of motion (\ref{MFTFSUEQM}) reduce to 
\begin{subequations}\label{FSUEQM}
	\begin{eqnarray}
			g_s\phi_0 &=& \frac{g_{s}^2}{m_s^2}\Big( 
			\frac{1}{\pi^2}\,\int^{k_p}_0 dk\,\frac{ k^2\,m^*}{\sqrt{{k}^2 + {m^*}^2}} 
			+ \frac{1}{\pi^2}\,\int^{k_n}_0 dk\,\frac{ k^2\,m^*}{\sqrt{{k}^2 + {m^*}^2}}\Big) \nonumber\\
			&& +\frac{g_{s}^2}{m_s^2}\Big( - 
			\frac{\kappa}{2}(g_s\phi_0)^2 - \frac{\lambda}{6}(g_s\phi_0)^3\Big)\label{FSUsigma}\\
			g_{v}V_0 &=& \frac{g_{v}^2}{m_\omega^2}\Big(\rho_p + \rho_n - 
			\frac{\zeta}{6}(g_vV_0)^3 -  2\Lambda_v(g_vV_0)(g_\rho b_0)^2\Big)\label{FSUomega}\\
			g_{\rho}\,b_0 &=&
			 \frac{g_\rho^2}{m_\rho^2}\Big(\frac{1}{2}\big(\rho_p - \rho_n \big) - 
			 2\Lambda_v(g_vV_0)^2(g_\rho b_0)\Big)\label{FSUrho}
	\end{eqnarray}	
\end{subequations}
The equations of motion of the fields in the Ring convention (\ref{MFTRingEQM}) are, after also considering the results of Sec.\,\ref{secFSUexp}, as well as the difference between the conventions (Table \ref{tab:convention}),
\begin{subequations}\label{FSURingEQM}
	\begin{eqnarray}
			m_s^2\phi_0 &=& -\,g_s\Big(\frac{1}{\pi^2}\,\int^{k_p}_0 dk\,\frac{ k^2\,m^*}{\sqrt{{k}^2 + {m^*}^2}} 
				+ \frac{1}{\pi^2}\,\int^{k_n}_0 dk\,\frac{ k^2\,m^*}{\sqrt{{k}^2 + {m^*}^2}}\Big)\nonumber\\
	 			&&-\,g_2\phi_0^2 - g_3\phi_0^3\label{PK1sigma}\\
			m_\omega^2\,V_0 &=& g_v\big(\,\rho_n + \rho_p\,\big) - c_3 V_0^3\label{PK1omega}\\
			m_{\rho}^2\,b_0 &=& g_\rho\big(\,\rho_n - \rho_p\,\big)\label{PK1rho}\,.
	\end{eqnarray}
\end{subequations}
Using the expression for $\left\langle {\cal L}\right\rangle$ (\ref{varL}) and the results of Sec.\,\ref{secFSUexp}, the energy density (\ref{FSUeps1}) and the pressure (\ref{FSUpres1}) in the Walecka convention yields
\begin{subequations}\label{FSUEoSWalecka}
	\begin{eqnarray}
		\epsilon &=&
		\ \ \frac{1}{2}m_s^{2}\phi_0^{2} + \frac{\kappa}{3!}(g_s\phi_0)^3 + \frac{\lambda}{4!}(g_s\phi_0)^4 -
		 \frac{1}{2}m_\omega^{2}V_{0}^2 - \frac{\zeta}{4!}(g_vV_0)^4 \nonumber\\
		&& - \frac{1}{2}m_{\rho}^{2}b_0^2 - \Lambda_v(g_vV_0)^2 (g_\rho b_0)^2 + g_vV_0(\rho_n + \rho_p) 
		+ \frac{1}{2}g_\rho b_0(\rho_p - \rho_n)\nonumber\\
		&&+\frac{1}{\pi^{2}}\int^{k_{p}}_{0}dk\,k^{2}\sqrt{k^{2}+{m^{*}}^{2}}
		  + \frac{1}{\pi^{2}}\int^{k_{n}}_{0}dk\,k^{2}\sqrt{k^{2}+{m^{*}}^{2}}\label{FSUeps}
	\\\nonumber\\
		P &=& - \frac{1}{2}m_s^{2}\phi_0^{2} - \frac{\kappa}{3!}(g_s\phi_0)^3 - \frac{\lambda}{4!}(g_s\phi_0)^4 + 
		\frac{1}{2}m_\omega^{2}V_{0}^2 + \frac{\zeta}{4!}(g_vV_0)^4 \nonumber\\
		&&+ \frac{1}{2}m_{\rho}^{2}b_0^2 + \Lambda_v(g_vV_0)^2 (g_\rho b_0)^2 \nonumber\\
		 &&+ \frac{1}{3\pi^{2}}\int^{k_{p}}_{0}dk\,\frac{k^{4}}{\sqrt{k^{2}+{m^{*}}^{2}}} + 
		 \frac{1}{3\pi^{2}}\int^{k_{n}}_{0}dk\,\frac{k^{4}}{\sqrt{k^{2}+{m^{*}}^{2}}}\label{FSUpres}\,.
	\end{eqnarray}
\end{subequations}	
In the Ring convention the energy density and the pressure are given by
\begin{subequations}
	\begin{eqnarray}
		\epsilon &=&
		\ \ \frac{1}{2}m_s^{2}\phi_0^{2} + \frac{g_2}{3}\phi_0^3 + \frac{g_3}{4}\phi_0^4 - \frac{1}{2}m_\omega^{2}V_{0}^2 -
		 \frac{c_3}{4}V_0^4 \nonumber\\
		&& - \frac{1}{2}m_{\rho}^{2}b_0^2 + g_vV_0(\rho_n + \rho_p) 
		+ g_\rho b_0(\rho_n - \rho_p) \nonumber\\
		&& +\frac{1}{\pi^{2}}\int^{k_{p}}_{0}dk\,k^{2}\sqrt{k^{2}+{m^{*}}^{2}}
		  + \frac{1}{\pi^{2}}\int^{k_{n}}_{0}dk\,k^{2}\sqrt{k^{2}+{m^{*}}^{2}}\label{PK1eps}
	\\\nonumber\\
		P &=& -\frac{1}{2}m_s^{2}\phi_0^{2} - \frac{g_2}{3}\phi_0^3 - \frac{g_3}{4}\phi_0^4 + \frac{1}{2}m_\omega^{2}V_{0}^2 +
		 \frac{c_3}{4}V_0^4 + \frac{1}{2}m_{\rho}^{2}b_0^2 \nonumber\\
		 && + \frac{1}{3\pi^{2}}\int^{k_{p}}_{0}dk\,\frac{k^{4}}{\sqrt{k^{2}+{m^{*}}^{2}}} + 
		 \frac{1}{3\pi^{2}}\int^{k_{n}}_{0}dk\,\frac{k^{4}}{\sqrt{k^{2}+{m^{*}}^{2}}}\label{PK1pres}\,.
	\end{eqnarray}
\end{subequations}
\section{Nuclear matter observables}
\subsection{Symmetry energy}
The symmetry energy coefficient for the different parameter sets is calculated in exactly the same way as it was done for QHD-I in Sec.\,\ref{QHD1symm}, except that in this case the energy density has more isospin dependent terms.\\
\\
The general expression for the energy density is given by equation (\ref{FSUeps}). By examining the equations of motion of the different meson fields (\ref{FSUEQM}), it is clear that the scalar meson field is isospin independent. The omega meson field has an implied isospin dependence through the $\Lambda_v$-coupling to the rho meson field (which is explicitly isospin dependent). Thus the only explicitly isospin dependent part of the energy density, defined as $\epsilon_s$,  is
\begin{eqnarray}
	\epsilon_s &=&
	- \frac{1}{2}m_{\rho}^{2}b_0^2 - \Lambda_v(g_vV_0)^2 (g_\rho b_0)^2 + \frac{1}{2}g_\rho b_0(\rho_p - \rho_n) \nonumber\\
	&&+\frac{1}{\pi^{2}}\int^{k_{p}}_{0}dk\,k^{2}\sqrt{k^{2}+{m^{*}}^{2}}
	  + \frac{1}{\pi^{2}}\int^{k_{n}}_{0}dk\,k^{2}\sqrt{k^{2}+{m^{*}}^{2}} \nonumber\\
	   \nonumber\\
	  &=&
	  - b_0^2\,\frac{1}{2}\left(m_{\rho}^{2} + 2\,\Lambda_v(g_vV_0)^2 g_\rho^2\right) 
	+ \frac{1}{2}g_\rho b_0(\rho_p - \rho_n) \nonumber\\
	&&+\frac{1}{\pi^{2}}\int^{k_{p}}_{0}dk\,k^{2}\sqrt{k^{2}+{m^{*}}^{2}}
	  + \frac{1}{\pi^{2}}\int^{k_{n}}_{0}dk\,k^{2}\sqrt{k^{2}+{m^{*}}^{2}}\,. \nonumber
\end{eqnarray}
From the equation of motion of the rho meson field (\ref{FSUrho}), $b_0$ can be written in terms of $t$ as
\begin{eqnarray}
	b_0 &=& \frac{1}{2}\,\frac{g_\rho(\rho_p - \rho_n)}{m_\rho^2 + 2\Lambda_vg_\rho^2(g_vV_0)^2}\nonumber\\
		&=& \frac{1}{2}\,\frac{g_\rho (-t)\rho}{m_\rho^2 + 2\Lambda_vg_\rho^2(g_vV_0)^2}\,.\nonumber
\end{eqnarray}
The expression for $\epsilon_s/\rho$ then becomes
\begin{eqnarray}
	\frac{\epsilon_s}{\rho} &=&
	- \left(\frac{1}{2}\,\frac{g_\rho (-t)\rho}{m_\rho^2 + 2\Lambda_vg_\rho^2(g_vV_0)^2}\right)^2
	\left(\frac{m_{\rho}^{2}+ 2\Lambda_v(g_vV_0)^2 g_\rho^2}{2\,\rho} \right) +
	\frac{g_\rho(-t)\rho }{2\, \rho} \left(\frac{1}{2}\,\frac{g_\rho (-t)\rho}{m_\rho^2 + 2\Lambda_vg_\rho^2(g_vV_0)^2}\right)
	 \nonumber\\ 
	&&  +
	 \frac{1}{\pi^{2}\rho}\int^{k_{p}}_{0}dk\,k^{2}\sqrt{k^{2}+{m^{*}}^{2}}
	  + \frac{1}{\pi^{2}\rho}\int^{k_{n}}_{0}dk\,k^{2}\sqrt{k^{2}+{m^{*}}^{2}} \nonumber\\
	  \nonumber\\
	   &=& \frac{t^2\,\rho}{8}\left(\frac{g_\rho^2}{m_\rho^2 + 2\Lambda_vg_\rho^2(g_vV_0)^2}\right)\nonumber\\
	    && +
	 \frac{1}{\pi^{2}\rho}\int^{k_{p}}_{0}dk\,k^{2}\sqrt{k^{2}+{m^{*}}^{2}}
	  + \frac{1}{\pi^{2}\rho}\int^{k_{n}}_{0}dk\,k^{2}\sqrt{k^{2}+{m^{*}}^{2}}. \label{epssFSU}
\end{eqnarray}
Thus the defining equation for the symmetry energy coefficient Eq. (\ref{a4}) reduces to, 
\begin{eqnarray}
	a_{4} &=& \frac{1}{2}\left(\frac{\partial^{2}}{\partial t^{2}}\frac{\epsilon_s}{\rho}\right)_{t=0}\nonumber\\
	& =& \frac{k_f^3}{12\pi^2}\left(\frac{g_\rho^2}{m_\rho^2 + 2\Lambda_vg_\rho^2(g_vV_0)^2}\right) + \frac{k_f^2}{6\sqrt{k_f^2 + m^{*2}}}\,,\nonumber
\end{eqnarray}
after some algebra and using the definition for the baryon density at saturation (\ref{rho0}). Note that the last two terms of Eq. (\ref{epssFSU}) are the same as the two terms in Eq. (\ref{a4QHD1}) and the answer can be directly substituted.\\
\\
In the Ring convention $\epsilon_s$ is given by
\begin{eqnarray}
	\epsilon_s &=&
	- \frac{1}{2}m_{\rho}^{2}b_0^2 + \frac{1}{2}g_\rho b_0(\rho_p - \rho_n) \nonumber\\
	&&+\frac{1}{\pi^{2}}\int^{k_{p}}_{0}dk\,k^{2}\sqrt{k^{2}+{m^{*}}^{2}}
	  + \frac{1}{\pi^{2}}\int^{k_{n}}_{0}dk\,k^{2}\sqrt{k^{2}+{m^{*}}^{2}}\,, \nonumber
\end{eqnarray}
and $b_0$ by
\begin{eqnarray}
	b_0 &=& \frac{1}{2}\,\frac{g_\rho(\rho_p - \rho_n)}{m_\rho^2}\nonumber\\
		&=& \frac{1}{2}\,\frac{g_\rho (-t)\rho}{m_\rho^2}\,.\nonumber
\end{eqnarray}
The expression for $a_4$ in the Ring convention yields
\begin{eqnarray}
	a_{4}	= \frac{k_f^3}{12\pi^2}\left(\frac{g_\rho}{m_\rho}\right)^2 + \frac{k_f^2}{6\sqrt{k_f^2 + m^{*2}}}\,.\nonumber
\end{eqnarray}
\section{Summary}
In this chapter the equation of state of nuclear matter was calculated using various parameter sets of the QHD model of nuclear matter in RMF approximation.\\
This information will be used to calculate the equation of state of neutron star matter.

%% file: chapNS.tex
\chapter{Description of neutron star matter}\label{chapNS}
\section{Introduction}
Neutron stars are extraordinary objects. Having approximately the same density as atomic nuclei but being about $10^{20}$ times larger, these objects have to be approached in a manner \mbox{different} from ``normal'' matter. If the densities of neutron stars are considered, one can conclude that neutron stars could consist, in some part, of normal nuclear matter (protons and neutrons). Thus an oversimplified description of a neutron star is that of a giant nucleus with mass number of $10^{57}$ \cite{csg}. As mentioned in Chapter \ref{intro}, the difference between a neutron star and a normal nucleus is that where the nucleus is bound by the nuclear strong force, the neutron star is bound by gravity. Thus the description of a neutron star must be anchored in general relativity to account for the gravitational effect of not only the mass, but also the energy of this ultra-dense object. Using the principles of the special and general theories of relativity the Tolman-Oppenheimer-Volkoff (TOV) equation describing a static, spherical symmetric neutron star was derived in Chapter \ref{chap:TOV}. To solve the TOV equation, the relationship between the energy density and the pressure (equation of state) in the interior of the star must be known. \\
\\
The neutron star must also be charge neutral: while the gravitational forces are very strong on the large scale, a nett charge would result in very disruptive Coulomb-forces in the neutron star \cite{csg}.\\
\\
The focus in this chapter will be on deriving the equation of state for matter in the neutron star interior by considering various physical effects, such as beta-equilibrium as well as the contribution of the neutron star crust.
\section{Crustal effects}
The crust of a neutron star is usually divided into the inner and the outer crust. The outer crust ranges from the density where atomic nuclei become fully ionised, at about $10^{4}$ g/cm$^{3}$ ($10^{-8}$ MeV/fm$^{3}$), up to the density at which the nucleus becomes so neutron-rich that neutrons start to drip from the nucleus (	``the neutron drip line'') \cite{Rus}.\\
\\ 
The inner crust covers the density range from the neutron drip line up to the density \mbox{(denoted} by $\rho_c$) at which nuclei cease to exist and nuclear matter undergoes a phase transition to a uniform liquid \cite{BPS}.
\subsection{Outer crust}
The BPS equation of state is commonly used to describe the outer crust of neutron stars. This equation of state was first described by G. Baym, C. Pethick and P. Sutherland in 1971 \cite{BPS}. The departure point of the BPS equation of state is that the nuclei in the crust of a neutron star are arranged in a body-centred cubic (bcc) lattice immersed in a sea of free electrons, and that the energy of the system can be lowered by the capture of energetic electrons by protons (inverse beta decay) \cite{BPS}. The beta-decay of neutrons in the outer crust is (at this point) blocked by the filled electron states \cite{BPS}. The lattice contributes to lower the energy of the system by minimising the Coulomb interaction energy of the nuclei \cite{Rus}. \\
\begin{figure}
	\centering
		\includegraphics[width=0.75\textwidth]{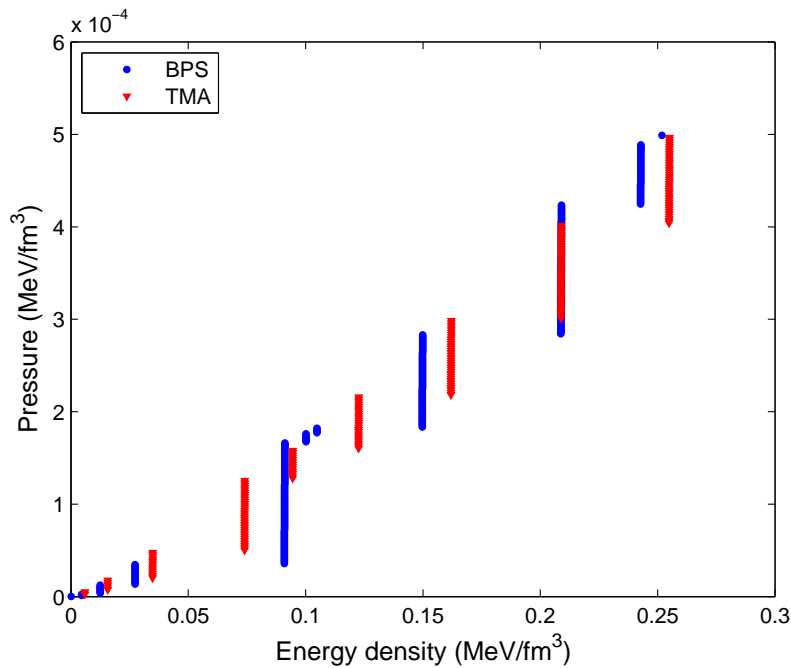}
	\caption{The BPS equation of state, calculated using the original sequence of equilibrium nuclei (labelled as BPS) and the sequence given by the TMA model (TMA). }
	\label{fig:TMABPS}
\end{figure}
\\
At a specific pressure the nucleus that minimises the chemical potential, $\mu$, of the system is deemed the equilibrium nucleus. The pressure range spanned by the outer crust is therefore defined by a sequence of equilibrium nuclei. Each equilibrium nucleus defines a certain pressure range which is characterised by the maximum mass (energy) density, $\epsilon_{\mbox{{\scriptsize max}}}$, at which the nucleus can be deemed the equilibrium nucleus. At any given pressure in the outer crust the properties of the equilibrium nucleus (either gathered from experimental data or from some \mbox{theoretical} model of the nucleus) together with the energy of the electrons determine the energy density. 
Since the pressure in the neutron star crust increases as a smooth function of depth, each transition from one equilibrium nucleus to another, will be accompanied by a discontinuity in the energy density \cite{BPS}. This can be clearly seen in Fig. \ref{fig:TMABPS} where the BPS equations of state represented by the sequence of equilibrium nuclei in Tables \ref{tab:BPS} and \ref{tab:TMA}, have been plotted. \\
\\
%
%
%
%
The BPS equation of state is described in detail in Refs \cite{Rus} and \cite{BPS}, and can be summarised as follows: the pressure, P, is given, for an equilibrium nucleus characterised by A (mass number) and Z (proton number), by
\begin{eqnarray}
	P = P_{e} + \frac{1}{3}w_{L}n_{N}\label{BPSpres}
\end{eqnarray}
where 
\begin{itemize}
	\item $\rho_{N}$ is the number density of nuclei, related to the baryon number density ($\rho_{b}$) by
					\[
							\rho_{b} = A \rho_{N},
					\]
 \item $P_{e}$ is the contribution from the free electrons with fermi momentum ($k_{e}$) and mass ($m_{e}$)
				\[
						P_{e} = \frac{1}{3\pi^{2}}\int^{k_{e}}_{0}dk\frac{k^{4}}{\sqrt{k^{2} + {m_{e}}^{2}}},
				\]
\item 	$w_{L}$ is the lattice energy per nucleus
				\[
					w_{L} = -1.819620\,Z^{2}e^{2}/a,
					\]
				where $e$ is the charge of an electron and $a$ is given for a bcc lattice by,
				\[
					n_{N}a^{3} = 2.
				\]
\end{itemize}
The energy density ($\epsilon$) is given by 
\begin{eqnarray}
	\epsilon = \epsilon_{e} + \rho_{N}(w_{L} + w_{N})\label{BPSeps}
\end{eqnarray}
where 
\begin{itemize}
	\item $w_{N}$ is the total energy of an isolated nucleus, and,
	\item $\epsilon_{e}$ the total electron energy density given by
\begin{eqnarray}
	\epsilon_{e} = \frac{1}{\pi^{2}}\int^{k_{e}}_{0}dk\,k^{2}\sqrt{k^{2} + {m_{e}}^{2}}\label{BPSepse}.
\end{eqnarray}
\end{itemize}
The equilibrium nucleus is determined by the $A$, $Z$ configuration that minimizes the chemical potential, $\mu$,
	\[ 
	\mu = \frac{w_{N} + \frac{4}{3} w_{L} + Z\sqrt{k_{e}^2 + m_e^2}}{A}\,.
\]
\\
The sequence of equilibrium nuclei used in the original BPS paper is shown in Table \ref{tab:BPS}, where the properties of the equilibrium nuclei were calculated in Ref.\,\cite{Rus}, using the nuclear data set that was used in the original BPS paper (Ref.\,\cite{BPS}, which was published in 1971, and used nuclear data from the 1960's). Since experimental techniques have undergone some improvement since the 1960's, the sequence of equilibrium nuclei between mass (energy) densities of $10^{4}$ g/cm$^{3}$ ($10^{-8}$ MeV/fm$^{3}$) and $10^{10}$ g/cm$^{3}$ ($10^{-2}$ MeV/fm$^{3}$) can be determined from experimental data \cite{Rus}. The sequence of equilibrium nuclei in Table \ref{tab:TMA} was calculated in Ref.\,\cite{Rus} using modern experimental nuclear data and nuclear models. The upper part of the table was determined from experimental data (compiled by G. Audi {\sl et al.}), while the lower part was  calculated using a modern relativistic nuclear field theory model by L.S. Geng {\sl et al.} called TMA (see Ref.\,\cite{Rus} and references therein).\\
\\
%
%
%
The BPS equation of state can be determined from either Table \ref{tab:BPS} or Table \ref{tab:TMA}. It can be done by either interpolating the energy density at a given pressure, or at a specific pressure (and hence equilibrium nucleus) the contribution of the electrons to the pressure can be determined from Eq. (\ref{BPSpres}) and the fermi momenta of the electrons ($k_e$) obtained.  If $k_e$ is known, $\epsilon_e$ can be determined from Eq. (\ref{BPSepse}). Using the value of $\epsilon_e$ and the properties of the equilibrium nuclei, the energy density can then be obtained from Eq. (\ref{BPSeps}).
%
%
%
\begin{table*}[tb]
	\centering
		\begin{tabular}{cccccccc}
			\hline\hline
			$\mu$ (Mev)&$\mu_e$ (Mev)&$\epsilon_{\mbox{\footnotesize max}}$(g$\cdot$cm$^{-3}$)&
			P (dyne$\cdot$cm$^{-2}$)&$\rho_b$(cm$^{-3}$)&Element&Z&N\\
			\hline
			930.60&0.95&$8.09\times10^6$&$5.29\times10^{23}$&$4.88\times10^{30}$&$^{56}$Fe&26&30\\
			931.31&2.60&$2.69\times10^{8}$&$6.91\times10^{25}$&$1.62\times10^{32}$&$^{62}$Ni&28&34\\
			932.00&4.24&$1.24\times10^{9}$&$5.20\times10^{26}$&$7.48\times10^{32}$&$^{64}$Ni&28&36\\
			933.33&7.69&$8.15\times10^{9}$&$5.78\times10^{27}$&$4.90\times10^{33}$&$^{84}$Se&34&50\\
			934.42&10.61&$2.23\times10^{10}$&$2.12\times10^{28}$&$1.34\times10^{34}$&$^{82}$Ge&32&50\\
			935.48&13.58&$4.88\times10^{10}$&$5.70\times10^{28}$&$2.93\times10^{34}$&$^{80}$Zn&30&50\\
			937.68&19.97&$1.63\times10^{11}$&$2.68\times10^{29}$&$9.74\times10^{34}$&$^{78}$Ni&28&50\\
			937.78&20.25&$1.78\times10^{11}$&$2.84\times10^{29}$&$1.07\times10^{35}$&$^{76}$Fe&26&50\\
			938.57&22.86&$2.67\times10^{11}$&$4.55\times10^{29}$&$1.60\times10^{35}$&$^{122}$Zr&40&82\\
			939.29&25.25&$3.73\times10^{11}$&$6.79\times10^{29}$&$2.23\times10^{35}$&$^{120}$Sr&38&82\\
			939.57&26.19&$4.32\times10^{11}$&$7.87\times10^{29}$&$2.59\times10^{35}$&$^{118}$Kr&36&82\\
			\hline
		\end{tabular}
	\caption{Sequence of equilibrium nuclei, from Ref.\,\cite{Rus}, that are used to reproduce the original equation of state of the outer crust of a neutron star as in Ref.\,\cite{BPS}.}
	\label{tab:BPS}
\end{table*}
\begin{table*}[tb]
	\centering
		\begin{tabular}{cccccccc}
			\hline\hline
			$\mu$ (Mev)&$\mu_e$ (Mev)&$\epsilon_{\mbox{\footnotesize max}}$(g$\cdot$cm$^{-3}$)&
			P (dyne$\cdot$cm$^{-2}$)&$\rho_b$(cm$^{-3}$)&Element&Z&N\\
			\hline
			930.60&0.95&$8.02\times10^6$&$5.22\times10^{23}$&$4.83\times10^{30}$&$^{56}$Fe&26&30\\
			931.32&2.61&$2.71\times10^{8}$&$6.98\times10^{25}$&$1.63\times10^{32}$&$^{62}$Ni&28&34\\
			932.04&4.34&$1.33\times10^{9}$&$5.72\times10^{26}$&$8.03\times10^{32}$&$^{64}$Ni&28&36\\
			932.09&4.46&$1.50\times10^{9}$&$6.44\times10^{27}$&$9.04\times10^{33}$&$^{66}$Ni&28&38\\
			932.56&5.64&$3.09\times10^{9}$&$1.65\times10^{27}$&$1.86\times10^{33}$&$^{86}$Kr&36&50\\
			933.62&8.38&$1.06\times10^{10}$&$8.19\times10^{27}$&$6.37\times10^{33}$&$^{84}$Se&34&50\\
			934.75&11.43&$2.79\times10^{10}$&$2.85\times10^{28}$&$1.68\times10^{34}$&$^{82}$Ge&32&50\\
			935.93&14.71&$6.21\times10^{10}$&$7.86\times10^{28}$&$3.73\times10^{34}$&$^{80}$Zn&30&50\\
			\\
			937.28&18.64&$1.32\times10^{11}$&$2.03\times10^{29}$&$7.92\times10^{34}$&$^{78}$Ni&28&50\\
			937.63&19.80&$1.68\times10^{11}$&$2.55\times10^{29}$&$1.01\times10^{35}$&$^{124}$Mo&42&82\\
			938.13&21.38&$2.18\times10^{11}$&$3.48\times10^{29}$&$1.31\times10^{35}$&$^{122}$Zr&40&82\\
			938.67&23.19&$2.89\times10^{11}$&$4.82\times10^{29}$&$1.73\times10^{35}$&$^{120}$Sr&38&82\\
			939.18&24.94&$3.73\times10^{11}$&$6.47\times10^{29}$&$2.23\times10^{35}$&$^{118}$Kr&36&82\\
			939.57&26.29&$4.55\times10^{11}$&$8.00\times10^{29}$&$2.72\times10^{35}$&$^{116}$Se&34&82\\
			\hline
		\end{tabular}
	\caption{Sequence of equilibrium nuclei, calculated by using modern atomic data and the relativistic TMA model of the nucleus, from Ref.\,\cite{Rus}, that are used to calculate the equation of state of the outer crust of a neutron star.}
	\label{tab:TMA}
\end{table*}
\subsection{Inner crust}
The BBP equation of state (by G. Baym, H. A. Bethe and C. J. Pethick \cite{BBP}) assumes that 
at densities above the neutron drip line, up to $\rho_c$, nuclear matter consists of a lattice of nuclei immersed in a gas of free neutrons. The equation of state is calculated by matching the pressure and the energy density of the lattice nuclei to that of a free neutron gas \cite{BBP}. The equation of state is given in Table \ref{tab:BBP} (at the end of the chapter).\\
\\
The value of the density at which the inner crust becomes a uniform liquid ($\rho_c$) depends on the model (and parameter set) used to describe the matter of the neutron star crust and interior. J. Carriere {\sl et al.} calculated the value of $\rho_c$ for different QHD parameter sets in Ref.\,\cite{Car}. In Ref.\,\cite{Car} the authors used the relativistic random-phase approximation to analyse if and when a system consisting of protons, neutrons and electron would undergo a second-order phase transition. For the inner crust they assumed a polytropic equation of state,
\begin{eqnarray}
	P(\epsilon) = A + B\epsilon^{4/3}, \label{polyeos}
\end{eqnarray}
where $A$ and $B$ are constants that are determined so that the above equation of state matches the equation of state for the outer crust (pressure and energy density) at the neutron drip line as well as the equation of state of the liquid interior at $\rho_c$.
\section{Equilibrium conditions}
Free neutrons are not stable particles and have a half-life of about 10 minutes \cite{firestone}. The neutron will decay into a proton, electron and an antineutrino (beta-decay):
	$$
		 n \rightarrow p + e^{-} + \bar{\nu}_{e}\,.
  $$
Therefore free neutrons in the neutron star will undergo beta-decay creating a proton and neutron. Since a neutron stars 
exists for a period longer than the half-life of a neutron, 
a significant number of neutrons in the star will undergo beta-decay. 
Since a neutron star consists of a huge amount of neutrons, the number of electrons (resulting from the beta-decay of neutrons) present in the star will also be quite significant. Due to the densities of the electrons, some electrons will attain relativistic energies, making the inverse process (capture of a electron by a proton to form a neutron) energetically favourable, i.e.
$$
	p + e^{-} \rightarrow n + \nu_{e}\,.
$$
The two decay processes mentioned above represent two competing processes and since the neutron star has a lifetime that is long compared to the timescale of these decay processes, the 
matter in the star has time to reach beta-equilibrium. 
The beta-equilibrated state is the state where, for a fixed baryon number density ($\rho = \rho_p + \rho_n$), the proton and neutron number densities ($\rho_p$ and $\rho_n$) are such that the energy density ($\epsilon$) of the system is at a minimum.  From Ref.\,\cite{csg}, the condition for beta-equilibrium can be expressed in terms of the chemical potential (Fermi energy), $\mu$. If $\mu_x$ refers to the chemical potential of a specific particle specie, $x$, then
$$
	\mu_x = \frac{\partial\epsilon}{\partial \rho_x}\,.
$$
The effects of the neutrinos are not considered in the beta-equilibrated state, for their mean-free path is longer than the radius of the star \cite{csg}, and therefore the beta-equilibrated state is described by
\begin{eqnarray}
	\mu_n = \mu_p + \mu_e\,.\label{beta}
\end{eqnarray}\\
The number density of particle $x$ ($\rho_x$) can be expressed in terms of the fermi momentum ($k_x$) as
$$
	\rho_x = \frac{k_x^3}{3\pi^2}\,,
$$ 
and from the above $k_x$ is 
$$
	k_x = \pi^2\,(3\pi^2\rho_x)^{-\,{2}/{3}}\,,
$$
which is a handy result when evaluating the chemical potentials of different particles.\\
\\
Using the expression for the energy density (\ref{NSeps}), the chemical potentials of the different particles can be expressed as
\begin{subequations}
	\begin{eqnarray}
		\mu_n &=& \sqrt{k_n^2 + {m^*}^2} + g_v V_0 - \frac{1}{2}g_\rho b_0 \nonumber\\
		\mu_p &=& \sqrt{k_p^2 + {m^*}^2} + g_v V_0 + \frac{1}{2}g_\rho b_0 \nonumber\\
		\mu_e &=& \sqrt{k_e^2 + {m_e}^2}\nonumber\,.
	\end{eqnarray}
\end{subequations}
The condition for beta-equilibrium (\ref{beta}) can therefore be generalised, for all parameter sets considered in this work, to
\begin{eqnarray}
	\sqrt{k_n^2 + {m^*}^2} = \sqrt{k_p^2 + {m^*}^2} + g_\rho b_0 + \sqrt{k_e^2 + {m_e}^2}\,.\label{waleckabeta}
\end{eqnarray}
In the Ring convention, the expression for beta-equilibrium is
\begin{eqnarray}
	\sqrt{k_n^2 + {m^*}^2} = \sqrt{k_p^2 + {m^*}^2} - 2g_\rho b_0 + \sqrt{k_e^2 + {m_e}^2}\,.\label{ringbeta}
\end{eqnarray}\\
In ultra-dense matter electrons can attain ultra-relativistic energies and therefore it may become energetically more favourable to populate muon states. Muons have the same charge as an electron, but a mass of 105.7 MeV \cite{muons}. In this work it is assumed that muon states will be populated once the electrons reach energies comparable to the mass of the muon. Muons are unstable particles and undergoes the following decay reaction \cite{csg}:
$$
	\mu^{-} \rightarrow e^{-} + \bar{\nu}_{e} + \nu_\mu\,.
$$
Equilibrium with regards to the above reaction implies that
\begin{eqnarray}
	\mu_e = \mu_\mu\,,\label{muoneqm}
\end{eqnarray}
where the muon chemical potential is given by 
$$
	\mu_\mu = \sqrt{k_\mu^2 + m_\mu^2}\,.
$$\\
Since the neutron star is assumed to be neutral, the number of protons must be equal to the number of electrons present. In terms of the fermi momenta of the species this means that
$$
	k_e = k_p,
$$
but if muon-states are also populated, the condition for charge equilibrium will have to be expressed as
\begin{eqnarray}
		\rho_\mu + \rho_e = \rho_p\,,\nonumber
\end{eqnarray}
which implies that
\begin{eqnarray}
	k_p = (k_e^3 + k_\mu^3)^{1/3}.\label{charge}
\end{eqnarray}\\
Neutron stars are in general equilibrium when the conditions of (\ref{beta}), (\ref{muoneqm}) and (\ref{charge}) are satisfied.
\section{Equation of state}
\subsection{Without a crust}
The energy density and the pressure of neutron star matter in general equilibrium, assuming a QHD description for nuclear matter that contains protons, neutrons, electrons, muons and mesons ($\sigma$, $\omega$ and $\rho$) is, from (\ref{FSUEoSWalecka}):
\begin{subequations}\label{NSEoS}
	\begin{eqnarray}
		\epsilon &=&
		\frac{1}{2}m_s^{2}\phi_0^{2} + \frac{\kappa}{3!}(g_s\phi_0)^3 + \frac{\lambda}{4!}(g_s\phi_0)^4 -
		 \frac{1}{2}m_\omega^{2}V_{0}^2 - \frac{\zeta}{4!}(g_vV_0)^4 \nonumber\\
		&-& \frac{1}{2}m_{\rho}^{2}b_0^2 - \Lambda_v(g_vV_0)^2 (g_\rho b_0)^2 + g_vV_0(\rho_n + \rho_p) 
		+ \frac{1}{2}g_\rho b_0(\rho_p - \rho_n)\nonumber\\
		&+&\frac{1}{\pi^{2}}\int^{k_{p}}_{0}dk\,k^{2}\sqrt{k^{2}+{m^{*}}^{2}}
		  + \frac{1}{\pi^{2}}\int^{k_{n}}_{0}dk\,k^{2}\sqrt{k^{2}+{m^{*}}^{2}}\nonumber\\
		  &+& \sum_{\lambda}\frac{1}{\pi^{2}}\int^{k_{\lambda}}_{0}dk\,k^{2}\sqrt{k^{2}+{m_\lambda}^{2}}\label{NSeps}\,,
	\end{eqnarray} 
	and,
	\begin{eqnarray}
		P &=& - \frac{1}{2}m_s^{2}\phi_0^{2} - \frac{\kappa}{3!}(g_s\phi_0)^3 - \frac{\lambda}{4!}(g_s\phi_0)^4 + 
		\frac{1}{2}m_\omega^{2}V_{0}^2 + \frac{\zeta}{4!}(g_vV_0)^4 \nonumber\\
		&+& \frac{1}{2}m_{\rho}^{2}b_0^2 + \Lambda_v(g_vV_0)^2 (g_\rho b_0)^2 \nonumber\\
		 &+& \frac{1}{3\pi^{2}}\int^{k_{p}}_{0}dk\,\frac{k^{4}}{\sqrt{k^{2}+{m^{*}}^{2}}} + 
		 \frac{1}{3\pi^{2}}\int^{k_{n}}_{0}dk\,\frac{k^{4}}{\sqrt{k^{2}+{m^{*}}^{2}}}\nonumber\\
		 &+& \sum_{\lambda}\frac{1}{3\pi^{2}}\int^{k_{\lambda}}_{0}dk\,k^{2}\sqrt{k^{2}+{m_\lambda}^{2}}\,,\label{NSpres}
	\end{eqnarray}
\end{subequations}
where $\lambda$ denotes the lepton (electrons and muons) state.
\begin{figure}
	\centering
		\includegraphics[width=0.70\textwidth]{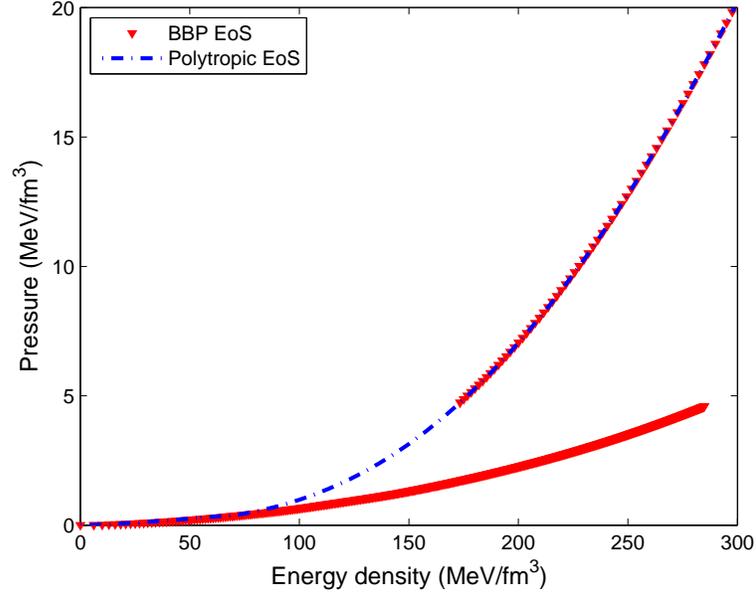}
	\caption{Behaviour of the equation of state (EoS) of the neutron star interior when different equations of state for the inner crust are used.}
	\label{fig:BBP}
\end{figure}

\subsection{Including a crust}
In this work the BPS equation of state is used for the outer crust. Both the original \mbox{sequence} of nuclei (of Table \ref{tab:BPS}) and modern calculations of the sequence of nuclei using the TMA model (as in Table \ref{tab:TMA}) are used. The difference between the resulting equations of state is shown in Fig. \ref{fig:TMABPS}. When the values of Table \ref{tab:BPS} are used for the sequence of equilibrium nuclei, it will be denoted by BPS. When those of Table \ref{tab:TMA} are used, it will be denoted by TMA. \\
\\
For the equation of state of the inner crust the BBP equation of state was initially used, but since only the energy density or the pressure can be matched at $\rho_c$ the equation of state has a rather large discontinuity, as can be seen in Fig. \ref{fig:BBP}. As such the BBP equation of state was not used in further calculations presented in this work. Rather the polytropic equation of state (\ref{polyeos}) for the inner crust is used, since the values of $A$ and $B$ can be calculated to 
the equation of state of the liquid interior at $\rho_c$. Due to time constraints, the calculation of 
$\rho_c$ for the different QHD parameter sets was not repeated. Rather the published values for $\rho_c$, where available, were used. 
For the parameter sets for which values of $\rho_c$ are not available, the value of the pressure at $\rho_c$ was estimated to be $4.982\times 10^{32}$dyne/cm$^{2}$ ($0.311$ MeV/fm$^{3}$). This pressure is simply the average pressure, corresponding to the published values for $\rho_c$, in the specific parameter sets. The published values for $\rho_c$, together with the corresponding pressures, are listed in Table \ref{tab:rhoc}.\\
\\
When crustal effects are included, the equation of state of the liquid interior of the star is given by the energy density and the pressure derived for the QHD description of the neutron star interior (\ref{NSEoS}).\\
The equation of state of the neutron star interior, when crustal effects are included, can therefore be summarised as
\begin{eqnarray}
	 P(\epsilon) = \left\{\begin{array}{ll}
													P_{\mbox{\scriptsize{BPS}}}(\epsilon)&\mbox{if}
													\ \ \epsilon_{\mbox{\footnotesize{min}}}\leq
													\epsilon\leq\epsilon_{\mbox{\footnotesize{outer}}}\\
													A + B\epsilon^{4/3}& \mbox{if}\ \ 
													 \epsilon_{\mbox{\footnotesize{outer}}}<\epsilon\leq\epsilon_c\\
													P_{\mbox{\scriptsize{QHD}}}(\epsilon)&\mbox{if}\ \ \epsilon>\epsilon_c
												\end{array}\right..
\end{eqnarray}
\begin{table*}[ttb]
	\centering
		\begin{tabular}{ccc}
			\hline\hline
			Parameter set & $\rho_c$ & Pressure at $\rho_c$ \\
			&(in MeV$\cdot$fm$^{-3}$)& (in dyne$\cdot$cm$^{-2}$)\\ 
			\hline
			FSUGold & 0.076& $6.56\times10^{32}$\\
			NL3 & 0.052&$3.42\times10^{32}$\\
			\hline
		\end{tabular}
	\caption{The values for transition (number) density from the inner crust to the liquid interior of the FSUGold and NL3 parameter sets, from Ref.\,\cite{FSU}. The corresponding pressures (at $\rho_c$) were calculated in this work, assuming that the neutron star is in general equilibrium. }
	\label{tab:rhoc}
\end{table*}
\section{Neutron star constraints}
As with any investigation into natural phenomena, experimental and observational inputs are crucial to the development of the description of a physical system. In the case of neutron stars these inputs could stem from 
experiments 
or astrophysical observations. Recently a testing scheme based on various different constraints derived from heavy-ion collisions and astrophysical observations has been proposed by T. Kl$\ddot{\mbox{a}}$hn {\sl et al.} \cite{Klaehn}. In this section the constraints proposed in Ref.\,\cite{Klaehn} and other publications, applicable to the scope of this work, will be discussed.
\subsection{General theory of relativity}
From the Tolman-Oppenheimer-Volkoff equation (\ref{TOV}) it is implied that the radius of the star ($R$) cannot be smaller or equal to $2GM$. If this condition is not obeyed, the change in the pressure is infinite or increases as one moves from the centre of the star to the boundary. $2GM$ defines a massive object's Schwarzschild radius and if the radius of the object lies within its Schwarzschild radius, i.e. $$ R < 2GM,$$ then the object has collapsed under its gravity and is a black hole  \cite{csg}.
\begin{figure}[ttb]
	\centering
		\includegraphics[width=0.70\textwidth]{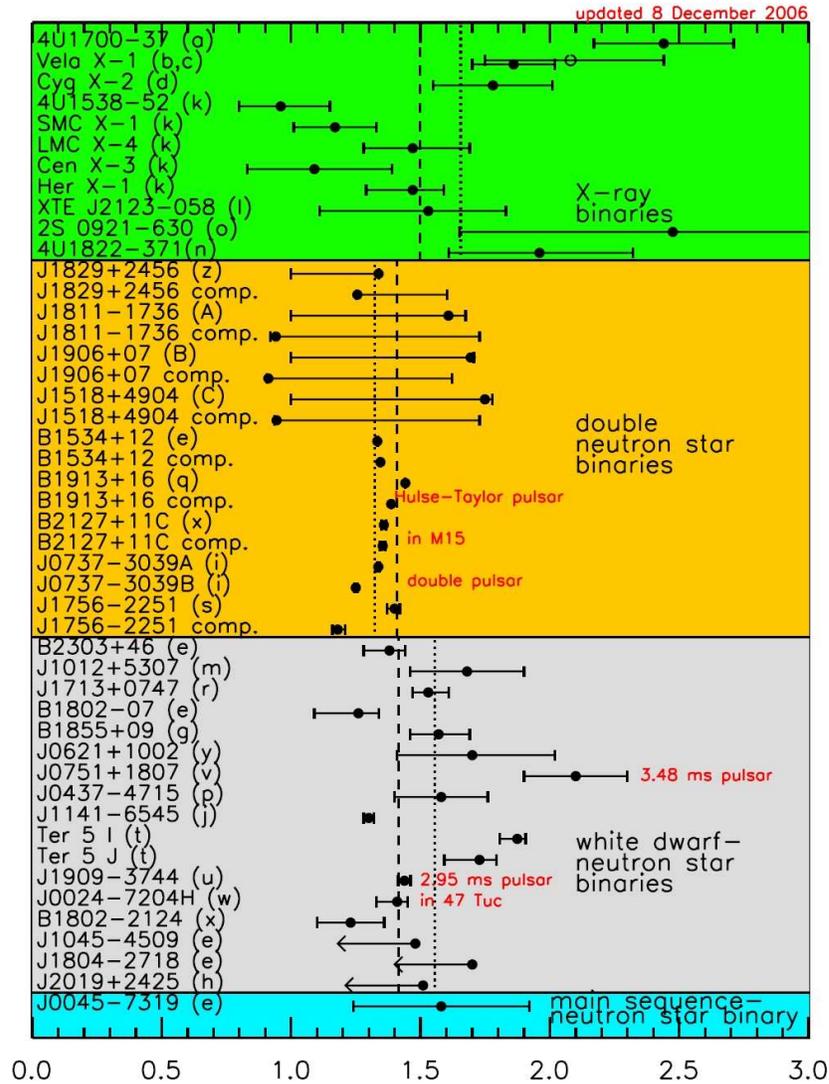}
	\caption{Table of masses (with error bars) of known neutron stars. 
	Taken from Ref.\,\cite{Lattimer07}.}
	\label{fig:NSmass}
\end{figure}
\subsection{Observational constraints}\label{sec:obs}
The mass of a neutron star in a binary system can be quite accurately measured \cite{Lattimer07} and the mass of various neutron stars have been determined. The list of known masses of neutron stars is given in Fig. \ref{fig:NSmass}.\\
\\
The known neutron star masses cover a narrow range, between about 1.4 and 1.5 $M_{\odot}$ (solar mass). This narrow mass range may be due to special circumstances that govern the evolution of binaries, but no thorough explanation has been put forward to date \cite{Lattimer07}. The heaviest known neutron star is PSR J0751+1807 with a mass of $2.1\pm 0.2 M_{\odot}$ \cite{Nice}. The reproduction of this mass by any model for the neutron star interior (equation of state) has been hailed as a very important constraint in determining whether certain models for the neutron star interior are applicable or not \cite{Klaehn}. J. M. Lattimer and M. Prakash have however warned in Ref.\,\cite{Lattimer07} that caution should be exercised  before the currently known heavy neutron stars are taken as firm evidence that such heavy neutron stars exists, due to uncertainties in the measurements.\\
\\
To determine the radius of a neutron star is much more difficult than the mass and there are large uncertainties in the measurements \cite{Links}. To date, the only neutron star whose mass and radius has been determined is EXO 0748-676 \cite{Ozel}. A. R. Villarreal and T. E. Strohmayer established that the mass of EXO 0748-676 must lie between 1.5 and 2.3 $M_{\odot}$ while the radius could lie between 9.5 and 15 km \cite{Villarreal}. F. $\ddot{\mbox{O}}$zel determined the lower limits of the mass and radius of EXO 0748-676 to be $2.10\pm 0.28\, M_{\odot}$ and $13.8\pm1.8\,$km \cite{Ozel}.\\
\\
These considerations are used in this work to evaluate different equations of state for the neutron star interior.
%
%
%
%
%
%
%
%
%
%
%
\begin{table}[htb]
	\centering
		\begin{tabular}{ccccc}
		\hline\hline
		$\epsilon$ (g/cm$^3$)&P (dyne$\cdot$cm$^{-2}$)&$\rho_b$(cm$^{-3}$)&Z&N\\
		\hline
		$4.46\times10^{11}$&$7.89\times10^{29}$&$2.67\times10^{35}$&126&40\\
		$5.23\times10^{11}$&$8.35\times10^{29}$&$3.13\times10^{35}$&128&40\\
		$6.61\times10^{11}$&$9.10\times10^{29}$&$3.95\times10^{35}$&130&40\\
		$7.96\times10^{11}$&$9.83\times10^{29}$&$4.76\times10^{35}$&132&41\\
		$9.73\times10^{11}$&$1.08\times10^{30}$&$5.81\times10^{35}$&135&41\\
		$1.20\times10^{12}$&$1.22\times10^{30}$&$7.14\times10^{35}$&137&42\\
		$1.47\times10^{12}$&$1.40\times10^{30}$&$8.79\times10^{35}$&140&42\\
		$1.80\times10^{12}$&$1.64\times10^{30}$&$1.08\times10^{36}$&142&43\\
		$2.20\times10^{12}$&$1.95\times10^{30}$&$1.31\times10^{36}$&146&43\\
		$2.93\times10^{12}$&$2.59\times10^{30}$&$1.75\times10^{36}$&151&44\\
		$3.83\times10^{12}$&$3.51\times10^{30}$&$2.29\times10^{36}$&156&45\\
		$4.93\times10^{12}$&$4.77\times10^{30}$&$2.94\times10^{36}$&163&46\\
		$6.25\times10^{12}$&$6.48\times10^{30}$&$3.73\times10^{36}$&170&48\\
		$7.80\times10^{12}$&$8.75\times10^{30}$&$4.65\times10^{36}$&178&49\\
		$9.61\times10^{12}$&$1.17\times10^{31}$&$5.73\times10^{36}$&186&50\\
		$1.25\times10^{13}$&$1.69\times10^{31}$&$7.42\times10^{36}$&200&52\\
		$1.50\times10^{13}$&$2.21\times10^{31}$&$8.91\times10^{36}$&211&54\\
		$1.78\times10^{13}$&$2.85\times10^{31}$&$1.06\times10^{37}$&223&56\\
		$2.21\times10^{13}$&$3.93\times10^{31}$&$1.31\times10^{37}$&241&58\\
		$2.99\times10^{13}$&$6.18\times10^{31}$&$1.78\times10^{37}$&275&63\\
		$3.77\times10^{13}$&$8.77\times10^{31}$&$2.24\times10^{37}$&311&67\\
		$5.08\times10^{13}$&$1.39\times10^{32}$&$3.02\times10^{37}$&375&74\\
		$6.19\times10^{13}$&$1.88\times10^{32}$&$3.67\times10^{37}$&435&79\\
		$7.73\times10^{13}$&$2.66\times10^{32}$&$4.58\times10^{37}$&529&88\\
		$9.83\times10^{13}$&$3.90\times10^{32}$&$5.82\times10^{37}$&683&100\\
		$1.26\times10^{14}$&$5.86\times10^{32}$&$7.47\times10^{37}$&947&117\\
		$1.59\times10^{14}$&$8.59\times10^{32}$&$9.37\times10^{37}$&1390&143\\
		$2.00\times10^{14}$&$1.29\times10^{33}$&$1.18\times10^{38}$&2500&201\\
		$2.52\times10^{14}$&$1.90\times10^{33}$&$1.48\times10^{38}$&&\\
		$2.76\times10^{14}$&$2.24\times10^{33}$&$1.62\times10^{38}$&&\\
		$3.08\times10^{14}$&$2.75\times10^{33}$&$1.81\times10^{38}$&&\\
		$3.43\times10^{14}$&$3.37\times10^{33}$&$2.02\times10^{38}$&&\\
		$3.89\times10^{14}$&$4.29\times10^{33}$&$2.28\times10^{38}$&&\\
		$4.64\times10^{14}$&$6.10\times10^{33}$&$2.71\times10^{38}$&&\\
		$5.09\times10^{14}$&$7.39\times10^{33}$&$2.98\times10^{38}$&&\\
		\hline	
		\end{tabular}
	\caption{The BBP equation of state of matter in the inner crust of a neutron star, as calculated in Ref.\,\cite{BBP}. 
	}
	\label{tab:BBP}
\end{table}
\section{Summary}
In this chapter a description of a neutron star was presented, based on certain assumptions about the crust of the neutron star as well as equilibrium conditions in the interior of the star.\\ The equation of state of neutron star matter was also derived. 

%% file: chapResults.tex
\chapter{Results and discussion}
The results in this chapter were generated by writing a computer program to solve the applicable equations in the \mbox{FORTRAN90} programming language. A description of the programs as well as the actual equations that were used in the computations, are given in Appendix \ref{ap:code}.
\section{Properties of saturated nuclear matter}
The properties of saturated nuclear matter were calculated using the different parameter sets discussed in Chapters \ref{chapQHD1} and \ref{chapFSUG} (QHD-I, NL-SH, TM1, TM2, NL3, PK1 and FSUGold). The calculated values are compared to the published ones for the different parameter sets. The results are listed in Tables \ref{tab:satprop1}, \ref{tab:satprop2} and \ref{tab:satprop3}. The properties of saturated nuclear matter calculated in this work shows good agreement with the published values. 
\\
\begin{table}[htb]
	\centering
		\begin{tabular}{|lllllll|}
		\hline
		Parameter set&\multicolumn{3}{c}{$\rho_{\mbox{\footnotesize sat}}$(fm$^{-3}$)}&
		\multicolumn{3}{c|}{E$_b$(MeV)}\\
		&Published&Calculated&Observed&Published&Calculated&Observed\\
		\hline\hline
		&&&0.153&&&-16.3\\
		QHD1 \cite{recentprogress}	&0.148	&0.148&	&-15.75&	-15.75&\\		
		NL-SH \cite{NLSH}&	0.146	&0.146&	&-16.328&	-16.346&\\
		TM1	\cite{TM1}&0.145	&0.145&	&-16.3	&-16.3	&\\
		TM2	\cite{TM1}&0.132	&0.132&	&-16.2	&-16.2	&\\
		NL3	\cite{NL3}&0.148	&0.148&	&-16.299&	-16.240&\\
		PK1	\cite{PK1}&0.148195	&0.148192	&&-16.268&	-16.268	&\\
		NL3	\cite{FSUL}&0.148	&0.148	&&-16.24	&-16.24&\\
		FSUGold	\cite{FSUL}&0.148	&0.148&&	-16.30&	-16.28	&\\
		\hline
		\end{tabular}
	\caption{Comparison between the published values and the ones calculated in this work for the saturation density of nuclear matter and the binding energy per nucleon at the saturation density. The references after the name of the parameter set indicate the source of the published values.}
	\label{tab:satprop1}
\end{table}
\begin{table}[htb]
	\centering
		\begin{tabular}{|lllllll|}
		\hline
		Parameter set&
		\multicolumn{3}{c}{K(MeV)}&\multicolumn{3}{c|}{$a_4$(MeV)}\\
		&Published&Calculated&Observed&Published&Calculated&Observed\\
		\hline\hline
		&&&234&&&32.5\\
		QHD1 \cite{recentprogress}	&	not available&546.55&&not available&19.3&\\
		NL-SH \cite{NLSH}&	354.95&355.34&	&36.1	&36.1	&\\
		TM1	\cite{TM1}&281&	281&	&36.9	&36.9	&\\
		TM2	\cite{TM1}&344&	344&	&35.8	&36.0&\\
		NL3	\cite{NL3}&	271.76	&271.52&&	37.4	&37.4	&\\
		PK1	\cite{PK1}&		282.644&	282.685&	&37.641	&37.640&\\
		NL3	\cite{FSUL}&	271	&272	&&37.3	&37.3&\\
		FSUGold	\cite{FSUL}&230	&230&	&32.6	&32.6	&\\
		\hline
		\end{tabular}
	\caption{Comparison between the published values and the ones calculated in this work for the compressibility and symmetry energy of nuclear matter at saturation. The references after the name of the parameter set indicate the source of the published values.}
	\label{tab:satprop2}
\end{table}
\begin{table}[tbbt]
	\centering
		\begin{tabular}{|lllll|}
		\hline
		Parameter set&
		\multicolumn{2}{c}{$m^*$/$M_n$}&\multicolumn{2}{c|}{$m^*$/$M_n$}\\
		&Published&Calculated&Published&Calculated\\
		\hline\hline
		QHD1 \cite{recentprogress}	&	0.54&	0.54&	0.54&	0.54\\
		NL-SH \cite{NLSH}&0.60&	0.60&	0.60&	0.60\\
		TM1	\cite{TM1}&0.634	&0.634&	0.634	&0.634\\
		TM2	\cite{TM1}&	0.571	&0.571&	0.571	&0.571\\
		NL3	\cite{NL3}&0.60&	0.60&	0.60	&0.60\\
		PK1	\cite{PK1}&	0.605525&	0.605526&	0.604981&	0.604983\\
		NL3	&	not available&	0.60	&not available&	0.60\\
		FSUGold	&not available&0.61&not available	&	0.61\\
		\hline
		\end{tabular}
	\caption{A comparison of the published values and the ones calculated in this work of the ration of the reduced (baryon) mass to the baryon mass at saturation. The references after the name of the parameter set indicate the source of the published values.}
	\label{tab:satprop3}
\end{table}
\begin{figure}[ttb]
	\centering
		\includegraphics[width=0.75\textwidth]{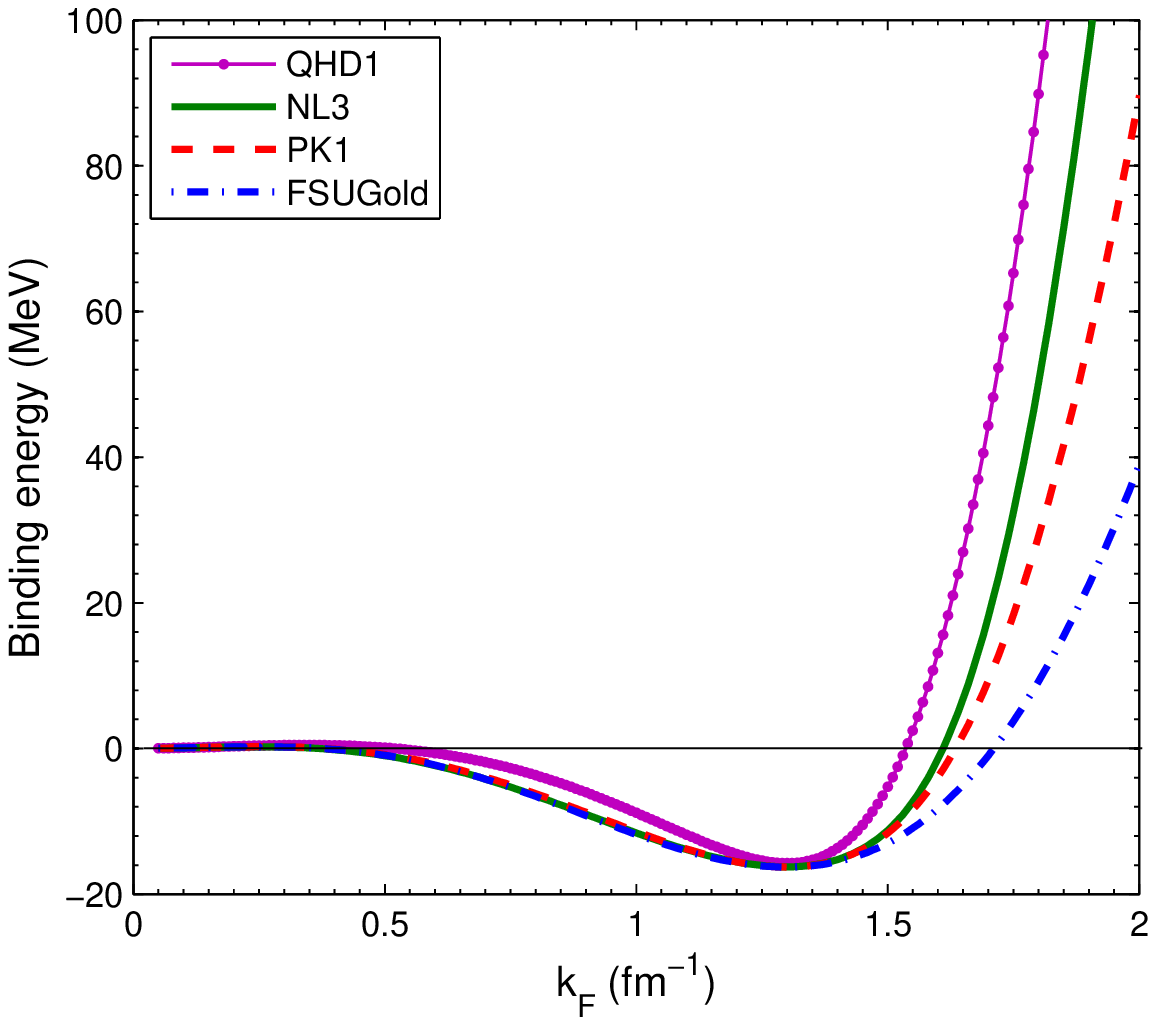}
	\caption{A comparison of the binding energy per nucleon of nuclear matter of different parameter sets.}
	\label{fig:BE}
\end{figure}\\
To determine the accuracy with which the FORTRAN90 program calculates the properties of saturated nuclear matter, these properties were also calculated using the NL-SH, TM1 and TM2 parameter sets. However, from this point onwards only the QHD-I, NL3, PK1 and FSUGold parameter sets will be considered.\\
\\
The binding energies around the saturation density (at a fermi momentum of 1.30 fm$^{-1}$) of the QHD-I, NL3, PK1 and FSUGold parameter sets are shown in Fig. \ref{fig:BE}. From this plot it is clear that the different parameters sets have very different high density behaviour. All the parameter sets give almost exactly the same values for the binding energies 
below saturation, but above saturation the values diverge. (The QHD1 parameter set was fitted to reproduce a binding energy at saturation of -15.75 MeV and therefore its description differs slightly from that given by the other parameter sets \cite{recentprogress}.) These discrepancies in the description of the binding energy at high densities emphasise that more information about dense matter is needed to better constrain the descriptions of nuclear matter.
\section{Neutron star properties}
In this work the NL3, PK1 and FSUGold parameter sets will be used to compare different calculated neutron star properties.
The FSUGold parameter set will be the default.
\begin{figure}
	\centering
		\includegraphics[width=0.70\textwidth]{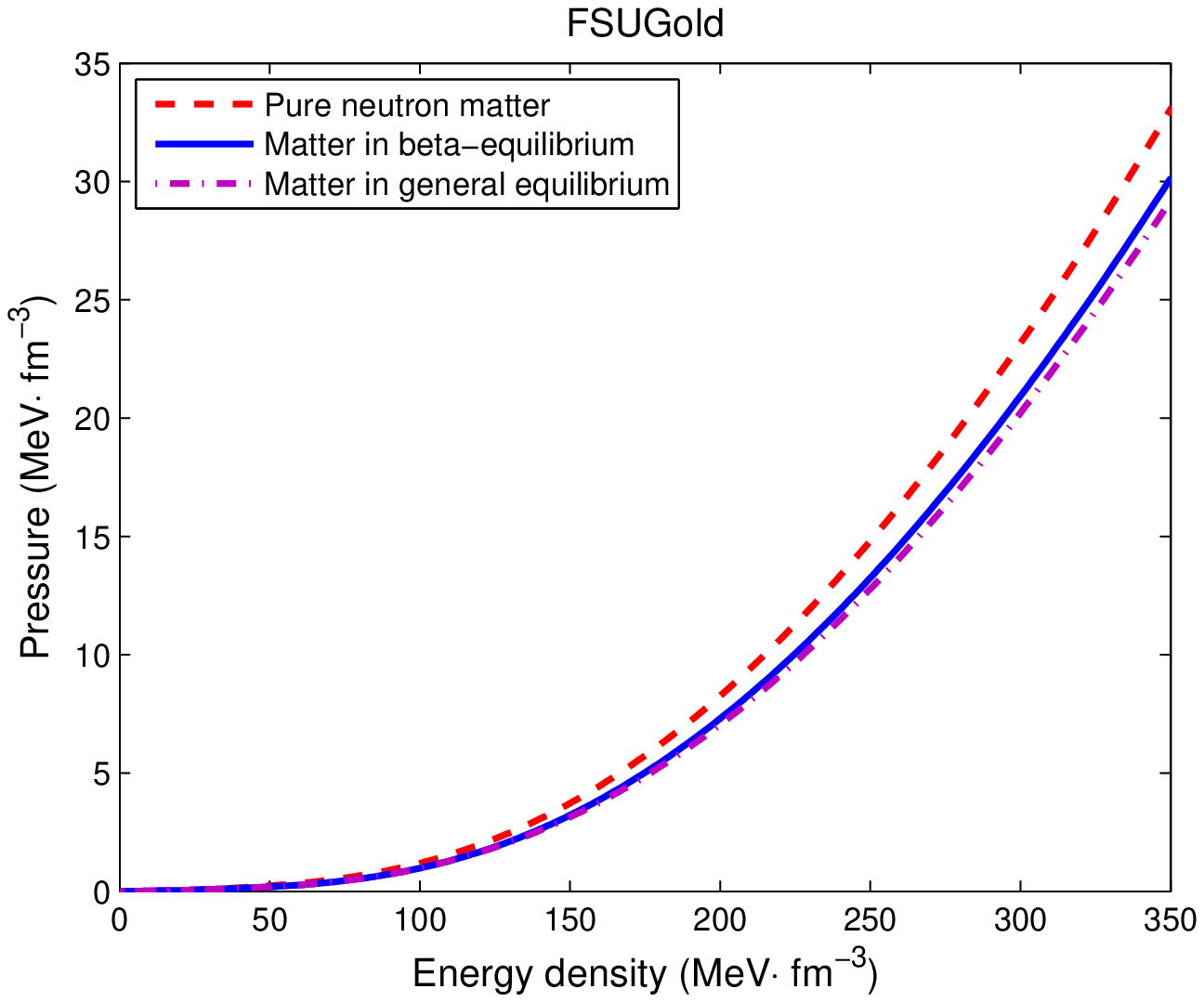}
	\caption{Softening of the equation of state due to the inclusion of different particles to the description of the neutron star interior.}
	\label{fig:softep}
\end{figure}
\begin{figure}[ttb]
	\centering
		\includegraphics[width=0.70\textwidth]{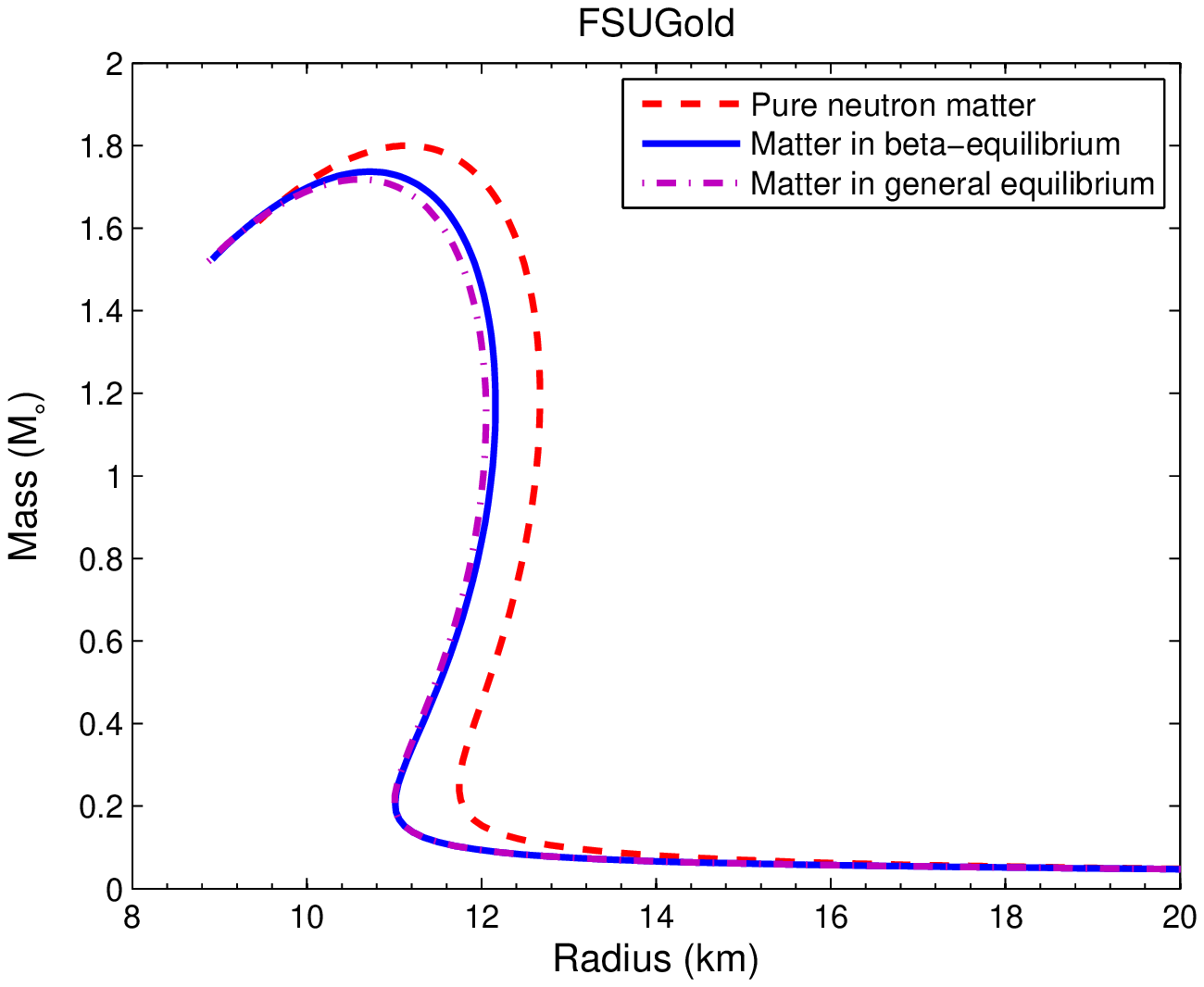}
	\caption{Influence, due to the inclusion of various particles in the description of the neutron star interior, on the
	mass-radius relationship of neutron stars.}
	\label{fig:softMR}
\end{figure}
\subsection{Inclusion of different particles}
As a first approximation the neutron star can be considered to be exclusively composed out of neutrons. This satisfies the condition of charge neutrality, but not the condition of beta-equilibrium, as expounded upon in Chapter \ref{chapNS}. By including the effects of beta-equilibrium the maximum mass of the neutron star sequence is reduced, i.e. the equation of state is softened. As was explained in Sec.\,\ref{comp}, a hard equation of state is when the energy density increases rapidly with an increase in pressure and a soft equation of state is when the increase in the energy density is more gradual. The equation of state is softened by the inclusion of different particles in the description of neutron star matter, because if it becomes energetically favourable to populate the states of, for example muons, the lowest energy muon states will be populated first, instead of the electron states, thereby reducing the increase in the energy density as the pressure increases. Fig. \ref{fig:softep} is an illustration of the softening of the equation of state by the inclusion of additional particles to the description of neutron star matter. If the neutron star is further considered to not only be in beta-equilibrium, but in general equilibrium as well (thus also including muons, instead of just protons, neutrons and electrons), the maximum mass of a particular sequence is further reduced.\\
\begin{table}[bbt]
	\centering
		\begin{tabular}{|l|ccc|}
			\hline
			Parameter set &\multicolumn{3}{c|}{Maximum mass (M$_\odot$)}\\
			\hline
			& \small Pure neutron matter & \small Beta-equilibrated matter& \small Matter in general equilibrium\\
			&\small (only neutrons)&\small (+ protons and electrons)&\small (+ muons)\\
			\hline\hline
			NL3		& 2.902 & 2.786	& 2.769 \\
			PK1		& 2.546 & 2.342	& 2.308 \\
			FSUGold& 1.800&	1.737 & 1.718 \\			
			\hline
		\end{tabular}
	\caption{Reduction of the maximum neutron star mass as different particles are included in the description of the neutron star interior.}
	\label{tab:soften}
\end{table}
\\
Table \ref{tab:soften} lists the maximum masses of a pure neutron neutron star, one in beta-equilibrium and a neutron star in general equilibrium, for different parameter sets. Fig. \ref{fig:softMR} shows the effect of the softening of the equation of state on the mass-radius relationship (equation of state described by the FSUGold parameter set).\\
\begin{figure}
	\centering
		\includegraphics[width=0.70\textwidth]{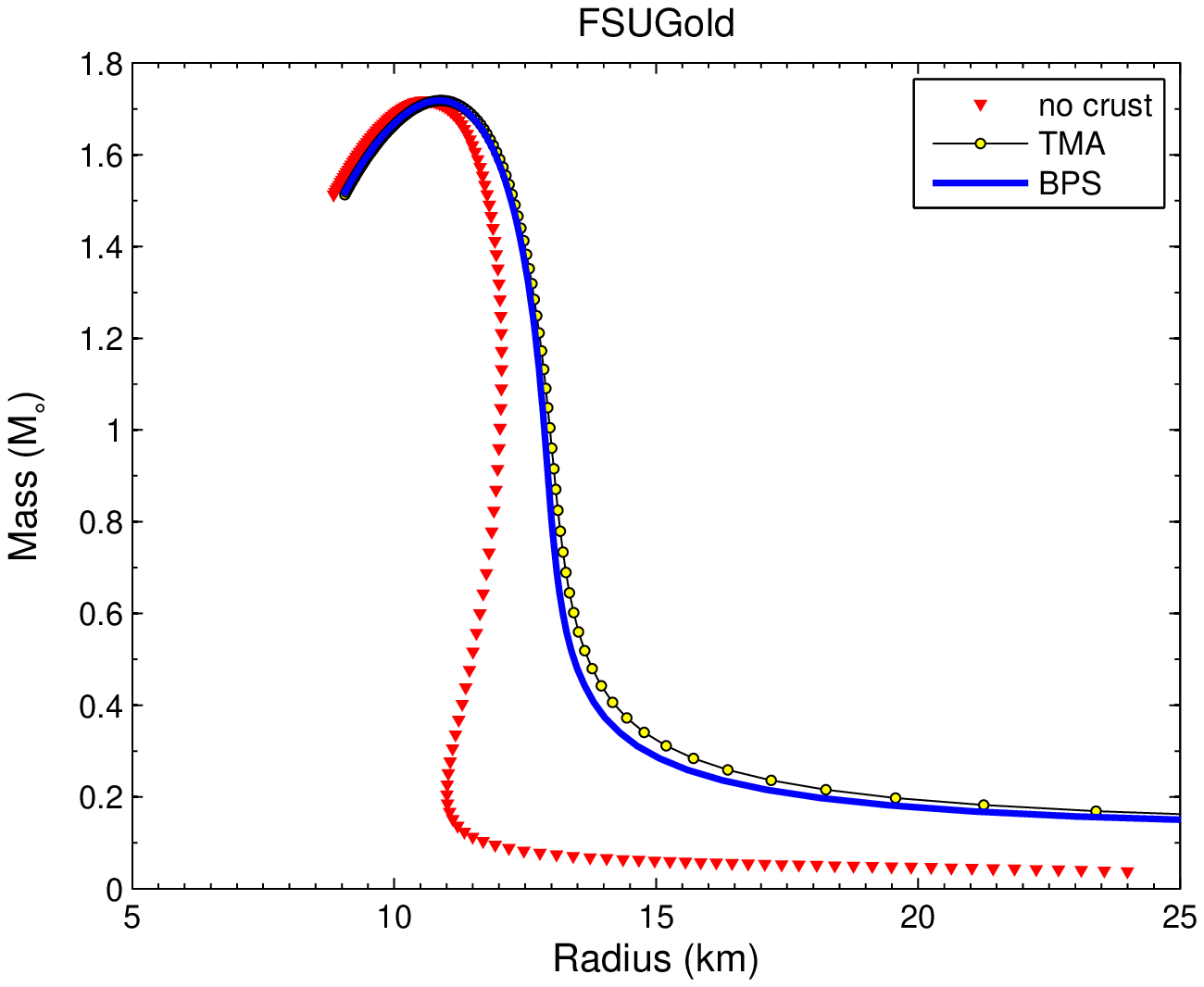}
	\caption{Sequence of neutron star mass-radius relationships, where outer and inner crustal effects have been 
	included in the description of the neutron star interior. The TMA plot refers to the BPS equation of state where the sequence of equilibrium nuclei is given by	Table \ref{tab:TMA}, while the BPS plot relates to that of Table \ref{tab:BPS}. }
	\label{fig:crustFSU}
\end{figure}
\subsection{Crustal effects}
\begin{figure}
	\centering
			\includegraphics[width=0.70\textwidth]{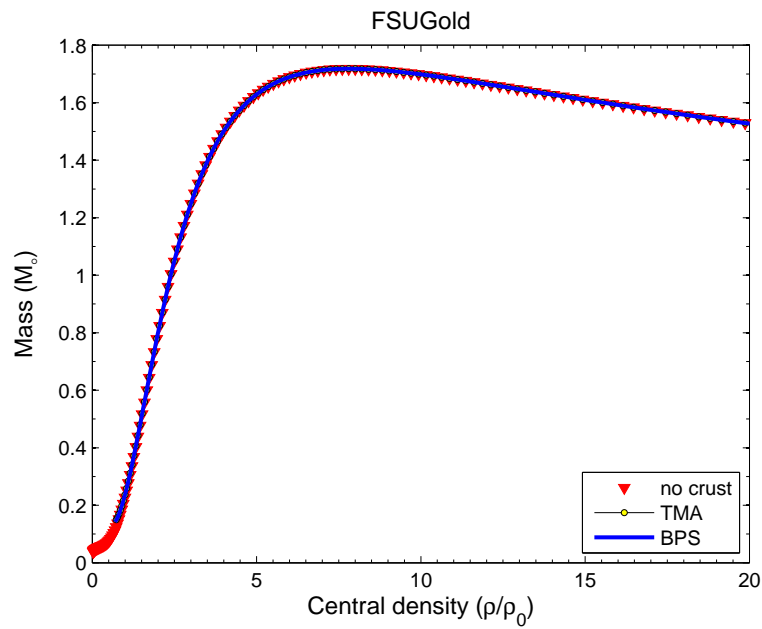}
			\caption{Plot of neutron star mass versus central density, where crustal effects have been included, corresponding to Fig. \ref{fig:crustFSU}. }
	\label{fig:crustFSUrho}
\end{figure}

The inclusion of crustal effects does not affect the mass of the neutron star, when compared to the case where 
crustal effects have not been included. However, the radii of neutron stars are in general larger when crustal effects are considered. 
\\\\
The reason that the inclusion of crustal effects do not influence the mass of the neutron star is that 
it does not affect the central pressure in the neutron star. The central pressure corresponds to the amount of overlying mass. The effect of the inclusion of crustal effects is rather to lower the pressure gradient near the boundary of the star, thereby enlarging the radius of the star. 
%
Fig. \ref{fig:crustFSU} shows the neutron star mass-radius sequence, for an equation of state based on the FSUGold parameter set, where different descriptions for the outer crust (TMA and BPS), as well as the polytropic equation of state for the inner crust have been included. 
Fig. \ref{fig:crustFSUrho} shows the plot of the neutron star mass against the central density of this calculation.\\
\\
%
%
%
\begin{figure}
	\centering
		\includegraphics[width=0.75\textwidth]{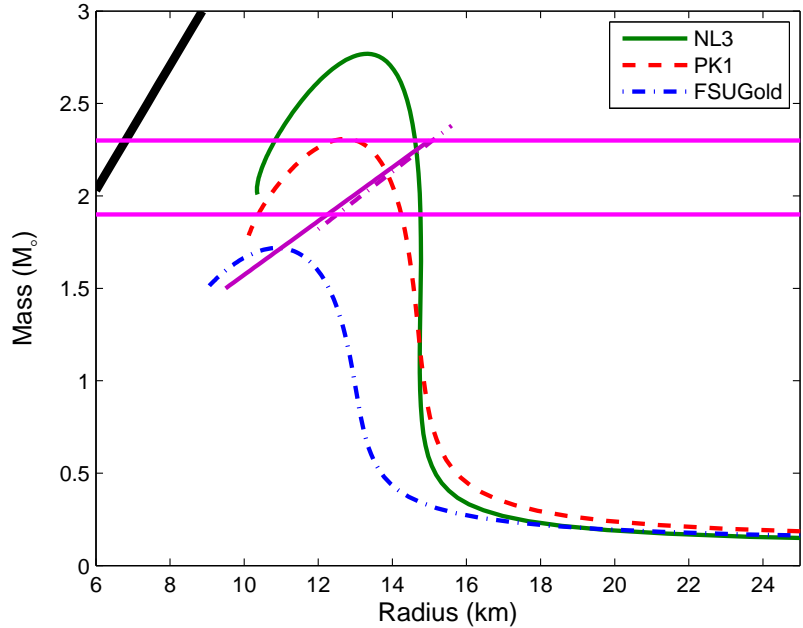}
	\caption{Sequences of neutron star mass-radius relationships corresponding to equations of state based on different parameter sets. All equations of state shown include a description of the inner crust (polytropic equation of state, matched at the appropriate densities for the transition to the liquid interior) and the BPS equation of state (specifically the TMA rendition) for the outer crust. The liquid interior of the star is assumed to consist of protons, neutrons, electrons and muons in general equilibrium. The horizontal (pink) lines correspond to the constraints imposed by the mass measurement of PSR J0751+1807. The diagonal dotted (purple) line represents the possible mass-radius relationships of EXO 0748-676, as calculated by F. $\ddot{\mbox{O}}$zel \cite{Ozel}, while the diagonal solid (purple) line corresponds to the calculation made by A. R. Villarreal and T. E. Strohmayer \cite{Villarreal}. The diagonal solid (black) line in the upper left-hand corner denotes the constraint imposed on the mass-radius relationship of a neutron star by the general theory of relativity. }
	\label{fig:obsconstrain}
\end{figure}
\begin{table}[bbt]
	\centering
		\begin{tabular}{|l|c|c|c|c|}
			\hline
			Parameter set & \multicolumn{2}{c}{TMA}&\multicolumn{2}{|c|}{BPS}\\
			\hline
			 &Maximum mass (M$_\odot$)& Radius (km)&Maximum mass (M$_\odot$)& Radius (km)\\
			\hline\hline
			NL3		& 2.768 &13.337 & 2.768& 13.331\\
			PK1		& 2.308 &12.761&  2.308 &12.752 \\
			FSUGold& 1.718&10.893& 1.718 &10.863  \\			
			\hline
		\end{tabular}
	\caption{Maximum masses and radii of different neutron star sequences. (The neutron star matter is considered to be in general equilibrium and crustal effects are included in the description.)}
	\label{tab:maxmass}
\end{table}
\subsection{Observational constraints}
As argued in Section \ref{sec:obs} observational evidence and theoretical constraints have to be taken into account to determine the appropriate descriptions of the neutron star interior. Fig. \ref{fig:obsconstrain} shows the plot of mass-radius sequences corresponding to different equations of state.\\
\\
From Fig. \ref{fig:obsconstrain} it can be seen that the NL3 and PK1 parameter sets provide a description which is in good agreement with current observational data, while the FSUGold description may seem to be too soft, if the maximum masses of each sequence (as given in Table \ref{tab:maxmass}) are considered.\\
\\
J. Piekarewicz has pointed out these apparent deficiencies of the FSUGold parameter set in Ref.\,\cite{FSUL}, but also stated that it would be rather naive to only consider neutron star constraints to determine the equations of state of dense matter. The difference between NL3 and PK1 parameter sets, is that the PK1 set includes an additional self-coupling in the omega meson field ($V^\mu$). Comparing the PK1 and FSUGold parameter sets, it will be seen that the FSUGold parameter set contains a further coupling between the two vector boson fields (${\bm b}^\mu$ and $V^\mu$), $\Lambda_v$, which was introduced to soften the symmetry at high density \cite{FSU}. However, the density dependence of the symmetry energy is currently unknown, so $\Lambda_v$ could not yet be firmly constrained \cite{FSUL}. However, the Parity Radius Experiment (PREX) aims to accurately measure the neutron radius of $^{208}$Pb, which would provide an accurate determination of the density dependence of the symmetry energy \cite{Links}. PREX will be conducted at Jefferson Laboratories in the United States and is currently scheduled for February 2009 \cite{PREX}.
%
%
%
%

%% file: chapConclu.tex
\chapter{Conclusions}
In this work the equations of state of neutron star matter were derived using the quantum hadrodynamics model of nuclei and nuclear matter by the application of the relativisitc mean-field approximation for the NL3, PK1 and FSUGold parameter sets. It was also shown (as was done in similar works, such as Refs\ \cite{Car, FSUL} and \cite{mengNS}) that these derived equations of state can be applied to calculate the mass-radius relationship of a neutron star and that these calculated values show good agreement with current observational data. However, further measurements of the properties of nuclear matter as well as that of neutron stars are necessary to obtain a complete description of dense nuclear matter.\\ 
\\
As argued by various authors, a description of dense matter should not be based on a single set of observables or a class of observables, but rather the most applicable description of dense matter would be the one that 
describes a plethora of different observables accurately \cite{Klaehn, FSUL}. Unfortunately our knowledge of dense matter, in whatever form, is currently limited. Fortunately it is foreseen that with upcoming experiments at current and new particle accelerator facilities, as well as the development of new telescopes, particularly radio-telescopes, much will be \mbox{uncovered} and discovered about terrestrial and extra-terrestrial dense matter in the near future. To gather a more complete picture of dense matter, works, such as this one, would be increasingly important in providing a link between laboratory physics and astrophysics. However, through the study of nuclear physics alone, this picture would not be easily attained and therefore the need for cooperation between astro- and nuclear physicists is greater than ever.\\
\\
The application of relativistic mean-field theory to the quantum hadrodynamics has therefore shown itself to be relevant to the description of finite nuclei and nuclear matter \cite{recentprogress}, as well as the description of neutron stars, based solely on a handful of coupling constants. 

%% file: Appendix.tex
\chapter{Code Documentation}\label{ap:code}
All numerical problems in this work were solved using programs that were written for this purpose in FORTRAN90 programming language. This appendix will give a brief overview of these computer programs.
\section{Introduction}
To obtain the mass-radius relationship of a neutron star using a specific parameter set in QHD is quite a complex physical and mathematical problem. Even after using RMF to approximate the description of nuclear matter, the equations that have to be solved are rarely exact, at most self-consisted. The main problem of applying a description of nuclear matter to neutron stars can be split into the following components: 
\begin{enumerate}
	\item obtaining the values of the unknown meson fields in the RMF approximation,
	\item calculating the equation of state of neutron star matter, and,
	\item using the equation of state to solve the TOV equation.
\end{enumerate}
In the following sections different parts of the code pertaining to the above-mentioned problem will be discussed. Whenever a function or subroutine is mentioned, it will be written in \textbf{bold} and subroutines are denoted by the prefix \textsl{sub\_} and functions by the prefix \textsl{func\_}.
\begin{figure}
	\centering
			\includegraphics[width=0.99\textwidth]{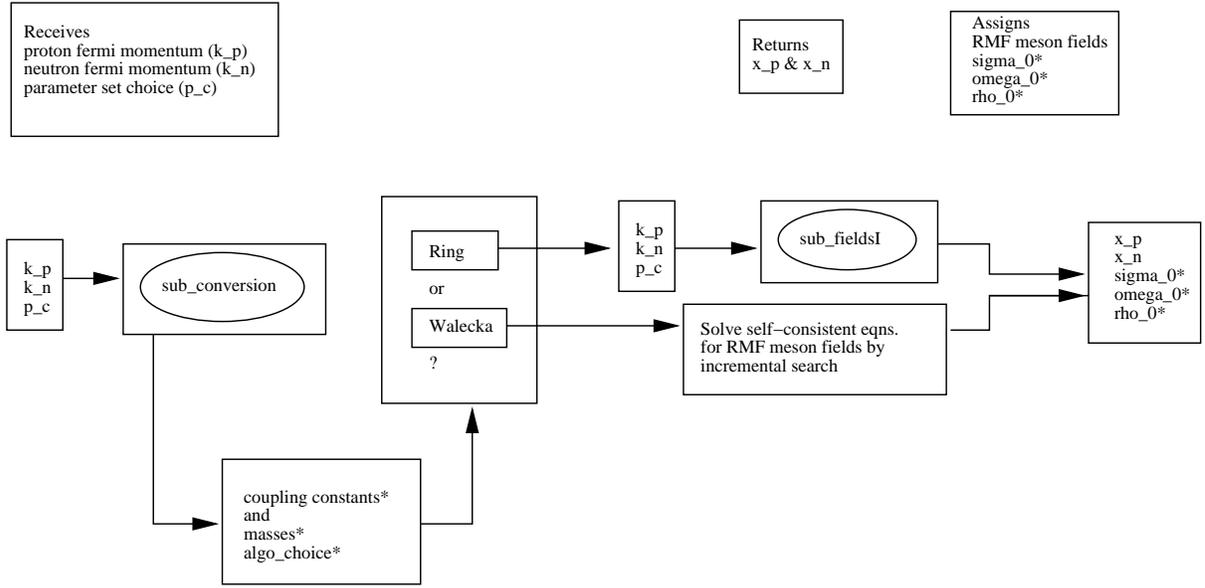}
		\caption{Illustration of the subroutine \textbf{sub\_fields} (global variables are denoted by an asterik).}
	\label{fig:subfields}
\end{figure}
\section{sub\_fields}
\textbf{Sub\_fields} is the subroutine used to calculate the RMF values of the meson fields ($\phi_0$, $V_0$ and $b_0$).\\\\A schematic representation of this subroutine is given in Fig. \ref{fig:subfields}. As imput \textbf{sub\_fields} \mbox{receives} values of the proton and neutron fermi momenta, $k_p$ (denoted by k\_p) and $k_n$ (k\_n), and the choice of the parameter set (para\_choice) for which the RMF meson field values have to calculated. For each specific value of k\_p and k\_n \textbf{sub\_conversion} is called. \textbf{Sub\_conversion} assigns all the coupling constants and masses 
as well as the baryon number density (density\_B) as global parameters for the specific choice of para\_choice. Once this is done the value of the RMF meson fields can be calculated. This is done by solving the self-consistent expressions for the different meson fields (\ref{FSUEQM}).\\
\\
The values of the $V_0$ and $b_0$ are determined by comparing the left- and right-hand sides of Eqs (\ref{FSUomega}) and (\ref{FSUrho}). The value of the field on the left-hand side of the equation is incrementally increased and then substituted into the expression on the right and the values on the left and right are compared. The RMF meson fields get assigned once the difference between the left and right sides are acceptability small (about $8.0\times10^{-7}$ fm$^{-4}$).\\
\\
In the case of $\phi_0$ the value of the reduced mass ($m^*$) is calculated, from which $\phi_0$ is then calculated using Eq. (\ref{mreduced1}). An expression for $m^*$ in terms of $\phi_0$ can be easily deduced from the expressions for $\phi_0$ (\ref{FSUsigma}) and $m^*$ (\ref{mreduced1}) and yields
\begin{eqnarray}
	m^* &=& M - \frac{g_{s}^2}{m_s^2}\Big( 
			\frac{1}{\pi^2}\,\int^{k_p}_0 dk\,\frac{ k^2\,m^*}{\sqrt{{k}^2 + {m^*}^2}} 
			+ \frac{1}{\pi^2}\,\int^{k_n}_0 dk\,\frac{ k^2\,m^*}{\sqrt{{k}^2 + {m^*}^2}}\Big) \nonumber\\
			&& -\frac{g_{s}^2}{m_s^2}\Big( - 
			\frac{\kappa}{2}(g_s\phi_0)^2 - \frac{\lambda}{6}(g_s\phi_0)^3\Big)\label{mredcode}
\end{eqnarray}
The integral in the above expression is evaluated using the analytical expression for the integral given in Appendix \ref{ap:sol}.\\
\\
After checking one final time if the self-consistent equations are satisfied to the desired accuracy (this is necessary, especially since in the FSUGold parameter set there is a coupling between $V_0$ and $b_0$ which complicates the calculation) the values of $\phi_0$ [denoted by sigma\_0 (\textsl{sic}) throughout the code], $m^*$ (m\_reduced), $V_0$ (omega\_0) and $b_0$ (rho\_0) are assigned (as global variables). 
$$
x_p\equiv\frac{k_p}{m^*}
$$
 and 
$$
 x_n\equiv \frac{k_p}{m^*}
$$ 
are also assigned. $x_p$ and $x_n$ are used in the calculation of the energy density and the pressure and stem from the analytical solution to the integrals Appendix \ref{ap:sol} in the expression for the energy density and the pressure (\ref{FSUEoSWalecka}).\\
\\
The differences between the Walecka and Ring conventions (see Sec. \ref{sec:convent}) necessitate the definition of a different subroutine to calculate the RMF meson field values in the Ring convention. \textbf{Sub\_fieldsI} is very similar to \textbf{sub\_fields}, but one of the few differences is that a numerical approximation is made to calculate the integral in the expression for $m^*$. 
\begin{figure}
	\centering
		\includegraphics[width=1.0\textwidth]{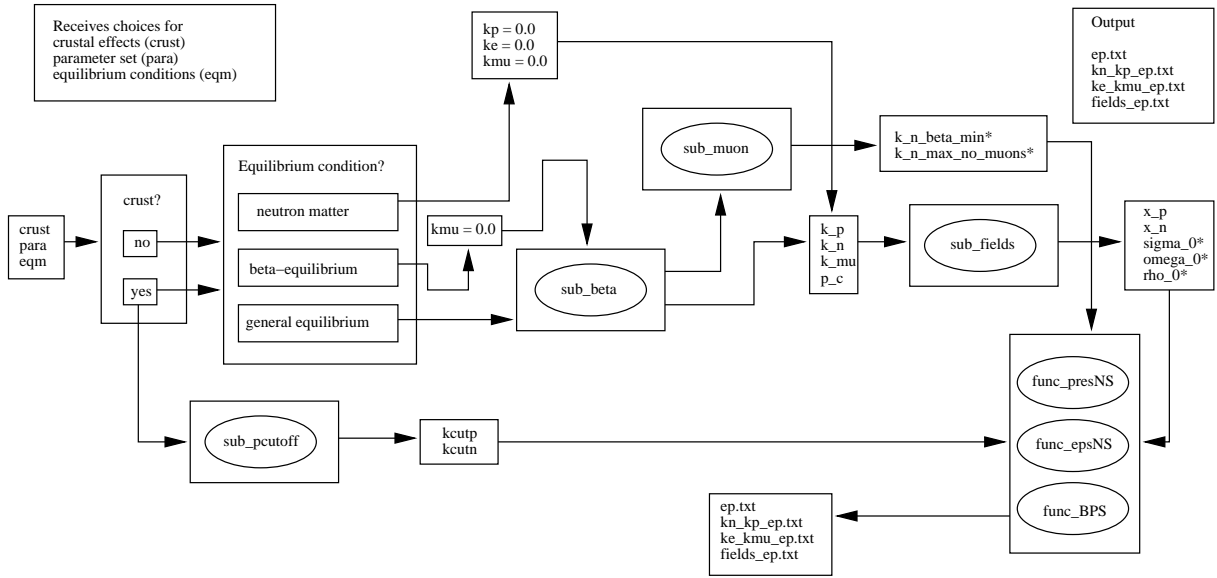}
	\caption{Illustration of the subroutine \textbf{sub\_epNS} (global variables are denoted by an asterik).}
	\label{fig:subepNS}
\end{figure}

\section{sub\_epNS}
This subroutine calculates the arrays containing corresponding values of the energy density and the pressure to be used in \textbf{sub\_TOV}, which solves the TOV equation. \textbf{Sub\_epNS} is \mbox{represented} graphically in Fig. \ref{fig:subepNS}.\\
\\
Specific choices of the crust (crust\_choice), equilibrium (beta of general, eqm\_choice) and the parameter set (para\_choice) serve as input to \textbf{sub\_epNS}.\\
\\ 
For a specific choice of the crust \textbf{sub\_pcutoff} determines the transition between the inner crust and the liquid interior and/or between the outer and inner crust. It returns a value of $k_n$ and $k_p$ (k\_cutn and k\_cutp) at the corresponding pressure or number density of the transition. \\
\\
If the description of matter in general equilibrium is required, \textbf{sub\_muon} returns the value of $k_n$ (k\_max\_no\_muons) above which the muons states will be populated. \textbf{Sub\_muon} also assigns the minimum values of $k_n$ (k\_n\_beta\_min) at which the matter can be in beta-equilibrium. The value of k\_n\_beta\_min is not calculated by \textbf{sub\_muon}, but was determined by hand and then hard coded to be $1.49\times10^{-1}$ fm$^{-1}$. \textbf{Sub\_beta} is used to determine the values of the proton, electron and muon fermi momenta (kp, ke and kmu) corresponding to a neutron fermi momentum, kn (which of course this calculation is parameter set specific). If only beta-equilibrated matter is desired, \textbf{sub\_beta} calls \textbf{sub\_beta2}, which does the necessary. \\
\\
If no crustal effects are considered, \textbf{sub\_fields} is called to determine the values of $\phi_0$ (sigma\_0), $\omega_0$ (omega\_0), $b_0$ (rho\_0), $x_p$ (x\_p) and $x_n$ (x\_n). These values, along with the values of $k_e$ (k\_e) and $k_\mu$ (k\_mu), are then used by \textbf{func\_epsNS} and \textbf{func\_presNS} to calculate the energy and the pressure of neutron star matter, using Eqs (\ref{FSUeps}) and (\ref{FSUpres}) respectively.\\
\\
The equation of state for the outer crust is calculated by \textbf{func\_bps}. For the choice of the \mbox{sequence} of equilibrium nuclei (either TMA from Table \ref{tab:TMA} or BPS from Table \ref{tab:BPS}) and \mbox{specific} value of the pressure in the outer crust, \textbf{func\_bps} returns the energy density. For the inner crust the BBP equation of state can be used or the polytropic equation of state [denoted by RPA in the program and given by Eq. (\ref{polyeos})]. The BBP equation of state is calculated by reading the values given in Table \ref{tab:BBP} from file (BBPE.txt and BBPP.txt for the energy density and the pressure respectively) and determined by interpolating the energy density at a specific pressure. The polytropic equation of state is calculated by first determining the values for $A$ and $B$. Using equation \ref{polyeos} the energy density is calculated for each input value of the pressure. The equation of state of the liquid interior of the neutron star is given by calling \textbf{func\_epsNS} and \textbf{func\_presNS} for values of $k_n$ (kn) greater than kcutn.\\
\\
The output of \textbf{sub\_epNS} is a file, ep.txt, which contains the pressure in the first column and the energy density in the second. The following data is also printed to file: The value of $k_n$ (kn) and $k_p$ (kp) to kn\_kp\_ep.txt; $k_e$ (ke), $k_\mu$ (kmu) and density\_B ($\rho_p + \rho_n$) to ke\_kmu\_ep.txt, while fields\_ep.txt contains the values of $\phi_0$ (sigma\_0), $\omega_0$ (omega\_0) and $b_0$ (rho\_0).
\section{sub\_TOV}
This subroutine solves the TOV equation. As input \textbf{sub\_TOV} receives the neutron fermi momentum (k\_fn) assumed to describe the matter at the centre of the star, as well as the choices for the parameter set, crust and equilibrium conditions. The equation of state have to be read into the pres and eps arrays as global variables, this has to be done before \textbf{sub\_TOV} is called.\\
\\
\textbf{Sub\_TOV} solves the TOV equation using the Runge-Kutta method to solve a coupled differential equation.\\
For a given coupled differential equation, such as Eqs (\ref{TOV}) and (\ref{TOVM}),	
\begin{subequations}\label{TOVsolve}
	\begin{eqnarray}
		\frac{dP}{dr} &=& F(r, P, M)\nonumber\\
			&=&	-\frac{G\epsilon M}{r^{2}}\Biggl[1+\frac{P}{\epsilon}\Biggr]
				\Biggl[1+\frac{4\pi r^{3}P}{M}\Biggr]
				\Biggl[1-\frac{2GM}{r}\Biggr]^{-1}\\
	\shoveleft{\mbox{and}}\ \  \ \ \ \ \ \ \ \ \ \ \ \ &&\nonumber\\
		\frac{dM}{dr} &=&G(r,P,M)\nonumber\\
		&=& 4\pi \epsilon r^{2},
	\end{eqnarray}
\end{subequations}
the Runge-Kutta method defines a single integration step of size $h$ by \cite{runge}
\begin{eqnarray}
	P_{n+1}&=& P_n + \frac{1}{6}k_1 + \frac{1}{3}k_2 + \frac{1}{3}k_3 + \frac{1}{6}k_4\nonumber\\
	M_{n+1}&=& M_n + \frac{1}{6}d_1 + \frac{1}{3}d_2 + \frac{1}{3}d_3 + \frac{1}{6}d_4\nonumber
\end{eqnarray}
with
\begin{eqnarray}
	k_1 &=& h F(r_n, P_n, M_n)\nonumber\\
	k_2 &=& h F(r_n + h/2, P_n + k_1/2, M_n + d_1/2)\nonumber\\
	k_3 &=& h F(r_n + h/2, P_n + k_2/2, M_n + d_2/2)\nonumber\\
	k_4 &=& h F(r_n + h, P_n + k_3, M_n + d_3)\nonumber\\\nonumber\\
	d_1 &=& h G(r_n, P_n, M_n)\nonumber\\
	d_2 &=& h G(r_n + h/2, P_n + k_1/2, M_n + d_1/2)\nonumber\\
	d_3 &=& h G(r_n + h/2, P_n + k_2/2, M_n + d_2/2)\nonumber\\
	d_4 &=& h G(r_n + h, P_n + k_3, M_n + d_3)\nonumber.
\end{eqnarray}
If crustal effects are taking into consideration \textbf{sub\_TOV} uses the above method to integrate the TOV equation until the pressure [$P(r)$] becomes $10^{-8}$ MeV/fm$^{3}$, which denotes the value of the pressure were the crust of a neutron star starts, from Ref. \cite{Rus}. If crustal effects are not considered \textbf{sub\_TOV} integrates until the pressure [$P(r)$] becomes $2.0\times10^{-5}$ MeV/fm$^{3}$ (the lowest pressure at which matter can be in beta-equilibrium). \\
\\
Due to the units of the variables care should be taken with the expressions for $F(r, P, M)$ and $G(r,P,M)$ as pointed out in Refs \cite{ns4u, piekWD}. $M$ and $r$ are of the order of a couple solar masses and kilometres respectively, while $P$ and $\epsilon$ are in MeV/fm$^{3}$. Thus to be able to handle these equations numerically, they have to be scaled. An appropriate choice would be to describe $M$ in terms of solar masses, i.e. $$M = M_{\odot}\bar{M}$$ where
\begin{itemize}
	\item $M_\odot$ is a solar mass (in MeV) as given in Table \ref{tab:constants}, and,
	\item $\bar{M}$ is a dimensionless quantity
\end{itemize}
as well as to express $r$ in units of metres.\\
\\
Defining the following quantities
\begin{itemize}
	\item $G = G'\cdot10^{-30}$ with G the gravitational constant in MeV/fm$^{3}$ from Table \ref{tab:constants}, 
	and,
	\item $M_{\odot} = M'_{\odot}\cdot10^{45}$,
\end{itemize}
(\ref{TOVsolve}) can be rewritten as
\begin{subequations}\label{TOVsolve2}
	\begin{eqnarray}
		\frac{dP}{dr} &=& F(r, P, M)\nonumber\\
			&=&	-\frac{G'\epsilon M'M'_{\odot}}{r^{2}}\Biggl[1+\frac{P}{\epsilon}\Biggr]
				\Biggl[1+\frac{4\pi r^{3}P}{M'_{\odot}M'}\Biggr]
				\Biggl[1-\frac{2G'M'_{\odot}M}{r}\Biggr]^{-1}\\
	\shoveleft{\mbox{and}}\ \  \ \ \ \ \ \ \ \ \ \ \ \ &&\nonumber\\
		\frac{dM'}{dr} &=&G(r,P,M)\nonumber\\
		&=& \frac{4\pi \epsilon r^{2}}{M'_{\odot}},
	\end{eqnarray}
\end{subequations}
The above equations are the ones used by \textbf{sub\_TOV} to calculate the mass-radius relationships of neutron stars in this work.
%
%
%
\section{List of modules}
For completeness sake the modules in the FORTRAN90 program will be listed. (If the subroutine or function has not been mentioned previously a quick explanation will be given.)\\
\\
The module \textbf{mod\_comp\_val\_RING} contains
\begin{itemize}
	\item \textbf{sub\_fieldsI},
	\item \textbf{func\_rhosI} , which calculates the integral in the expression for $\phi_0$ numerically, and, 
	\item \textbf{func\_intrhos} calculates the integrand of the numerical integration in \textbf{func\_rhosI}, while,
	\item \textbf{func\_sigmaI},
	\item \textbf{func\_omegaI}, and,
	\item \textbf{func\_rhoI} are the self-consistent expressions for the meson fields called by \textbf{sub\_fields}.
	
\end{itemize}
The module \textbf{mod\_comp\_val\_WALECKA} contains
\begin{itemize}
	\item \textbf{sub\_fields},
	\item \textbf{func\_rho\_s}, which calculates the integral in the expression for $\phi_0$ numerically, and,
	\item \textbf{func\_int\_rho} the integrand of the numerical integration in \textbf{func\_rho\_s}, while,
	\item \textbf{func\_sigma},
	\item \textbf{func\_omega}, 
	\item \textbf{func\_rho} are the self-consistent expressions for the meson fields called by \textbf{sub\_fields}, and,
	\item \textbf{print\_parameters} that prints the value of the coupling constants and masses of a \mbox{particular} parameter set to the screen.
\end{itemize}
The two modules mentioned above contains all the subroutines and functions used in calculating 
the RMF values of the meson fields and refers to the two different conventions 
used in this work.\\
\\
The module \textbf{mod\_EoSFSU} contains
\begin{itemize}
	\item \textbf{func\_epsNS},
	\item \textbf{func\_presNS},
	\item \textbf{func\_epsFSU},
	\item \textbf{func\_presFSU},
	\item \textbf{func\_inteps} which contains the integrand for the numerical integration  
	of the integral in the expression for the energy density (this function was not used),
	\item \textbf{func\_bps}, and,
	\item \textbf{func\_getke} which calculates the electron fermi momentum for a given pressure in \textbf{func\_bps};
\end{itemize}
whereas the module \textbf{mod\_gauleg} contains 
\begin{itemize}
	\item \textbf{sub\_gauleg}, which calculates the roots and weights for the numerical integration,
\end{itemize}
and \textbf{mod\_splint} contains the interpolation functions 
\begin{itemize}
	\item \textbf{sub\_csplin} which calculates the coefficients for the numerical interpolation, and,
	\item \textbf{func\_cseval} which does the actual interpolation using the coefficients generated by 
	\textbf{sub\_csplin}. 
\end{itemize}
The module \textbf{mod\_globals} contains all the global parameters as well as
\begin{itemize}
	\item \textbf{sub\_choice} a simple subroutine that prints the choices made to the screen, and,
	\item \textbf{sub\_conversion,} 
\end{itemize}
while \textbf{mod\_obs} contains
\begin{itemize}
	\item \textbf{sub\_sat}, which calculates the properties of saturated nuclear matter for a choice of \mbox{parameter} set and prints the results to the screen and to the file sat\_prop.txt,
	\item \textbf{sub\_symm\_ep} calculates the equation of state of symmetric nuclear matter, used in \textbf{sub\_sat}, and, 
	\item \textbf{sub\_epNS}.
\end{itemize}
\textbf{mod\_procedures} houses all kinds of functions and subroutines used by other modules, such as
\begin{itemize}
	\item \textbf{sub\_beta},
	\item \textbf{sub\_beta2},
	\item \textbf{sub\_muon}, and,
	\item \textbf{sub\_pcutoff},
\end{itemize}
which were all introduced previously.\\
\\
Finally \textbf{mod\_TOVFSU} contains
\begin{itemize}
	\item \textbf{sub\_TOV},
	\item \textbf{func\_TOV1Fus}, and,
	\item \textbf{func\_TOV1Gus} which contains the expressions for the right hand side of the equations used to solve the TOV equation (\ref{TOVsolve2}).
\end{itemize}
\section{Codes for different choices}
\begin{figure}
	\centering
		\includegraphics[width=0.75\textwidth]{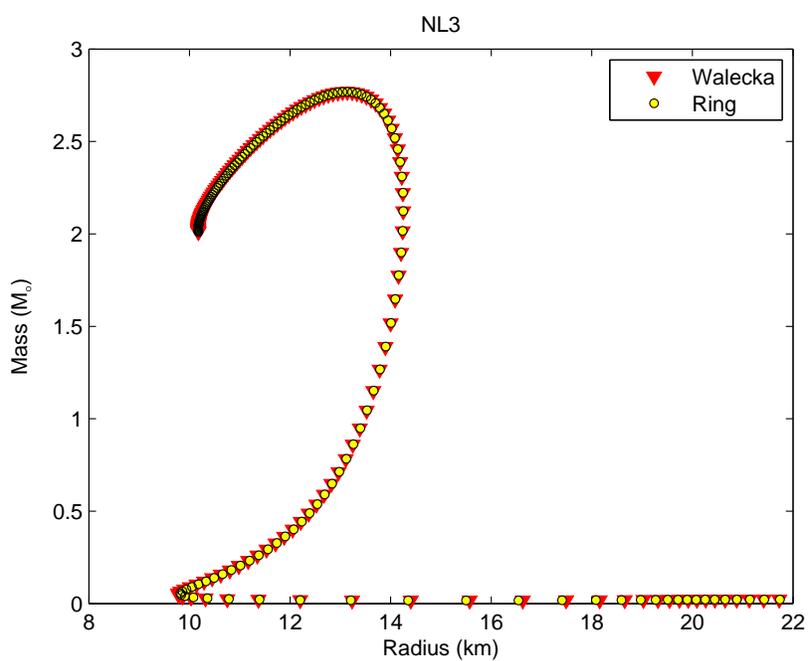}
	\caption{Mass-radius relationship for neutron star matter in general equilibrium, no crustal effects included, for the NL3 parameter set.}
	\label{fig:NL3MRWvsR}
\end{figure}

The codes for choices of the parameters sets, crustal effects and equilibrium condition are:
\textbf{Parameter sets}
\begin{itemize}
	\item 1 FSUGold
	\item 2 QHD1
	\item 3 NL3 (Walecka convention)
	\item 31 NL3 (Ring convention)
	\item 51 PK1
	\item 61 NL-SH
	\item 71 TM1
	\item 81 TM2
\end{itemize}
\textbf{Crustal choices}
\begin{itemize}
	\item 0 if no crustal effects included
	\item 1 if only outer crust, BPS equation of state based on TMA sequence of equilibrium nuclei,
	\item 2 if only outer crust, BPS equation of state based on BPS sequence of equilibrium nuclei,
	\item 31 if the BPS (TMA) equation of state for outer crust and BBP equation of state for the inner crust are required,
	\item 32 if the BPS (BPS) equation of state for outer crust and BBP equation of state for the inner crust are required,
	\item 11 if the BPS (TMA) equation of state for the outer crust and polytropic equation of state for inner crust are required, and,
	\item 12 if the BPS (BPS) equation of state for the outer crust and polytropic equation of state for inner crust are required.
\end{itemize}
\textbf{Equilibrium conditions}
\begin{itemize}
	\item 1 for neutron star matter in general equilibrium,
	\item 2 for neutron star matter in beta-equilibrium only, and, 
	\item 3 for pure neutron matter
\end{itemize}
To illustrate the accuracy of the different subroutines for the different conventions the mass-radius relationship for the NL3 parameter set, assuming neutron star matter in general equilibrium and no crustal effects, is shown in Fig. \ref{fig:NL3MRWvsR}.
\chapter{Solutions to integrals}\label{ap:sol}
\section{Energy density}
Any expressions for the energy density of fermi-gas type model contains the integral
\begin{eqnarray}
	\frac{1}{\pi^{2}}\int^{k_{F}}_{0}dk\,k^{2}\sqrt{k^{2}+{m}^{2}}\nonumber
\end{eqnarray}
where
\begin{itemize}
	\item $k_F$ is the fermi momentum of the specific particle, and
	\item $m$ is the mass.
\end{itemize}
This integral can be solved analytically by making the following substitution
$$
	\frac{k}{m}\rightarrow u,
$$
and defining
$$\frac{k_F}{m} \equiv x_F$$thereby
\begin{eqnarray}
	\frac{1}{\pi^{2}}\int^{k_{F}}_{0}dk\,k^{2}\sqrt{k^{2}+{m}^{2}}\;
	&=&\frac{m^4}{\pi^{2}}\int^{k_{F}/m}_{0}du\,u^{2}\,\sqrt{u^{2}+1}\;\nonumber\\
	&=&\frac{m^4}{\pi^{2}}\int^{x_{F}}_{0}du\,u^{2}\,\sqrt{u^{2}+1}\;\nonumber\,.
\end{eqnarray}
Using integration by parts, the last integral reduces to:
\begin{eqnarray}
	\frac{m^4}{\pi^{2}}\int^{x_{F}}_{0}du\,u^{2}\,\sqrt{u^{2}+1}\;
	\ =\ 
	\frac{m^4}{8\pi^{2}}\Big[\left(2x_F^3+x_F\right)
	\left(1+x_F^2\right)^{1/2}-\sinh^{-1}\left(x_F\right)\Big]\,.
\end{eqnarray}\\
\section{Pressure}
As with the energy density, any expression for the pressure contains the integral
\begin{eqnarray}
	\frac{1}{3\pi^{2}}\int^{k_{F}}_{0}dk\,\frac{k^{4}}{\sqrt{k^{2}+{m}^{2}}}\nonumber\,.
\end{eqnarray}
By making the same substitution as with the energy density integral, the integral reduces to
\begin{eqnarray}
	\frac{1}{3\pi^{2}}\int^{k_{F}}_{0}dk\,\frac{k^{4}}{\sqrt{k^{2}+{m}^{2}}}\,
	&=&\frac{m^4}{3\pi^{2}}\int^{k_{F}/m}_{0}du\,\frac{u^{4}}{\sqrt{u^{2}+1}}\,\nonumber\\
	&=&\frac{m^4}{3\pi^{2}}\int^{x_{F}}_{0}du\,\frac{u^{4}}{\sqrt{u^{2}+1}}\,.\nonumber
\end{eqnarray}
Which can be solved analytically as
\begin{eqnarray}
	\frac{m^4}{3\pi^{2}}\int^{x_{F}}_{0}du \,\frac{u^{4}}{\sqrt{u^{2}+1}}\ = \ 
	\frac{m^4}{3\pi^{2}}\frac{1}{8}
	\Big[\left(2x_F^3-3x_F\right)\left(1+x_F^2\right)^{1/2}+3\sinh^{-1}\left(x_F\right)\Big]\,.
\end{eqnarray}\\
\section{MFT scalar meson field}
The expression  $\phi_0$ (\ref{FSUsigma}) contains the integral
\begin{eqnarray}
	\frac{1}{\pi^2}\,\int^{k_F}_0 dk\,\frac{ k^2\,m^*}{\sqrt{{k}^2 + {m^*}^2}}\nonumber\,,
\end{eqnarray}
and as shown in Ref. \cite{walecka} this integral can be expressed as
\begin{eqnarray}
	\frac{1}{\pi^2}\,\int^{k_F}_0 dk\,\frac{ k^2\,m^*}{\sqrt{{k}^2 + {m^*}^2}} \ = \ 
	\frac{1}{\pi^2} \Biggl[k_F\left(k_F^2 + {m^*}^2\right)^{1/2}-
	{m^*}^2\,\ln\left(\frac{k_F+\left(k_F^2 + {m^*}^2\right)^{1/2}}{m^*}\right)\Biggr]\,.
\end{eqnarray}

%% file: biblio.tex
\specialhead{BIBLIOGRAPHY}

%% file: tesisJPWD.bbl
\markboth{}{}
\begin{thebibliography}{xx}
	\bibitem{PHENIX} http://www.phenix.bnl.gov/
	\bibitem{ALICE} http://aliceinfo.cern.ch/Public
	\bibitem{Links} C. J. Horowitz, Eur. Phys. J. A, {\bf 30}, 303 (2006).
	\bibitem{csg} N. K. Glendenning, {\em Compact stars}, 2\raisebox{1.5mm}{\footnotesize{nd}} edition, Springer, (2000).
	\bibitem{Lattimer07} J. M. Lattimer and M. Prakash, Phys. Rep. {\bf 442}, 1-6, 109 (2007).
	\bibitem{webertxt} F. Weber, {\em Pulsars as Astrophysical Laboratories for Nuclear and Particle Physics},
	IOP Publishing, (1999).
	\bibitem{shapiro} S. L. Shapiro and S. A. Teukolsky, {\em Black Holes, White Dwarfs, and Neutron stars},
	John Wiley \& Sons, (1983).
	\bibitem{weberSQM} F. Weber, Prog. Pat. Nucl. Phys. {\bf 54}, 193 (2005).
	\bibitem{hewish} A. Hewish, S. J. Bell, J. D. H. Pilkington, and R. A. Collins, Nature, {\bf 217}, 709 (1968).
	\bibitem{NS1} F. Pacini, Nature, {\bf 216}, 567 (1967).
	\bibitem{NS2} T. Gold, Nature, {\bf 221}, 25 (1969).
	\bibitem{SKA1} http://www.skatelescope.org/
	\bibitem{SKA} http://www.ska.ac.za/
	\bibitem{schutz} B. F. Schutz, {\em A first course in General Relativity}, Cambridge University Press, (1990).
	\bibitem{griff} D. J. Griffiths, {\em Introduction to Electrodynamics}, 
	3\raisebox{1.5mm}{\footnotesize{rd}} edition, Prentice-Hall, (1999).
	\bibitem{stephani} H. Stephani, {\em Relativity}, 
	3\raisebox{1.5mm}{\footnotesize{rd}} edition, Cambridge University Press, (2004).
	\bibitem{toeg} J. P. du Plessis, {\em Algemene Tensor Analise}, 
	class notes for Tensor Analysis course, Department of Mathematics, Applied Mathematics and Computer Science, 
	University of Stellenbosch, (unpublished).
	\bibitem{tolman} R. C. Tolman, Phys. Rev. {\bf 55}, 364 (1939).
	\bibitem{OV} J. R. Oppenheimer and G. M. Volkoff, Phys. Rev. {\bf 55}, 374 (1939).
	\bibitem{ns4u} R. R. Silbar and S. Reddy,  Am.\, J.\, Phys. {\bf 72}, 892 (2004).
	\bibitem{walecka} B. D. Serot and J. D. Walecka, Adv. Nuc. Phys. \textbf{16}, 1 (1986).
	\bibitem{recentprogress} B. D. Serot and J. D. Walecka, Int. J. Mod. Phys. E \textbf{6}, 515 (1997).
	\bibitem{machleidt} R. Machleidt, in  \textsl{Relativistic Dynamics and Quark 
	Nuclear Physics}, edited by M. B. Johnson and A. Picklesimer, John Wiley \& Sons, (1986).
	\bibitem{walecka1} J. D. Walecka, Ann. Phys. \textbf{83}, 491 (1974).
	\bibitem{waleckatext} J. D. Walecka, {\em Theoretical Nuclear and Subnuclear Physics}, Oxford University Press, 
	(1995).	
	\bibitem{A+H} I. J. R. Aitchinson and A. J. G. Hey, {\em Gauge Theories in Particle Physics},
	Adam Hilger LTD., (1982).	
	\bibitem{drell} J. D. Bjorken and S. D. Drell, {\em Relativistic Quantum Mechanics},  McGraw-Hill, (1964).
	\bibitem{capri} A. Z. Capri, {\em Relativistic Quantum Mechanics and Introduction to Quantum Field Theory},
	World Scientific, (2002).
	\bibitem{guidry} M. Guidry, {\em Gauge Field Theories},	John Wiley \& Sons, (1991).
	\bibitem{landau} R. H. Landau, {\em Quantum Mechanics II},  John Wiley \& Sons, (1989).
	\bibitem{goldstein} H. Goldstein, C. Poole and J. Safko, {\em Classical mechanics},
	 3\raisebox{1.5mm}{\footnotesize{th}} edition, Addison Wesley, (2002).
	\bibitem{jose} J. V. Jos$\acute{\mbox{e}}$ and E. J. Saletan, {\em Classical dynamics}, 
	Cambridge University Press, (1998).
	\bibitem{P+S} M. E. Peskin and D. V. Schroeder, {\em An introduction to Quantum Field Theory}, Perseus Books, (1995)
	\bibitem{M+S} F. Mandl and G. Shaw, {\em Quantum Field Theory}, Revised edition, John Wiley \& Sons, (1993).
	\bibitem{greiner} W. Greiner and J. Reinhardt, {\em Field Quantization}, Springer-Verlag, (1996).
	\bibitem{Lorenzpaper} J. D. Jackson and L. B. Okun, Rev. Mod. Phys. {\bf 73}, 663 (2001).
	\bibitem{FSU1} B. G. Todd and J. Piekarewicz, Phys. Rev. C {\bf 67}, 044317 (2003).
	\bibitem{griff2} D. J. Griffiths, {\em Introduction to Quantum Mechanics}, Prentice-Hall, (1995).
	\bibitem{B+B} J. Boguta and A. R. Bodmer, Nuc. Phys. {\bf A292}, 413 (1977).
	\bibitem{NLSH} M. M. Sharma, M. A. Nagarajan, and P. Ring, Phys. Lett. {\bf B312}, 337 (1993).
	\bibitem{TM1} Y. Sugahara and H. Toki, Nuc. Phys. {\bf A579}, 557 (1994).
	\bibitem{NL3} G. A. Lalazissis, J. K$\ddot{\mbox{o}}$nig, and P. Ring, Phys. Rev. C {\bf 55}, 540 (1997).
	\bibitem{PK1} W. Long, J. Meng, N. Van Giai, and S. Zhou, Phys. Rev. C {\bf 69}, 034319 (2004).
	\bibitem{FSU} B. G. Todd-Rutel and J. Piekarewicz, Phys. Rev. Lett. {\bf 95}, 122501 (2005).
	\bibitem{Rus} S. B. R$\ddot{\mbox{u}}$ster, M. Hempel, and J. Schaffner-Bielich, Phys. Rev. C {\bf 73},
	 035804 (2006).
	\bibitem{BPS} G. Baym, C. Pethick, and P. Sutherland,  Astrophys.\, J. {\bf 70}, 299 (1971).
	\bibitem{BBP} G. Baym, H. A. Bethe, and C. J. Pethick, Nuc.\, Phys. {\bf A175}, 225 (1971).
	\bibitem{Car} J. Carriere, C. J. Horowitz, and J. Piekarewicz,  Astrophys.\, J. {\bf 593}, 463 (2003).
	\bibitem{firestone} R. B. Firestone, {\em Table of Isotopes}, 8\raisebox{1.5mm}{\footnotesize{nd}} edition, 
	John Wiley \& Sons, (1999).
	\bibitem{muons} http://pdg.lbl.gov/
	\bibitem{Klaehn} T. Kl$\ddot{\mbox{a}}$hn, D. Blaschke, S. Typel, E. N. E. van Dalen, A. Faessler, C. Fuchs, 
	T. Gaitanos, H. Grigorian, A. Ho, E. E. Kolomeitsev, M. C. Miller, G. R$\ddot{\mbox{o}}$pke, 
	J. Tr$\ddot{\mbox{u}}$mper, D. N. Voskresensky, F. Weber, and H. H. Wolter,
	Phys. Rev. C {\bf 74}, 035802 (2006).
	\bibitem{Nice} D. J. Nice, E. M. Splaver, I. H. Stairs, O. L$\ddot{\mbox{o}}$hmer, A. Jessner, M. Kramer,
	 and J. M. Cordes, Astrophys. J. {\bf 634}, 1242 (2005).
	\bibitem{Ozel} F. $\ddot{\mbox{O}}$zel, Nature, {\bf 441}, 1115 (2006).
	\bibitem{Villarreal} A. R. Villarreal and T. E. Strohmayer, Astrophys. J. {\bf 614}, L121 (2004).
	\bibitem{FSUL} J. Piekarewicz, Phys. Rev. C {\bf 76}, 064310 (2007), arXiv:0709.2699v1 [nucl-th].
	\bibitem{PREX} http://hallaweb.jlab.org/parity/prex/
	\bibitem{mengNS} S. F. Ban, J. Li, S. Q. Zhang, H. Y. Jia, J. P. Sang, and, J. Meng, Phys. Rev. C {\bf 69}, 
	045805 (2004).
	\bibitem{piekWD} C. B. Jackson, J. Taruna, S. L. Pouliot, B. W. Ellison, D. D. Lee, and J. Piekarewicz,
	 Eur. J. Phys.,
	 {\bf 26}, 695 (2005).
	\bibitem{runge} M. Abramowitz and I. A. Stegun, {\em Handbook of Mathematical Functions}, Dover Publications, 
	(1972). 
\end{thebibliography}
